\documentclass[preprint, aps,prd,amsmath,amssymb,floatfix]{revtex4}
\usepackage{graphicx}
\usepackage{color}
\usepackage{graphicx}% Include figure files
\usepackage{dcolumn}% Align table columns on decimal point
\usepackage{bm}% bold math
\usepackage{amsmath}
\usepackage{amsfonts}
\usepackage{bbm}
\usepackage{subfigure}
\usepackage{setspace}
\usepackage{slashed}
\newcommand{\beq}{\begin{eqnarray}}
\newcommand{\eeq}{\end{eqnarray}}

\newcommand{\bmp}{\noindent\begin{minipage}{16cm}}
\newcommand{\emp}{\end{minipage}\vskip 7mm} % 7mm untightened

\def\drawbox#1#2{\hrule height#2pt
        \hbox{\vrule width#2pt height#1pt \kern#1pt
              \vrule width#2pt}
              \hrule height#2pt}

\def\Asym#1#2{\vcenter{\vbox{\drawbox{#1}{#2}
              \kern-#2pt % line up boxes
              \drawbox{#1}{#2}}}}

\def\met{{\slash\!\!\!\!\!\:E}_T}
\def\mpt{{\slash\!\!\!\!\!\:p}_T}
\def\mptv{{\slash\!\!\!\!\!\:\vec{p}}_T}

\begin{document}
%%%%%%%%%%%%%%%%%%%%%%%%%%%%%%%%%%%%%%%%%%%%%%%%%%%%%%%%%%%%%%%%%%%%%%%%%%%
\title{\Large  Technicolor Walks at the LHC }
\author{Alexander {\sc Belyaev}}
\email{a.belyaev@phys.soton.ac.uk}
\affiliation{
\mbox{NExT Institute:
School of Physics \& Astronomy} \\ \mbox{University of Southampton,
Highfield, Southampton SO17 1BJ, UK} \\
\mbox{Particle Physics Department, Rutherford Appleton Laboratory,} \mbox{Chilton,
Didcot, Oxon OX11 0QX, UK}}
\author{Roshan {\sc Foadi}}
\email{roshan@fysik.sdu.dk}
\author{Mads T. {\sc Frandsen}}
\email{toudal@nbi.dk}
\author{Matti  {\sc J\"arvinen}}
\email{mjarvine@ifk.sdu.dk}
\affiliation{High Energy Physics Center, University of Southern Denmark, Campusvej 55, DK-5230 Odense M, Denmark}
\author{Alexander {\sc Pukhov}}
\email{pukhov@lapp.in2p3.fr}
\affiliation{\mbox{Skobeltsyn Inst. of Nuclear Physics, Moscow State Univ., Moscow 119992, Russia}}
\author{Francesco {\sc Sannino}}
\email{sannino@ifk.sdu.dk}
\affiliation{High Energy Physics Center, University of Southern Denmark, Campusvej 55, DK-5230 Odense M, Denmark}
%%%%%%%%%%%%%%%%%%%%%%%%%%%%%%%%%%%%%%%%%%%%%%%%%%%%%%%%%%%%%%%%%%%%%%%%%%%%%%%%%%%%%%%%%%%%%%%%%%%%%%%%%%%%%%%%%%%%%%%%%%%%%%%%%%%%%%%%%%%%%%

\begin{abstract}

We analyze the potential of the Large Hadron Collider (LHC) to observe 
signatures of phenomenologically viable Walking Technicolor  models. 
We study and compare the Drell-Yan (DY) and
Vector Boson Fusion (VBF)  mechanisms for the production of composite
heavy vectors. We find that the heavy vectors are most easily produced and detected via the DY processes.
The composite Higgs phenomenology is also studied. If Technicolor walks at the LHC its footprints will be visible and our analysis will help uncovering them. 
\end{abstract}

%%%%%%%%%%%%%%%%%%%%%%%%%%%%%%%%%%%%%%%%%%%%%%%%%%%%%%%%%%%%%%%%%%%%%%%%
\maketitle

\section{Introduction}

Dynamical electroweak symmetry breaking (DEWSB) has a fair chance to constitute 
the correct extension of the Standard Model (SM). However, electroweak 
precision data (EWPD) and constraints from
flavor changing neutral currents (FCNC) both disfavor underlying gauge dynamics
resembling too closely a scaled-up version of Quantum Chromodynamics (QCD)
(see \cite{Sannino:2008ha,Hill:2002ap} for recent reviews).  With QCD-like dynamics ruled out, what kind of 
four dimensional gauge theory can be a realistic candidate for
DEWSB?

Based on recent progress 
\cite{Sannino:2004qp,Dietrich:2006cm,Ryttov:2007sr,Ryttov:2007cx,Dietrich:2005jn,Sannino:2008ha} in the
understanding of Walking Technicolor (WT)  dynamics \cite{Holdom:1981rm,Holdom:1984sk,Eichten:1979ah,Lane:1989ej} various phenomenologically viable models have been proposed. Primary examples are: i) the $SU(2)$ theory with two techniflavors in the adjoint representation,  known as Minimal Walking Technicolor (MWT); ii) the $SU(3)$ theory with
two flavors in the two-index symmetric representation which is called Next to Minimal
Walking Technicolor (NMWT). These gauge theories have remarkable 
properties~\cite{Sannino:2004qp,Hong:2004td,Dietrich:2005jn,Dietrich:2006cm,Ryttov:2007sr,Ryttov:2007cx} and alleviate the tension with the EWPD when used for
DEWSB~\cite{Sannino:2004qp,Dietrich:2005jn,Foadi:2007ue,Foadi:2007se}.  First principle lattice simulations already started~\cite{Catterall:2007yx,Catterall:2008qk,Shamir:2008pb,DelDebbio:2008zf,DelDebbio:2008wb} giving preliminary
support to the claim that these theories are indeed (near) conformal. The finite temperature properties of these models  have been recently studied in \cite{Cline:2008hr} in connection with the order of the electroweak phase transition.

We focus the present analysis on NMWT since this theory possesses the simplest global symmetry (SU(2)$_{\rm L}\times$SU(2)$_{\rm R}$) yielding fewer composite particles than MWT 
(with its SU(4) global symmetry)~\footnote{The
technibaryon number is assumed to be broken via new interactions beyond the
electroweak sector}. {}Following our construction in Ref.~\cite{Foadi:2007ue} we provide a
comprehensive Lagrangian for this model. Key ingredients are (i) the global
symmetries of the underlying gauge theory, (ii) vector meson dominance,
(iii) walking dynamics, and (iv) the ``minimality'' of the theory, 
that is the small number of flavors and thus a small $S$ parameter. Based
on (i) and (ii) we use for the low-energy physics a chiral 
resonance model containing spin zero and spin one fields. Some of the coefficients 
of the corresponding Lagrangian are then constrained using (iii) and (iv) 
through the modified Weinberg's sum rules (WSR's)~\cite{Appelquist:1998xf}. 
Given that we cannot compute the entire set of the coefficients of the 
effective Lagrangian directly from the underlying gauge theory we use the 
practical approach of studying the various LHC observables for different 
values of the unknown parameters. In this respect our low-energy theory 
can also be seen as a template for any strongly coupled theory which may 
emerge at the LHC. An analysis of unitarity of the longitudinal WW 
scattering versus precision measurements, within the effective Lagrangian 
approach, can be found in Ref.~\cite{Foadi:2008ci}, and shows that it 
is possible to pass the precision tests while simultaneously delay the 
onset of unitarity violation.

Clean signatures of the NMWT model come from the production of spin one resonances and the composite Higgs, followed by their decays to SM fields. In particular in this work we focus on Drell-Yan (DY) and vector boson fusion (VBF) 
production of the vector resonances. We also study the 
associate Higgs production together with a $W$ or a $Z$ boson. This channel is interesting due to the interplay among the SM gauge bosons, the heavy vectors and the composite Higgs. 

In Section II we introduce the model and impose constraints on its parameter space from LEP and Tevatron. We also use information from the underlying gauge dynamics in the form of the generalized WSRs. The LHC phenomenology is studied in Section III. More specifically we investigate the heavy vector production as well as the associate composite Higgs production. We summarize our results in Section IV. 

\section{The Simplest Model of Walking Technicolor}
We have explained that NMWT has the simplest chiral symmetry, $SU(2)_{\rm L} \times SU(2)_{\rm R} $since it is expected to be near walking with just two Dirac flavors.  The low energy
description of this model can be encoded in a chiral Lagrangian including spin one resonances. {}Following  Ref.~\cite{Foadi:2007ue} and \cite{Appelquist:1999dq} we write:
\begin{eqnarray}
{\cal L}_{\rm boson}&=&-\frac{1}{2}{\rm Tr}\left[\widetilde{W}_{\mu\nu}\widetilde{W}^{\mu\nu}\right]
-\frac{1}{4}\widetilde{B}_{\mu\nu}\widetilde{B}^{\mu\nu}
-\frac{1}{2}{\rm Tr}\left[F_{{\rm L}\mu\nu} F_{\rm L}^{\mu\nu}+F_{{\rm R}\mu\nu} F_{\rm R}^{\mu\nu}\right] \nonumber \\
&+& m^2\ {\rm Tr}\left[C_{{\rm L}\mu}^2+C_{{\rm R}\mu}^2\right]
+\frac{1}{2}{\rm Tr}\left[D_\mu M D^\mu M^\dagger\right]
-\tilde{g^2}\ r_2\ {\rm Tr}\left[C_{{\rm L}\mu} M C_{\rm R}^\mu M^\dagger\right] \nonumber \\
&-&\frac{i\ \tilde{g}\ r_3}{4}{\rm Tr}\left[C_{{\rm L}\mu}\left(M D^\mu M^\dagger-D^\mu M M^\dagger\right)
+ C_{{\rm R}\mu}\left(M^\dagger D^\mu M-D^\mu M^\dagger M\right) \right] \nonumber \\
&+&\frac{\tilde{g}^2 s}{4} {\rm Tr}\left[C_{{\rm L}\mu}^2+C_{{\rm R}\mu}^2\right] {\rm Tr}\left[M M^\dagger\right]
+\frac{\mu^2}{2} {\rm Tr}\left[M M^\dagger\right]-\frac{\lambda}{4}{\rm Tr}\left[M M^\dagger\right]^2
\label{eq:boson}
\end{eqnarray}
where $\widetilde{W}_{\mu\nu}$ and $\widetilde{B}_{\mu\nu}$ are the ordinary electroweak field strength tensors, $F_{{\rm L/R}\mu\nu}$ are the field strength tensors associated to the vector meson fields $A_{\rm L/R\mu}$~\footnote{In Ref.~\cite{Foadi:2007ue}, where the chiral symmetry is SU(4), there is an additional term whose coefficient is labeled $r_1$. With an SU($N$)$\times$SU($N$) chiral symmetry this term is just identical to the $s$ term.}, and the $C_{{\rm L}\mu}$ and $C_{{\rm R}\mu}$ fields are
\begin{eqnarray}
C_{{\rm L}\mu}\equiv A_{{\rm L}\mu}-\frac{g}{\tilde{g}}\widetilde{W_\mu}\ , \quad
C_{{\rm R}\mu}\equiv A_{{\rm R}\mu}-\frac{g^\prime}{\tilde{g}}\widetilde{B_\mu}\ .
\end{eqnarray}
The 2$\times$2 matrix $M$ is
\begin{eqnarray}
M=\frac{1}{\sqrt{2}}\left[v+H+2\ i\ \pi^a\ T^a\right]\ ,\quad\quad  a=1,2,3
\end{eqnarray}
where $\pi^a$ are the Goldstone bosons produced in the chiral symmetry breaking, $v=\mu/\sqrt{\lambda}$ is the corresponding VEV, $H$ is the composite Higgs, and $T^a=\sigma^a/2$, where $\sigma^a$ are the Pauli matrices. The covariant derivative is
\begin{eqnarray}
D_\mu M&=&\partial_\mu M -i\ g\ \widetilde{W}_\mu^a\ T^a M + i\ g^\prime \ M\ \widetilde{B}_\mu\ T^3\ . 
\end{eqnarray}
When $M$ acquires its VEV, the Lagrangian of Eq.~(\ref{eq:boson}) contains mixing matrices for the spin one fields. The mass eigenstates are the ordinary SM bosons, and two triplets of heavy mesons, of which the lighter (heavier) ones are denoted by $R_1^\pm$ ($R_2^\pm$) and $R_1^0$ ($R_2^0$). These heavy mesons are the only new particles, at low energy, relative to the SM.

Some remarks should be made about the Lagrangian of Eq.~(\ref{eq:boson}).
First, the new strong interaction preserves parity, which implies
invariance under the transformation
\begin{eqnarray}
\quad M \leftrightarrow M^\dagger \ , \quad C_{\rm L} \leftrightarrow C_{\rm R} \ .
\end{eqnarray}
Second, we have written the Lagrangian in a ``mixed'' gauge. As explained
in the appendix of Ref.~\cite{Foadi:2007ue}, the Lagrangian for this model
can be rewritten by interpreting the vector meson fields as gauge fields
of a ``mirror'' gauge group SU(2)$^\prime_{\rm L}\times$SU(2)$^\prime_{\rm
R}$. This is equivalent to the idea of Hidden Local
Symmetry~\cite{Bando:1984ej,Bando:1987br}, used in a similar context for
the BESS models~\cite{Casalbuoni:1995qt}. In Eq.~(\ref{eq:boson}) the
vector mesons have already absorbed the corresponding pions, while the
SU(2)$\times$U(1) {gauge} fields are still transverse. Finally,
Eq.~(\ref{eq:boson}) contains all ${\cal O}(p^2)$ operators of dimension
two and four.

Now we must couple the SM fermions. The interactions with the Higgs and the spin one mesons are mediated by an unknown ETC sector, and can be parametrized at low energy by Yukawa terms, and mixing terms with the $C_{\rm L}$ and $C_{\rm R}$ fields. Assuming that the ETC interactions preserve parity, the most general form for the quark Lagrangian is~\footnote{The lepton sector works out in a similar way, the
only difference being the possible presence of Majorana neutrinos.}
\begin{eqnarray}
{\cal L}_{\rm quark}&=&\bar{q}^i_L\ i \slashed{D} q_{iL} + \bar{q}^i_R\ i \slashed{D} q_{iR} \nonumber \\
&+&\tilde{g}\  \bar{q}^i_L\ K_i^j\  \slashed{C}_{\rm L} \ q_{jL}
+\tilde{g}\  \bar{q}^i_R\ K_i^j\ \slashed{C}_{\rm R}\  q_{jR} \nonumber \\
&-&\left[\bar{q}^i_L\ (Y_u)_i^j\ M\ \frac{1+\tau^3}{2}\ q_{jR}
+\bar{q}^i_L\ (Y_d)_i^j\ M \ \frac{1-\tau^3}{2}\ q_{jR} + {\rm h.c.}\right] \ ,
\label{eq:quark}
\end{eqnarray}
where $i$ and $j$ are generation indices, $i=1,2,3$, $q_{iL/R}$ are electroweak doublets, $K$ is a 3$\times$3
Hermitian matrix, $Y_u$ and $Y_d$ are 3$\times$3 complex matrices. The covariant derivatives are the ordinary
electroweak ones,
\begin{eqnarray}
\slashed{D}q_{iL}&=&\left(\slashed{\partial}-i\ g\ \slashed{\widetilde{W}}^a\ T^a
-i\ g^\prime \slashed{\widetilde{B}} Y_{\rm L}\right)q_{iL} \ ,\nonumber \\
\slashed{D}q_{iR}&=&\left(\slashed{\partial}-i\ g^\prime \slashed{\widetilde{B}} Y_{\rm R}\right)q_{iR} \ ,
\end{eqnarray}
where $Y_{\rm L}=1/6$ and $Y_{\rm R}={\rm diag}(2/3,-1/3)$. One can exploit the global symmetries of the kinetic and $K$-terms to reduce the number of physical parameters in the Yukawa matrices. Thus we can take
\begin{eqnarray}
Y_u={\rm diag}(y_u,y_c,y_t) \ , \quad Y_d= V\ {\rm diag}(y_d,y_s,y_b) \ ,
\end{eqnarray}
and
\begin{eqnarray}
q^i_L=\left(\begin{array}{c} u_{iL} \\ V_i^j d_{jL} \end{array}\right) \ , \quad
q^i_R=\left(\begin{array}{c} u_{iR} \\ d_{iR} \end{array}\right) \ ,
\end{eqnarray}
where $V$ is the CKM matrix. In principle one could also have a mixing matrix for the right-handed fields, due to the presence of the $K$-terms. However at this point this is an unnecessary complication, and we set this mixing matrix equal to the identity matrix. Finally, we also set
\begin{eqnarray}
K=\kappa \ {\bf 1}_{3\times 3} \ ,
\label{eq:kappa}
\end{eqnarray} 
to prevent flavor changing neutral currents (FCNC) to show up at tree-level. A more precise approach would require taking the experimental bounds on FCNC and using these to constrain $K-\kappa \ {\bf 1}_{3\times 3}$.

\subsection{Weinberg Sum Rules}
In its general form Eq.~(\ref{eq:boson}) describes any model of DEWSB with a spontaneously broken SU(2)$_{\rm L}\times$SU(2)$_{\rm R}$ chiral symmetry. In order to
make contact with the underlying gauge theory, and discriminate between different classes of models, we make
use of the WSRs. In Ref.~\cite{Appelquist:1998xf} it was argued that the zeroth WSR -- which is nothing but
the definition of the $S$ parameter --
\begin{equation}
S=4\pi\left[\frac{F_V^2}{M_V^2}-\frac{F_A^2}{M_A^2}\right] \ , \label{eq:WSR0} 
\end{equation}
 and the first WSR,  
 \begin{equation}
 F_V^2 - F_A^2 = F_\pi^2 \ , \label{eq:WSR1} 
\end{equation}
 do not receive significant contributions from the
near conformal region, and are therefore unaffected. 
In these equations $M_V$ ($M_A$) and $F_V$ ($F_A$) are mass and decay constant of the vector-vector (axial-vector) meson, respectively, in the limit of zero electroweak gauge couplings. $F_\pi$ is the decay constant of the pions: since this is a model of DEWSB, $F_\pi=246$ GeV. The Lagrangian of Eq.~(\ref{eq:boson}) gives
\begin{eqnarray}
M_V^2 &=& m^2 + \frac{\tilde{g}^2\ (s-r_2)\ v^2}{4} \nonumber \\
M_A^2 &=& m^2 + \frac{\tilde{g}^2\ (s+r_2)\ v^2}{4} \ ,
\label{eq:masses}
\end{eqnarray}
and
\begin{eqnarray}
F_V & = & \frac{\sqrt{2}M_V}{\tilde{g}} \ ,  \nonumber \\
F_A & = & \frac{\sqrt{2}M_A}{\tilde{g}}\chi \ , \nonumber \\
F_\pi^2 & = & \left(1+2\omega\right)F_V^2-F_A^2 \ ,
\label{eq:FVFAFP}
\end{eqnarray}
where
\begin{eqnarray}
\omega \equiv \frac{v^2 \tilde{g}^2}{4 M_V^2}(1+r_2-r_3) \ , \quad \quad 
\chi \equiv 1-\frac{v^2\ \tilde{g}^2\ r_3}{4 M_A^2} \ . \label{eq:chi}
\end{eqnarray}
Then Eqs.~(\ref{eq:WSR0}) and (\ref{eq:WSR1}) give
\begin{eqnarray}
\label{eq:s_of_chi}
& & S=\frac{8\pi}{\tilde{g}^2}\left(1-\chi^2\right) \ , \label{eq:S} \\
& & r_2 = r_3-1 \ .
\end{eqnarray}
The second WSR does receive important contributions from the near conformal region, and is modified to
\begin{eqnarray}
F_V^2 M_V^2 - F_A^2 M_A^2 = a \frac{8\pi^2}{d(R)} F_\pi^4 \ , \label{eq:WSR2}
\end{eqnarray}
where $a$ is expected to be positive and ${\cal O}(1)$, and $d(R)$ is the dimension of the representation of
the underlying fermions~\cite{Appelquist:1998xf}. For each of these sum rules a more general spectrum would involve a sum over vector
and axial states. 

In the effective Lagrangian we codify the walking behavior in $a$ being positive and ${\cal O}(1)$,
and the minimality of the theory in $S$ being small. A small $S$ is both due to the small number of flavors in
the underlying theory and to the near conformal dynamics, which reduces the contribution to $S$ relative to a
running theory~\cite{Appelquist:1998xf,Sundrum:1991rf,Kurachi:2006mu}. In NMWT (three colors in the two-index 
symmetric representation) the {naive} one-loop  $S$ parameter is $S=1/\pi\simeq 0.3$: this is a reasonable input for 
$S$ in Eq.~(\ref{eq:WSR0}).

\begin{figure}
{\includegraphics[height=8cm,width=8cm]{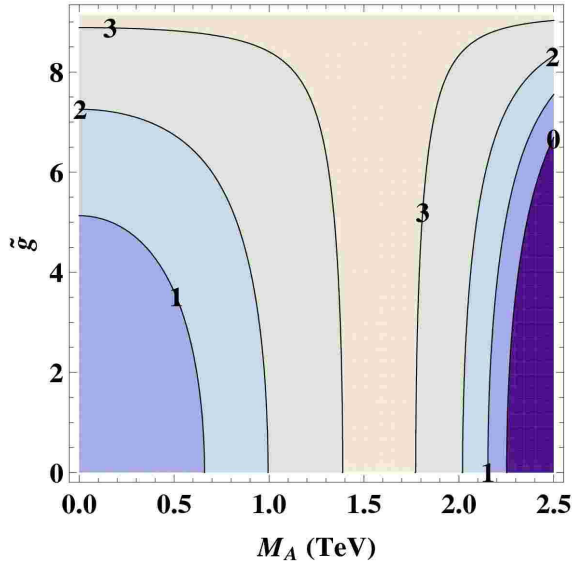}}
\caption{Contour plot for $a$ in the $(M_A,\tilde{g})$ plane, for $S=0.3$ in NMWT ($d({\rm R})=6$). We plot contours for $a=0,1,2,3$, and $3\leq a\leq a_{\rm max}=d({\rm R})/(2\pi S)\simeq 3.18$ (central region). The running regime corresponds to the $a=0$ contour, which is on the lower right of the parameter space. Walking dynamics requires $a={\cal O}(1)>0$, which is achieved for a large portion of the parameter space.}\label{fig:a}
\end{figure}
Fig.~\ref{fig:a} displays a contour plot of $a$ in the $(M_A,\tilde{g})$ plane for $S=0.3$, in NMWT ($d({\rm R})=6$). This plot is obtained after imposing Eqs.~(\ref{eq:WSR0}) and (\ref{eq:WSR1}). Notice that $a={\cal O}(1)$ for a large portion of the parameter space, since the maximum value of $a$ is found to be $a_{\rm max}=d({\rm R})/(2\pi S)$, and this gives 3.18 for $S=0.3$ and $d({\rm R})=6$. The running regime, $a=0$, is only attained for large values of $M_A$. However walking regimes, $a={\cal O}(1)$, are also compatible with smaller values of $M_A$. For example, if we require $1<a<2$ with $S=0.3$ in NMWT, from Fig~(\ref{fig:a}) we see that this is both possible for $M_A\gtrsim$ 2.0 TeV and $M_A\lesssim$ 1.0 TeV. Although a walking regime with large values of $M_A$ is more plausible, since this is more naturally achieved by moving away from a running regime, a walking scenario with small values of $M_A$ cannot be excluded based solely on the WSRs analysis.

\subsection{Electroweak Parameters}\label{sec:EWP}
If the $\kappa$ parameter of Eq.~(\ref{eq:kappa}) is negligibly small, then the fermion Lagrangian of Eq.~(\ref{eq:quark}) describes a ``universal'' theory, in the sense that all the corrections to the electroweak observables show up in gauge current correlators. If this is the case the new physics effect on the low-energy observables are fully accounted for by the Barbieri {\em et. al.} parameters~\cite{Barbieri:2004qk}. In our model these are
\begin{eqnarray}
\hat{S}&=&\frac{g^2(1-\chi^2)}{2\tilde{g}^2+g^2(1+\chi^2)}  \ ,\nonumber \\
\hat{T}&=&0 \ ,\nonumber \\
W&=&M_W^2\frac{g^2\left(M_A^2+M_V^2\ \chi^2\right)}
{\left(2\tilde{g}^2+g^2(1+\chi^2)\right)M_A^2 M_V^2} \ , \nonumber \\
Y&=&M_W^2\frac{{g^\prime}^2\left(M_A^2+ M_V^2\ \chi^2\right)}
{\left(2\tilde{g}^2+{g^\prime}^2(1+\chi^2)\right) M_A^2 M_V^2} \ , \nonumber \\
\hat{U}&=&0 \ , \nonumber \\
V&=&0 \ , \nonumber \\
X&=&g g^\prime \frac{M_W^2}{M_A^2 M_V^2}
\frac{M_A^2-M_V^2\ \chi^2}
{\sqrt{\left(2\tilde{g}^2+g^2(1+\chi^2)\right)
\left(2\tilde{g}^2+{g^\prime}^2(1+\chi^2)\right)}} \ .
\end{eqnarray}

It is important to notice that these are the electroweak parameters from the
pure technicolor sector only. Important negative contributions to $\hat{S}$ (or
$S$) and positive contributions to $\hat{T}$ (or $T$) can arise from a mass
splitting between the techniup and the technidown fermions (which can arise from
the ETC sector) or from new nondegenerate lepton doublets, with either Majorana
or Dirac neutrinos. A new lepton doublet is actually required in MWT, where it
is introduced to cure the SU(2) Witten anomaly, and suffices to bring $S$ and
$T$ to within 1$\sigma$ of the experimental expectation value \cite{Foadi:2007ue}. Without these
extra contributions, if the underlying gauge theory is NMWT (with three colors
in the two-index symmetric representation), the {naive} one-loop contribution to
$S$ is $S=1/\pi\simeq 0.3$. Taking this as the true value of $S$, the prediction
for $\hat{S}$ is almost everywhere in the parameter space within 2$\sigma$ for a
light Higgs and 3$\sigma$ for a heavy Higgs. If NMWT is very close to the
conformal window $S$ can even be smaller: this scenario was considered in
Ref.~\cite{Foadi:2007se}, where $\hat{S}$ was taken as a viable input (within
1$\sigma$), and the other electroweak parameters, like $Y$ and $W$, were shown
to impose lower bounds on $M_A$ and $\tilde{g}$.

\subsection{Parameter Space of NMWT}
In our analysis we will take zero $\kappa$, since it affects the tree-level anomalous couplings highly constrained by experiments.  We take $S=0.3$ corresponding to its naive value in NMWT.

The  remaining parameters
are $M_A,\ \tilde{g},\ s$ and $M_H$, with $s$ and $M_H$ having a sizable effect in processes involving the composite Higgs \footnote{{
The information on the spectrum
alone is not sufficient to constrain $s$, but it can be measured studying
other physical processes.}}.

\begin{figure}
\includegraphics[height=8cm,width=10cm]{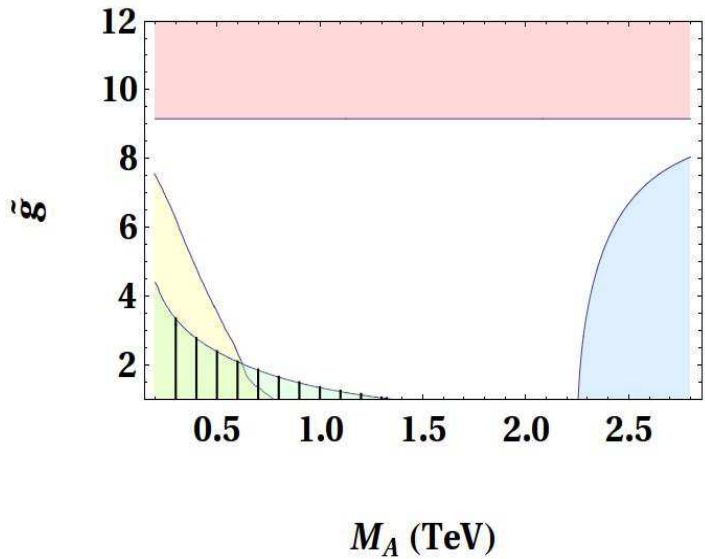}
\caption{Bounds, for $S=0.3$, in the $(M_A,\tilde{g})$ plane from: (i) CDF
direct searches of $R_1^0$ at Tevatron, in $p\bar{p}\rightarrow e^+e^-$, for
$s=1$ and  $M_{H}=200$~GeV.
The forbidden regions is the uniformly shaded one in the left corner.
The parameters $M_H$ and $s$ affect indirectly the Tevatron bounds by
changing the BR
of the $Z$ boson decay to two composite Higgsses.  However, we have checked that
the effects on the constraints coming from varying the parameter $s$ and
$M_H$ are small. 
(ii) Measurement of the electroweak parameters W and Y at
95\% confidence level. The forbidden region is the striped one in
the left corner. (iii) The constraint $a>0$, where $a$
is defined in Eq.~(\ref{eq:WSR2}). The corresponding  limiting curve
is given by Eq.~(\ref{eq:MAbound}). The forbidden region is
the shaded one in the right corner. (iv) Consistency of the theory: no
imaginary numbers for physical quantities like $F_V$ and $F_A$ . The
forbidden region is the horizontal stripe in the upper  part
of the figure. The limiting curve here is given by
Eq.~(\ref{eq:gtbounds}). 
%The bounds from {(ii), (iii), and  (iv)} do not depend on $s$. 
We repeat that %, in each plot, 
the shaded regions are excluded.}
\label{fig:bounds}
\end{figure}

CDF imposes lower bounds on $M_A$ and $\tilde{g}$ from direct searches of
$R_1^0$ in  the $p\bar{p}\rightarrow e^+e^-$ process, as shown by {the uniformly shaded region} 
in the lower left of Fig~\ref{fig:bounds}. {To present this bound we have applied the
CDF public results of Ref.~\cite{CDF-limit}
to our model.} Additional lower bounds on $M_A$ and
$\tilde{g}$ come from the electroweak parameters $W$ and $Y$, as explained
in Ref.~\cite{Foadi:2007se}.
The measurements of $W$ and $Y$ exclude the striped region 
on the lower left in Fig~\ref{fig:bounds} at 95\% confidence level 
(which corresponds roughly to the $2\sigma$ limit of a one-dimensional distribution).
%As explained in Sec.~\ref{sec:EWP} {the $2\sigma$ level} is
%motivated by the fact that new leptons or a techniup-technidown mass
%splitting can give important contributions to the electroweak parameters,
%and we need not assume the Technicolor contribution alone to be within
%1$\sigma$. Here for simplicity, and in order to focus exclusively on the
%Technicolor sector, we exclude the presence of new sectors, and accept a
%2$\sigma$ agreement as viable.

The upper bound for $\tilde{g}$,
\begin{eqnarray}
\tilde{g} < \sqrt{\frac{8\pi}{S}} \ ,
\label{eq:gtbounds}
\end{eqnarray}
is dictated by the internal consistency of the model. For $S=0.3$ this gives
$\tilde{g}\lesssim 9.15$, and is shown by the upper horizontal line in
Fig~\ref{fig:bounds}. The upper bound for $M_A$ corresponds to the value for
which both WSR's are satisfied in a running regime, and above which $a$ in
Eq.~(\ref{eq:WSR2}) becomes negative:
\begin{eqnarray}
M_A^2 < \frac{4\pi F_\pi^2}{S}\left(1+\frac{1}{\sqrt{1-\frac{\tilde{g}^2
S}{8\pi}}}\right) \ .
\label{eq:MAbound}
\end{eqnarray}
This is shown by the lower right curve in Fig~\ref{fig:bounds}.

\section{Phenomenology}\label{sec:pheno}
We  use
the CalcHEP package~\cite{Pukhov:2004ca} since it is a convenient tool to investigate collider phenomenology. The LanHEP package~\cite{Semenov:2008jy} has been used to derive the Feynman rules for the model.  

We tested the CalcHEP model implementation  in different ways. We have
implemented the model in both unitary and t'Hooft-Feynman gauge, and checked the gauge
invariance of the physical output. We investigated the Custodial Technicolor
(CT) limit~\cite{Foadi:2007se} of the model, corresponding to $r_2=r_3=0$, for
which $S=0$ and $M_A=M_V$. If we further require $s=0$ this model is then
identical to the degenerate BESS model (D-BESS)~\cite{Casalbuoni:1995qt} for
which results are available in the literature~\cite{Casalbuoni:2000gn}. We find agreement with the latter for the widths and BR's. 

New physics signals are expected from the vector meson and the
composite Higgs sectors. Here we focus on the production at LHC of the vector
mesons through DY and VBF channels, as well as the production of the composite Higgs
in association
with a weak gauge boson. We compare our results with the ones for Higgsless models \cite{Birkedal:2005yg,He:2007ge} and on the  associate Higgs production  with the analysis done by Zerwekh \cite{Zerwekh:2005wh}.

\subsection{Heavy Vectors: Masses, Decay Widths and Branching Ratios}\label{sec:decay}

\begin{figure}
{
\includegraphics[width=0.45\textwidth]{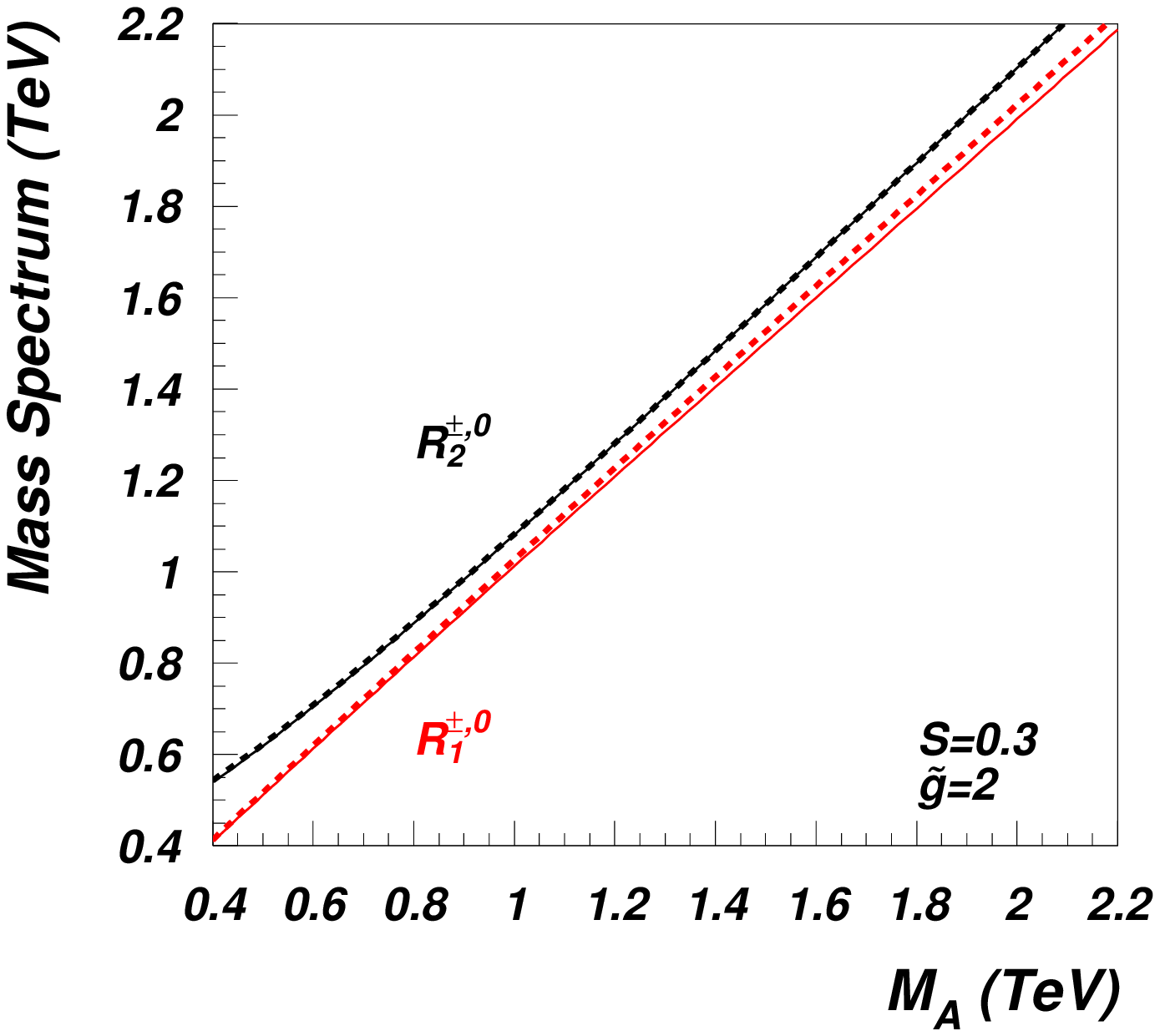}
\includegraphics[width=0.45\textwidth]{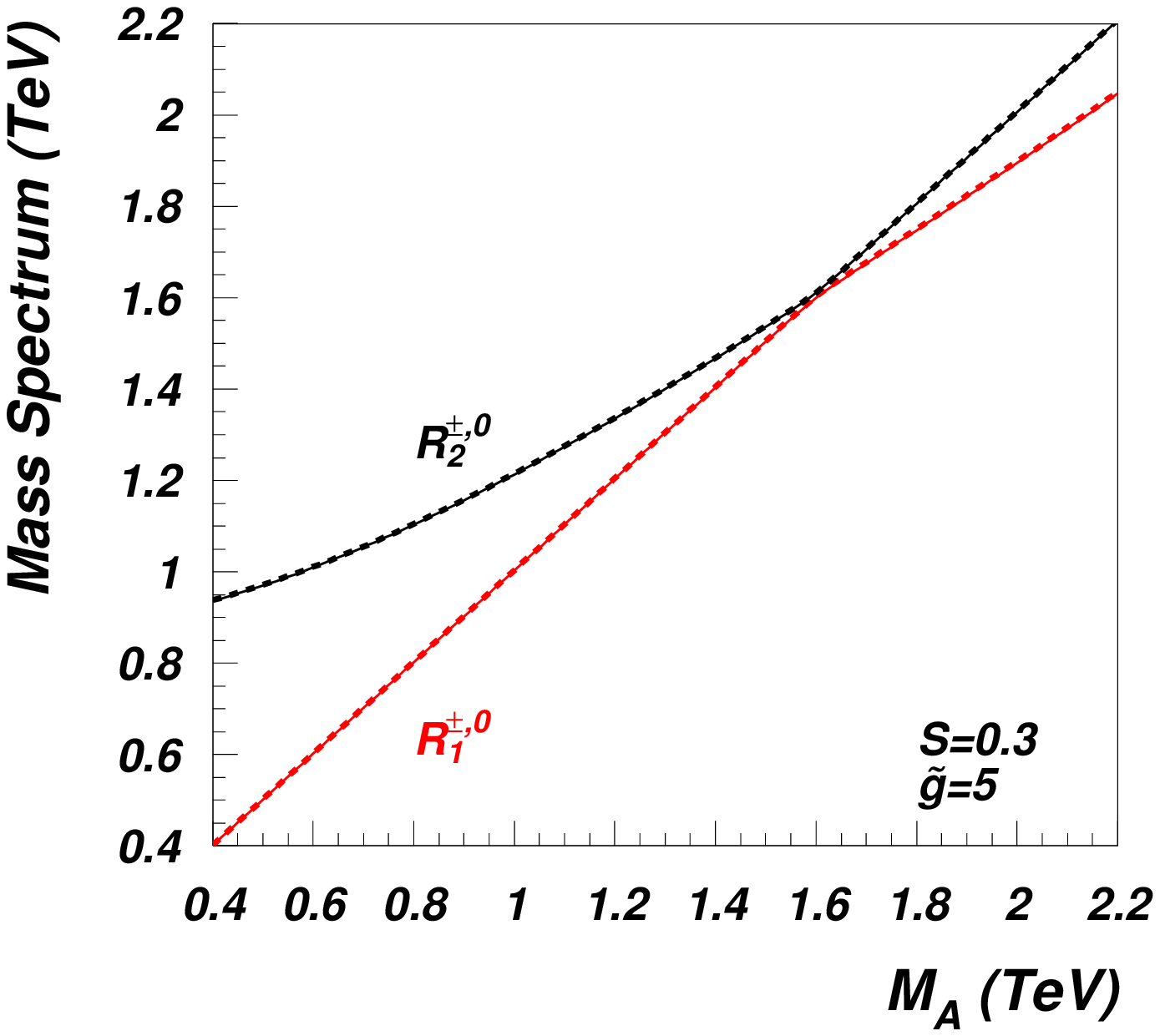}%
} 
 \caption{The mass spectrum of  the $M_{R^{\pm,0}_{1,2}}$
        vector mesons versus $M_A$
	for $\tilde{g}=2$ (left)
	and $\tilde{g}=5$ (right).
	The masses of the charged vector mesons are denoted by solid lines,
	while the masses of the neutral mesons are denoted by dashed lines.
	\label{fig:spectrum}}
\end{figure}

One important consequence of the failure of the second WSR \cite{Appelquist:1998xf,Foadi:2007ue} is the
possible mass spectrum inversion of the vector and axial spin one  mesons. \begin{figure}
{\includegraphics[height=6.5cm,width=7.82cm]{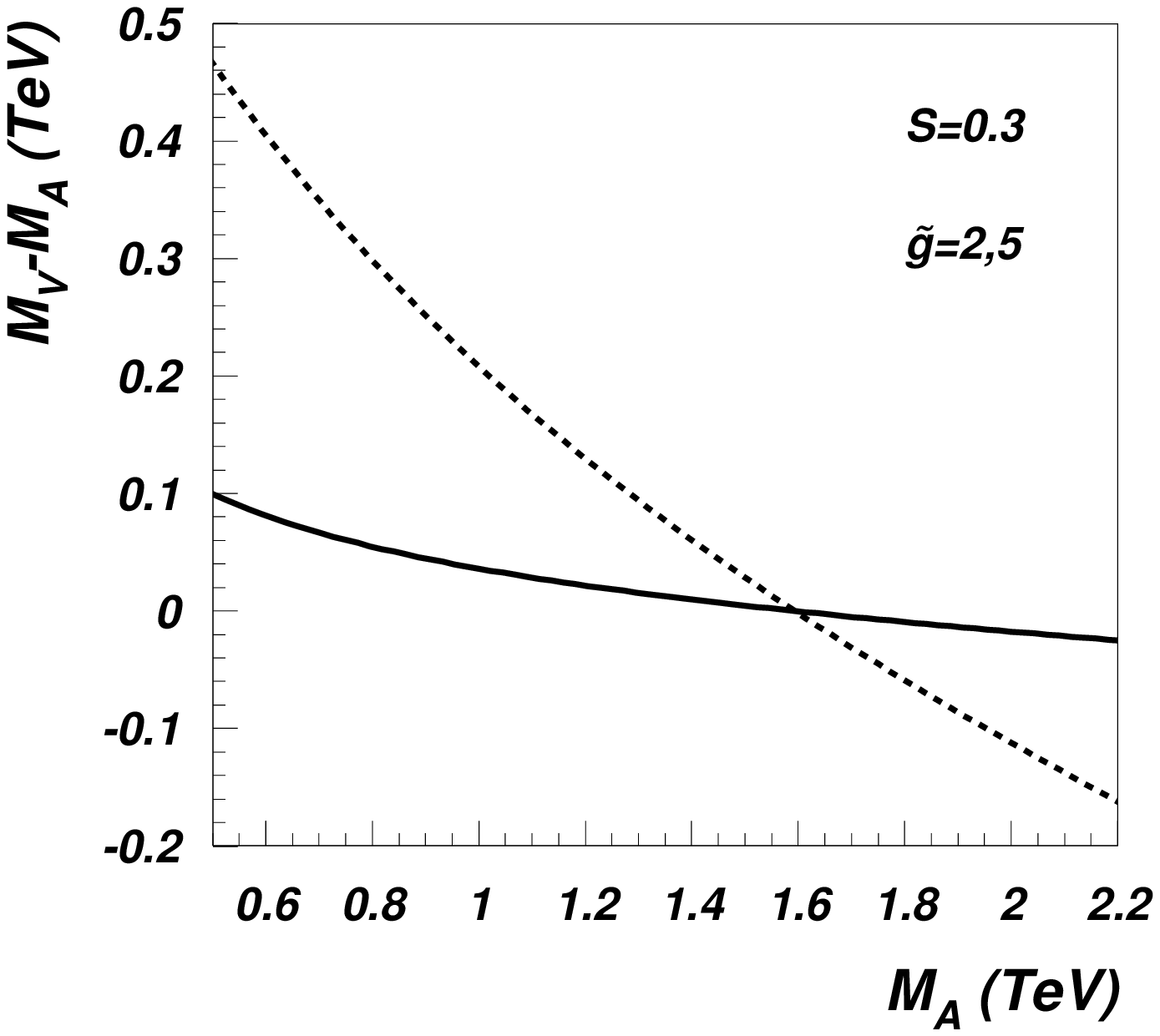}
 \includegraphics[height=6.5cm,width=7.82cm]{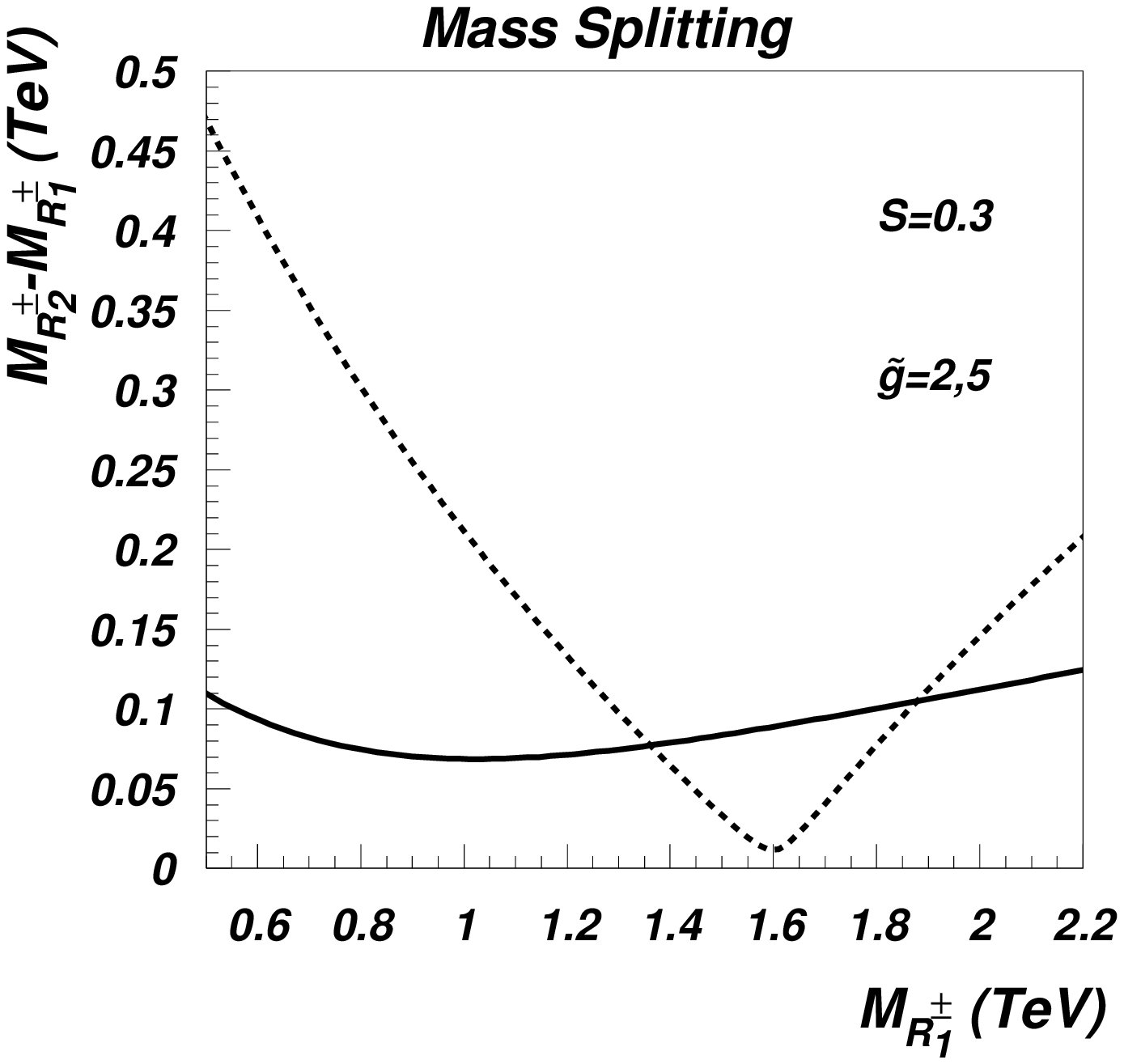}}
\caption{Mass splittings $M_V-M_A$ (left) and $M_{R_2^\pm}-M_{R_1^\pm}$ (right). The dotted lines are for $\tilde{g}=5$ while the solid lines are for $\tilde{g}=2$.}\label{fig:masses}
\end{figure}
In
Fig.~\ref{fig:masses} (left) we plot $M_V-M_A$ as a function
of $M_A$ for two reference values of $\tilde{g}$ and $S=0.3$. For generic values of $S$ the
inversion occurs for
\begin{eqnarray}
{M}^{\rm inv} = \sqrt{\frac{4\pi}{S}}F_\pi \ .
\label{eq:inv}
\end{eqnarray}
This gives ${M}^{\rm inv}\simeq 1.6$ TeV for $S=0.3$, as clearly shown in the plot. 
Fig.~\ref{fig:masses}  (right) shows $M_{R_2^\pm}-M_{R_1^\pm}$ as a function of $M_{R_1^\pm}$, where $R_1^{\pm,0}$ ($R_2^{\pm,0}$) are the lighter (heavier) vector resonances, with tree-level electroweak corrections included. This mass difference is always positive by definition, and the mass inversion becomes a kink in the plot. Away from ${M}^{\rm inv}$ $R_1$ ($R_2$) is 
an axial (vector) meson for $M_A<{M}^{\rm inv}$, and a vector (axial) meson for $M_A>{M}^{\rm inv}$. 
{The mass difference in Fig.~\ref{fig:masses} is proportional to $\tilde g^2$, and becomes relatively small for $\tilde g=2$. The effects of the electroweak corrections are larger for 
small $\widetilde{g}$ couplings. {}For example, the minimum of $M_{R_2^\pm}-M_{R_1^\pm}$ is shifted from  ${M}^{\rm inv}\simeq 1.59$~TeV to about 1~TeV for $\tilde g=2$. To help the reader we plot in Fig.~\ref{fig:spectrum}
 the actual spectrum for the vector boson masses
 versus $M_A$.  } Preliminary studies on the lattice of MWT and the mass inversion issue appeared in Ref.~\cite{DelDebbio:2008zf}.
\begin{figure}[tbhp]
 \vskip -0.4cm
 \includegraphics[width=0.45\textwidth,height=0.35\textwidth]{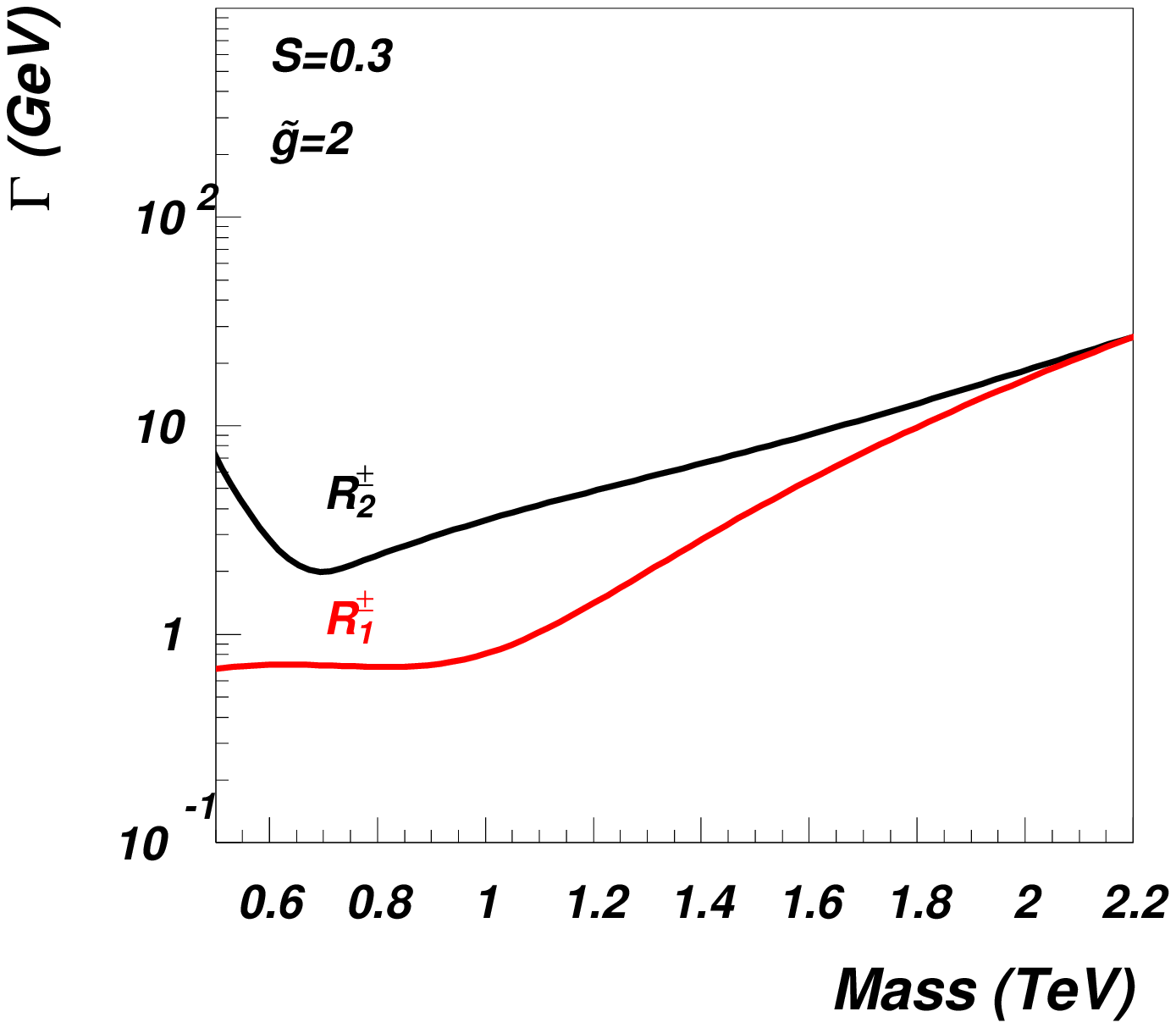}%
 \includegraphics[width=0.45\textwidth,height=0.35\textwidth]{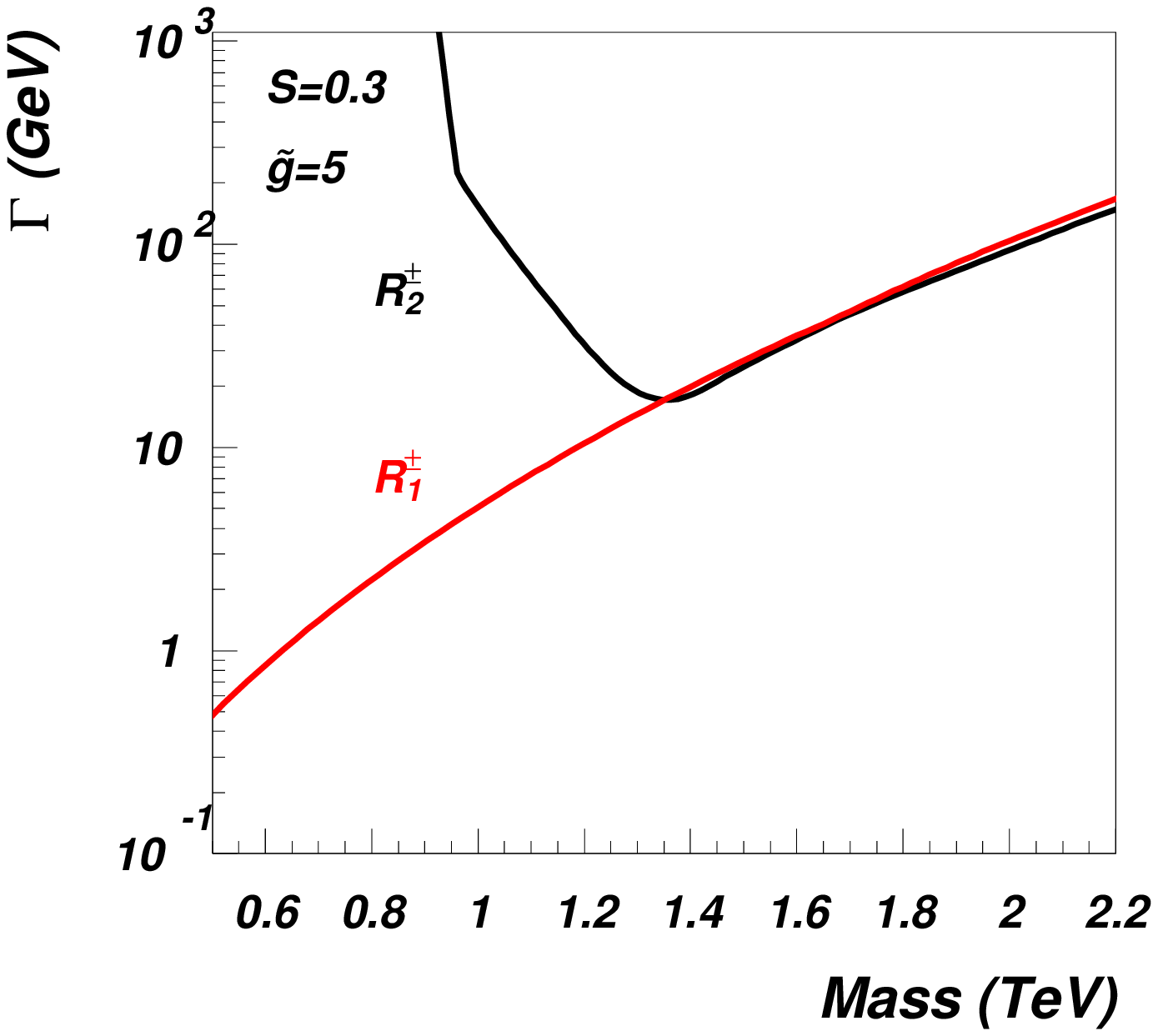}
 \vskip -0.4cm
 \includegraphics[width=0.45\textwidth,height=0.35\textwidth]{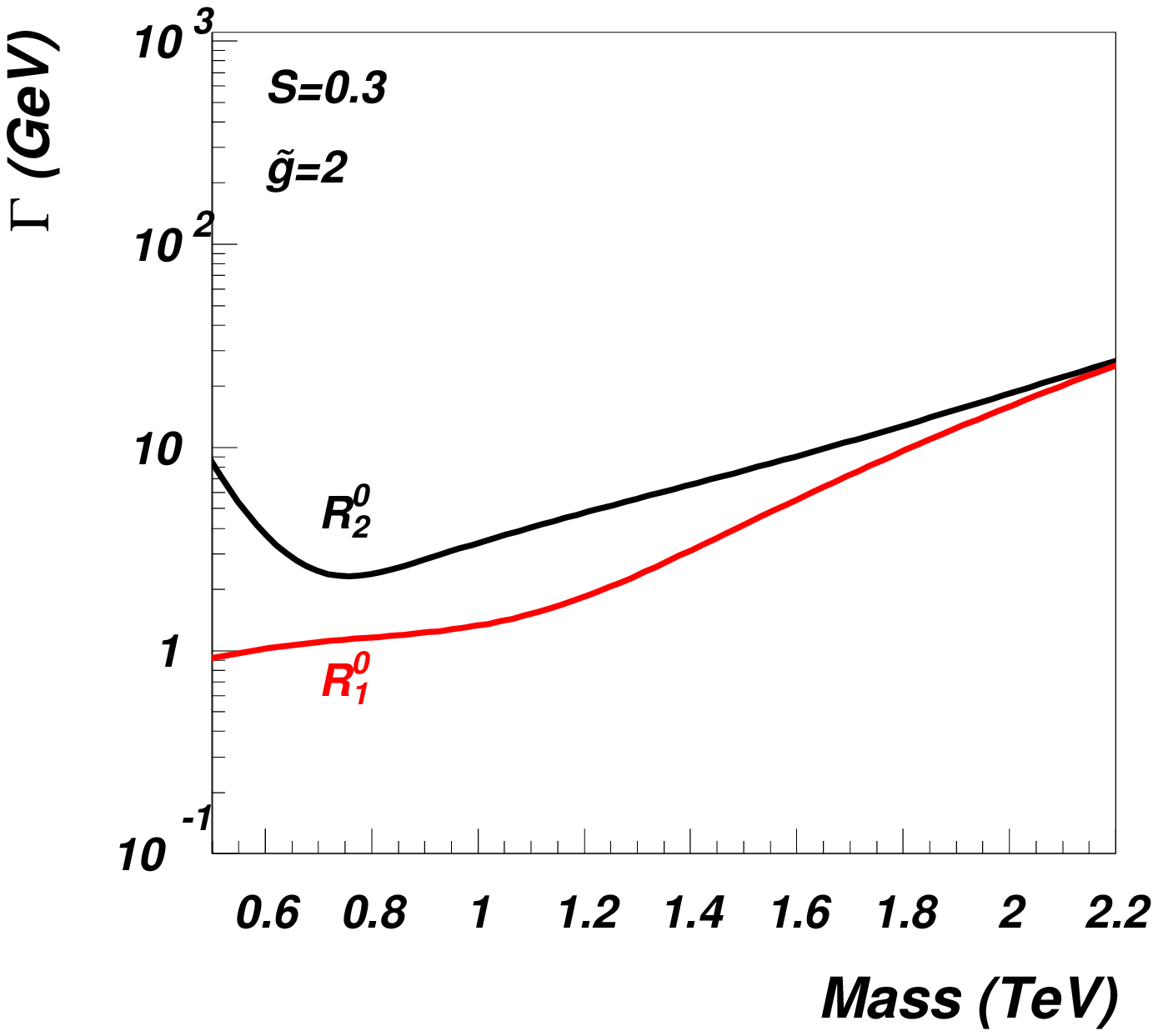}%
 \includegraphics[width=0.45\textwidth,height=0.35\textwidth]{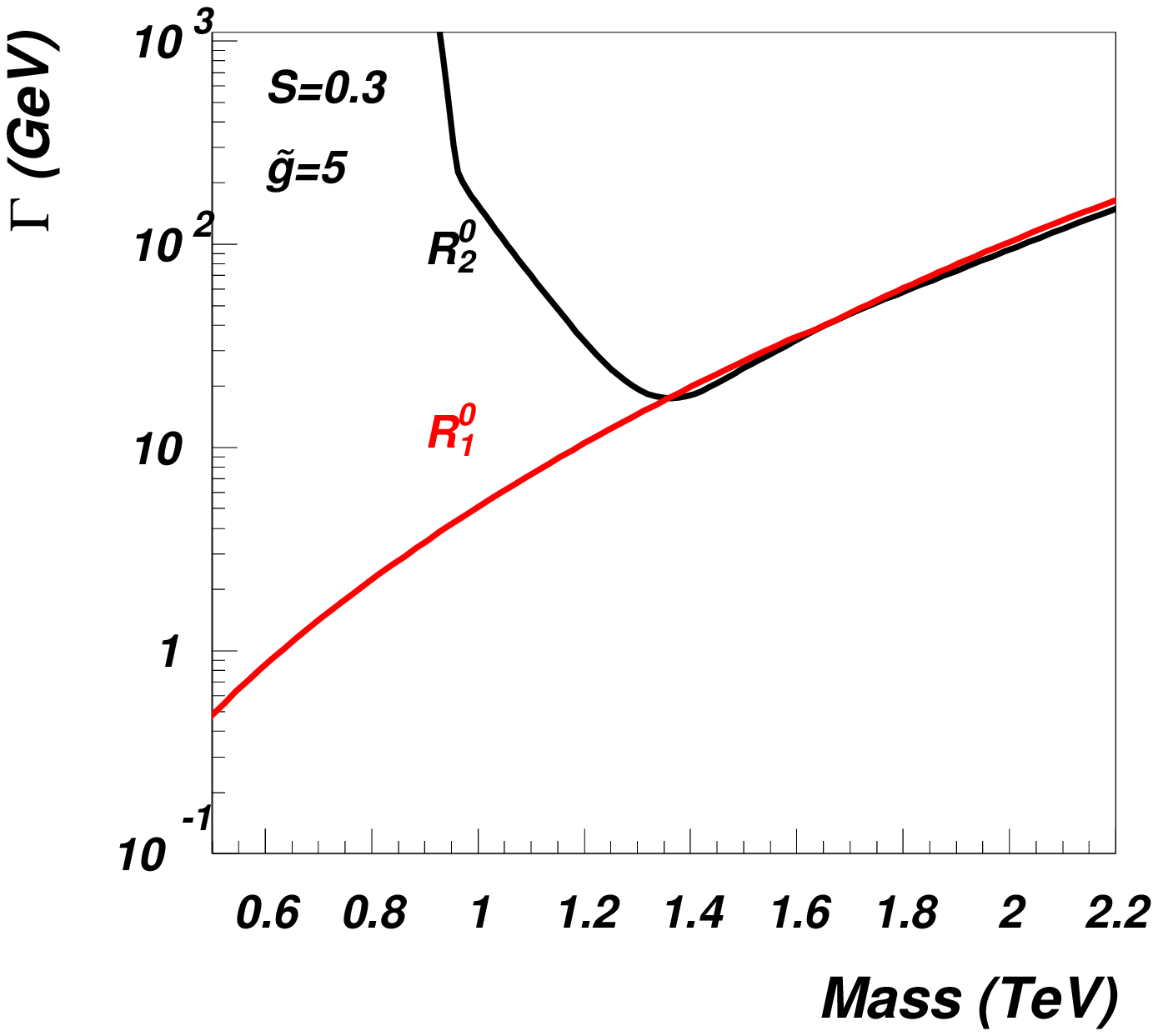}
 \vskip -0.4cm
\caption{Decay width of the charged (first row) and neutral (second row) 
vector mesons for $S=0.3$ and $\tilde{g}=2,5$.  We take $M_H = 0.2 \ \textrm{TeV},\ s=0$.}
\label{fig:decay2}
\end{figure}

 The widths of the heavy vectors are displayed
 in Fig.~\ref{fig:decay2}.  The lighter meson, $R_1$, is very narrow.
 The heavier meson, $R_2$, is very narrow for small values of
 $\tilde{g}$. In fact in this case $M_{R_2}\simeq M_{R_1}$, forbidding decays of $R_2$ to $R_1$ (+anything). For large
 $\tilde{g}$, $R_2$ is very narrow  for large masses, but then becomes
 broader when the $R_2\to R_1,X$ channels open up, where $X$ is a SM 
 gauge boson.  It becomes very broad when the $R_2\to 2R_1$ decay channel opens up.
  The former are only important below the inversion point,
 where $R_1$ is not too heavy. The latter is only possible when $R_2$
 is essentially a spin one vector  and $M_{R_2}>2 M_{R_1}$.

The narrowness of $R_1$ { (and $R_2$, when the $R_2\to R_1,X$ channels are forbidden)} 
is essentially due to the
small value of the $S$ parameter. In fact for $S=0$ the trilinear
couplings of the  vector mesons to two scalar fields of the strongly
interacting sector vanish. This can be understood as follows:  the
trilinear couplings with a vector resonance contain a derivative of
either the Higgs or the technipion,  and this can only come from $r_3$
in Eq.~(\ref{eq:boson}). Since $r_3=0$ implies $S=0$, as
Eqs.~(\ref{eq:S})  and (\ref{eq:chi})  show explicitly, it follows
that the decay width of $R_1$ and $R_2$ to two scalar fields  vanishes
as $S\rightarrow 0$. As a consequence, for $S=0$ the vector meson
decays to the longitudinal SM  bosons are highly suppressed,
because the latter are nothing but the eaten technipions. (The
couplings to the SM  bosons do not vanish exactly because of the mixing with the spin one resonances.)
A known scenario in which the widths of $R_1$ and $R_2$ are  highly
suppressed is provided by the D-BESS model~\cite{Casalbuoni:2000gn},
where the spin one  and the spin zero  resonances do not interact.
Therefore, in D-BESS {\em all} couplings involving one or more vector
resonances and  one or more scalar fields vanish, not just the
trilinear coupling with one vector field. The former scenario requires 
$r_2=r_3=s=0$, the latter only requires $r_3=0$. A somewhat
intermediate scenario is provided by CT, in  which $r_2=r_3=0$ but
$s\neq 0$. Narrow spin one  resonances seems to be a common feature in
various models of dynamical electroweak symmetry breaking. (see for
example Ref.~\cite{Brooijmans:2008se}).  Within our effective Lagrangian~(\ref{eq:boson}) this property is linked to having a
small $S$ parameter.  If it turns out that broader spin one resonances are observed at the LHC this fact can be accounted for by including operators of mass dimension greater than four, as shown in Sec.~\ref{sec:narrow}.

\bigskip

\begin{figure}[tbhp]
\vskip -0.2cm
\includegraphics[width=0.45\textwidth,height=0.35\textwidth]{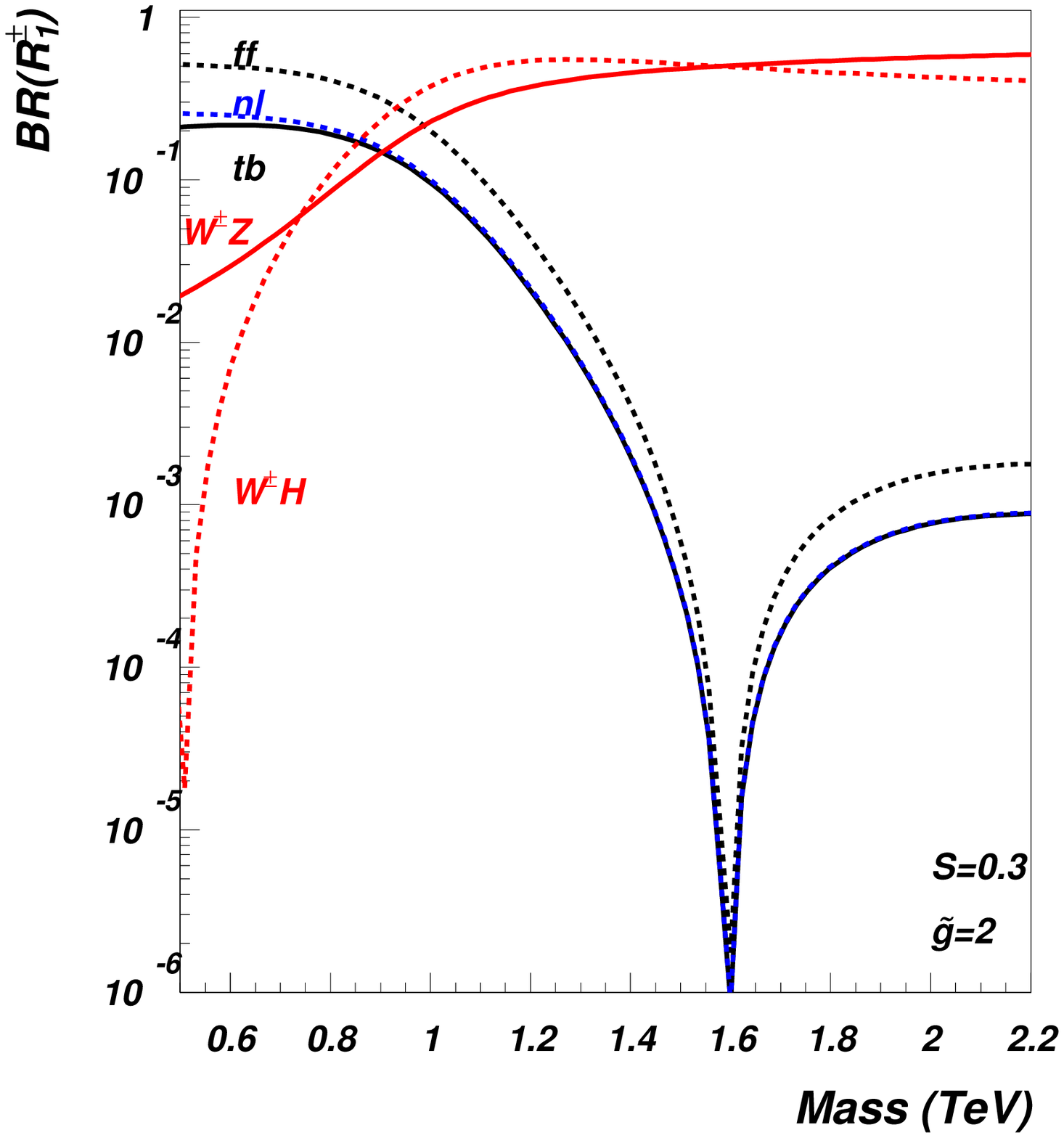}%
\includegraphics[width=0.45\textwidth,height=0.35\textwidth]{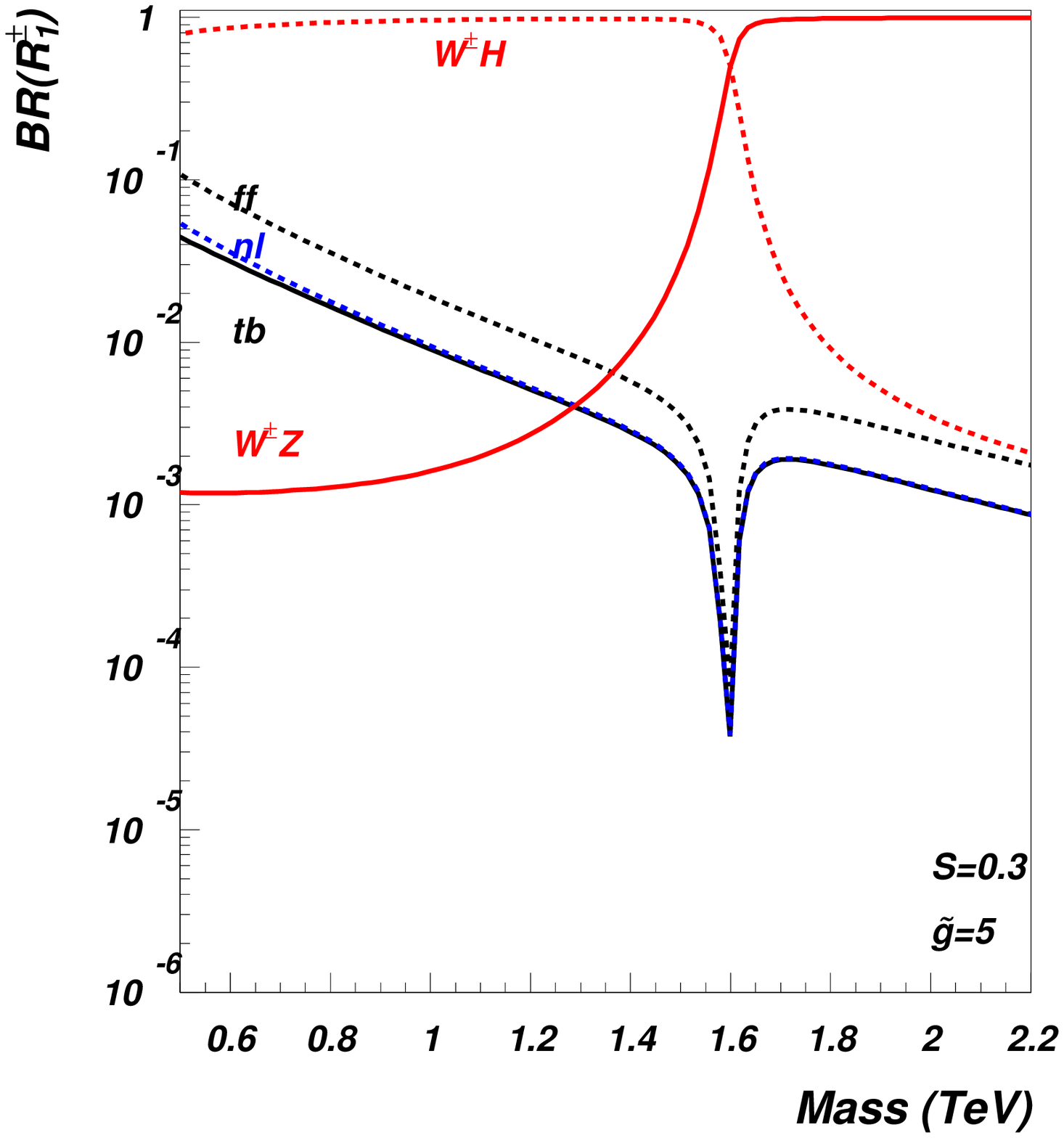}
 \vskip -0.2cm
 \includegraphics[width=0.45\textwidth,height=0.35\textwidth]{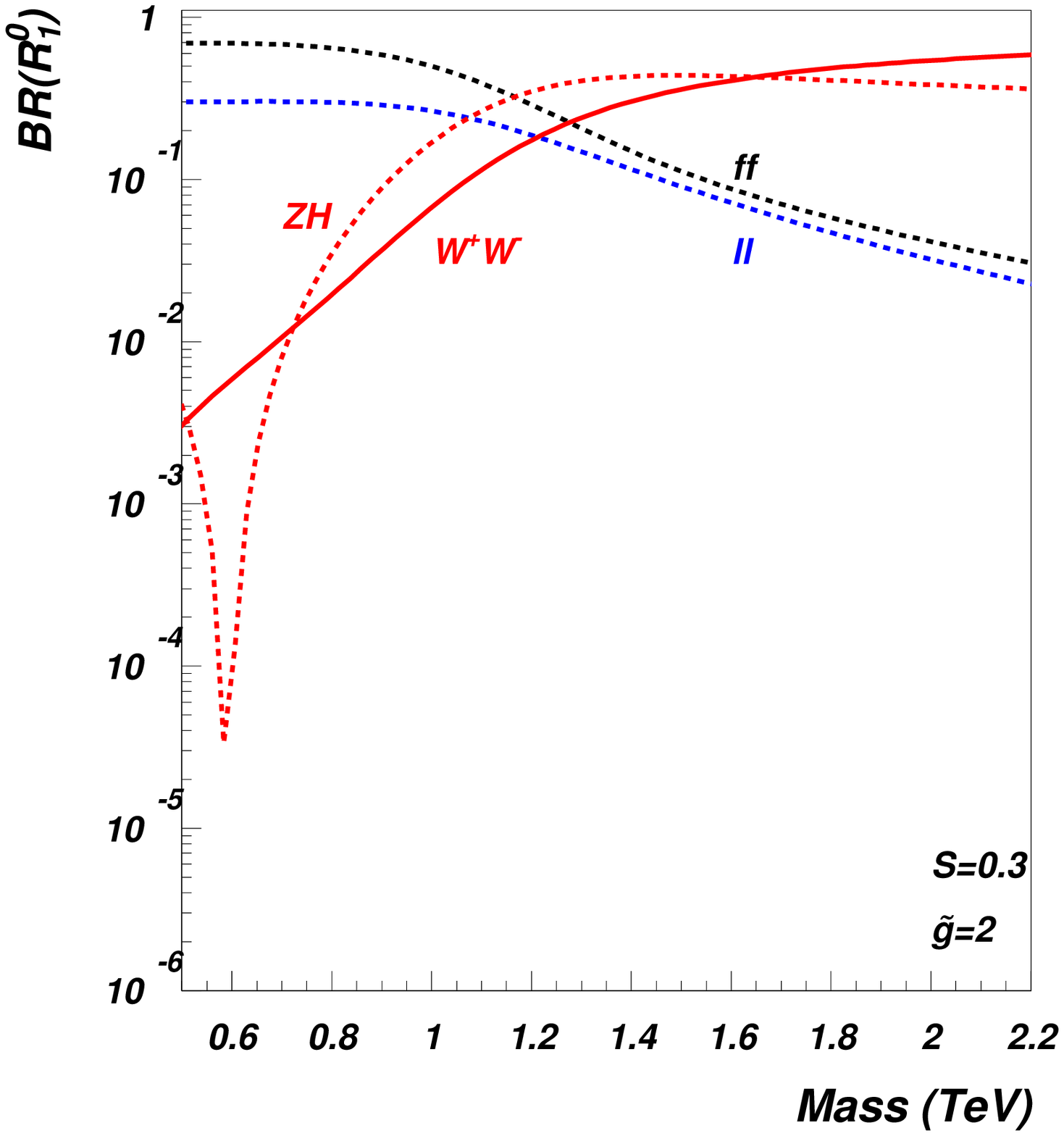}%
 \includegraphics[width=0.45\textwidth,height=0.35\textwidth]{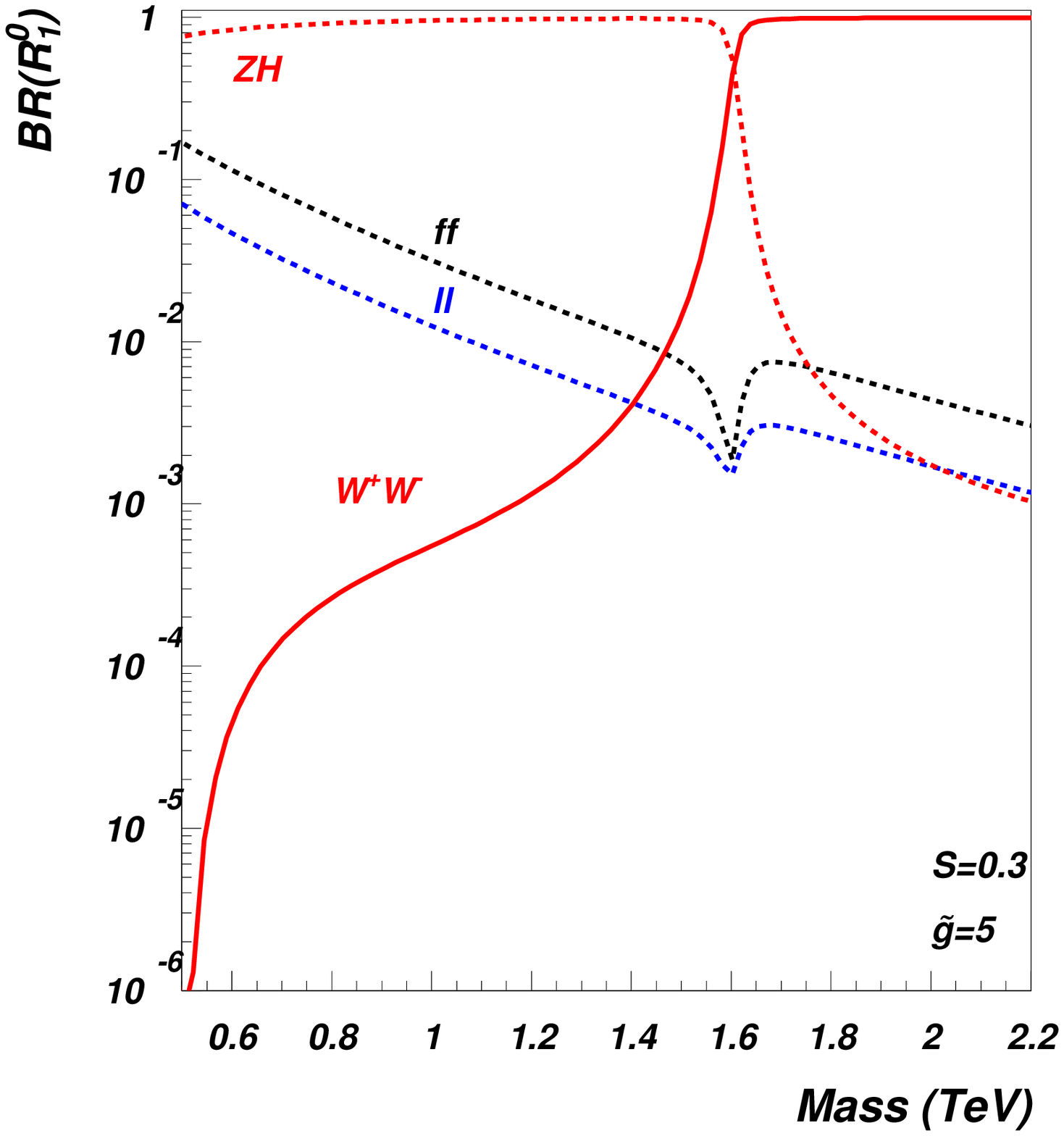}
 \vskip -0.2cm
\caption{Branching ratios of the charged (first row)
and neutral (second row)  $R_1$ resonance for $S=0.3$ and $\tilde{g}=2,5$.  We take $M_H = 0.2 \ \textrm{TeV}, s=0$}\label{fig:BRR1}
\end{figure}
The  $R_1$ branching ratios are shown in
Fig.~\ref{fig:BRR1}. 
The wild variations observed in the plots around $1.6$~{TeV} reflects the mass inversion discussed earlier. Here the mixing between $R_1^a$ and $\widetilde{W}^a$,  with $a=0,\pm$,  vanishes, suppressing the decay to SM fermions. 

The other observed structure for the decays in $ZH$ and $WH$, at low masses, is due to the opposite and competing contribution coming from the technicolor and electroweak sectors. This is technically possible since the coupling of the massive vectors to the longitudinal component of the gauge bosons and the composite Higgs is suppressed by the small value of S. 

{\begin{figure}[tbhp]
 \vskip -0.2cm
 \includegraphics[width=0.45\textwidth,height=0.35\textwidth]{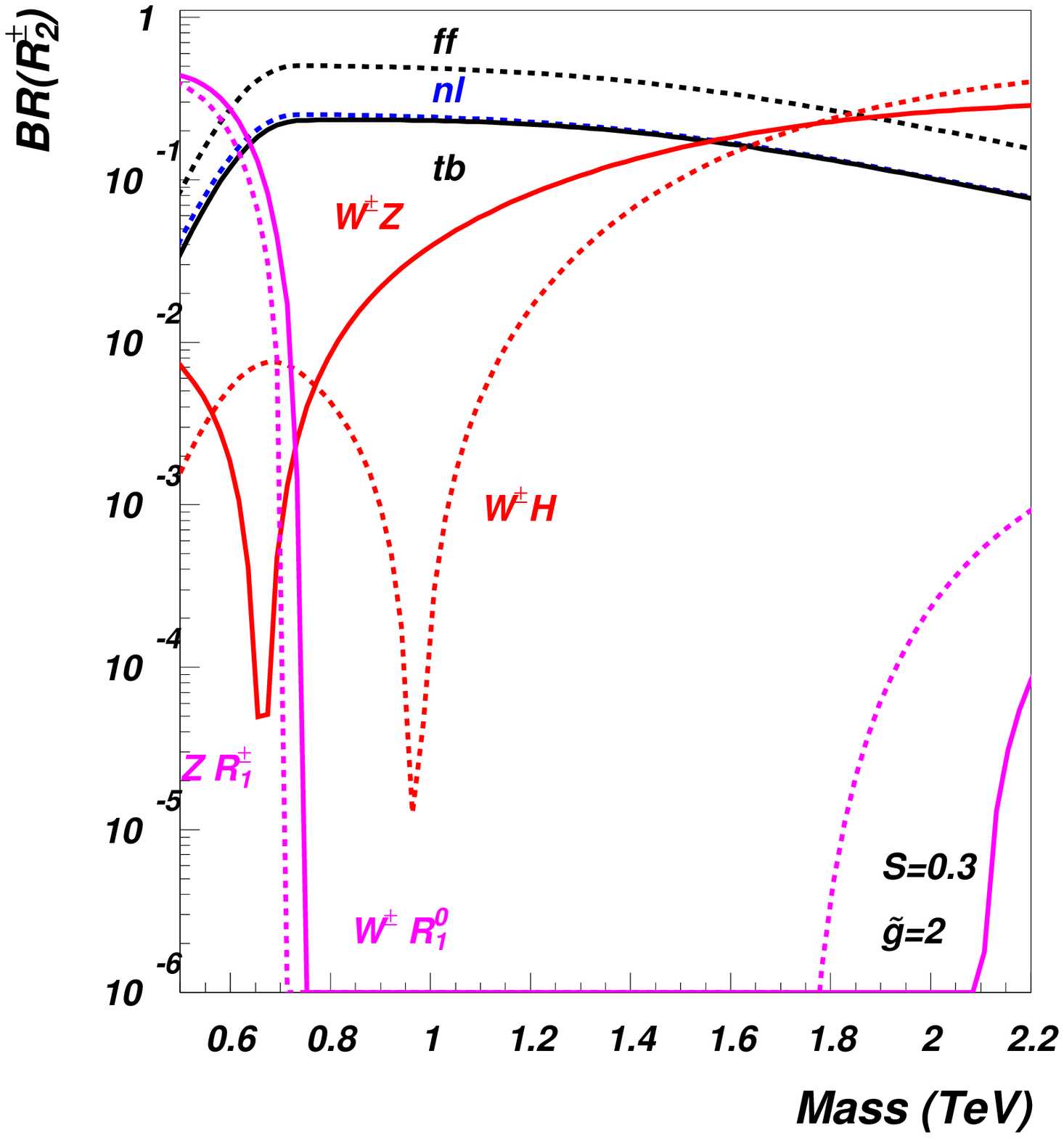}%
 \includegraphics[width=0.45\textwidth,height=0.35\textwidth]{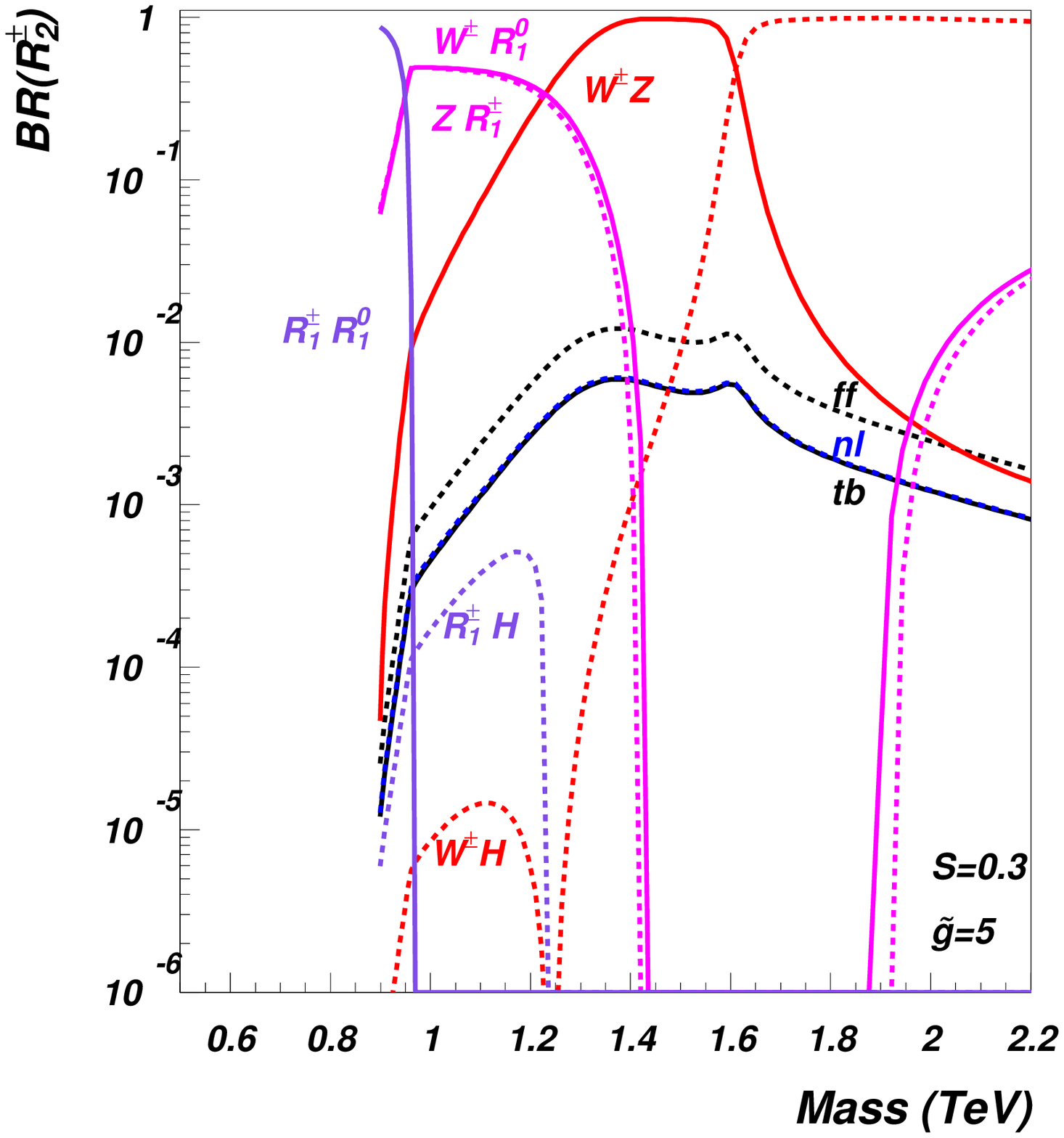}
 \vskip -0.2cm
 \includegraphics[width=0.45\textwidth,height=0.35\textwidth]{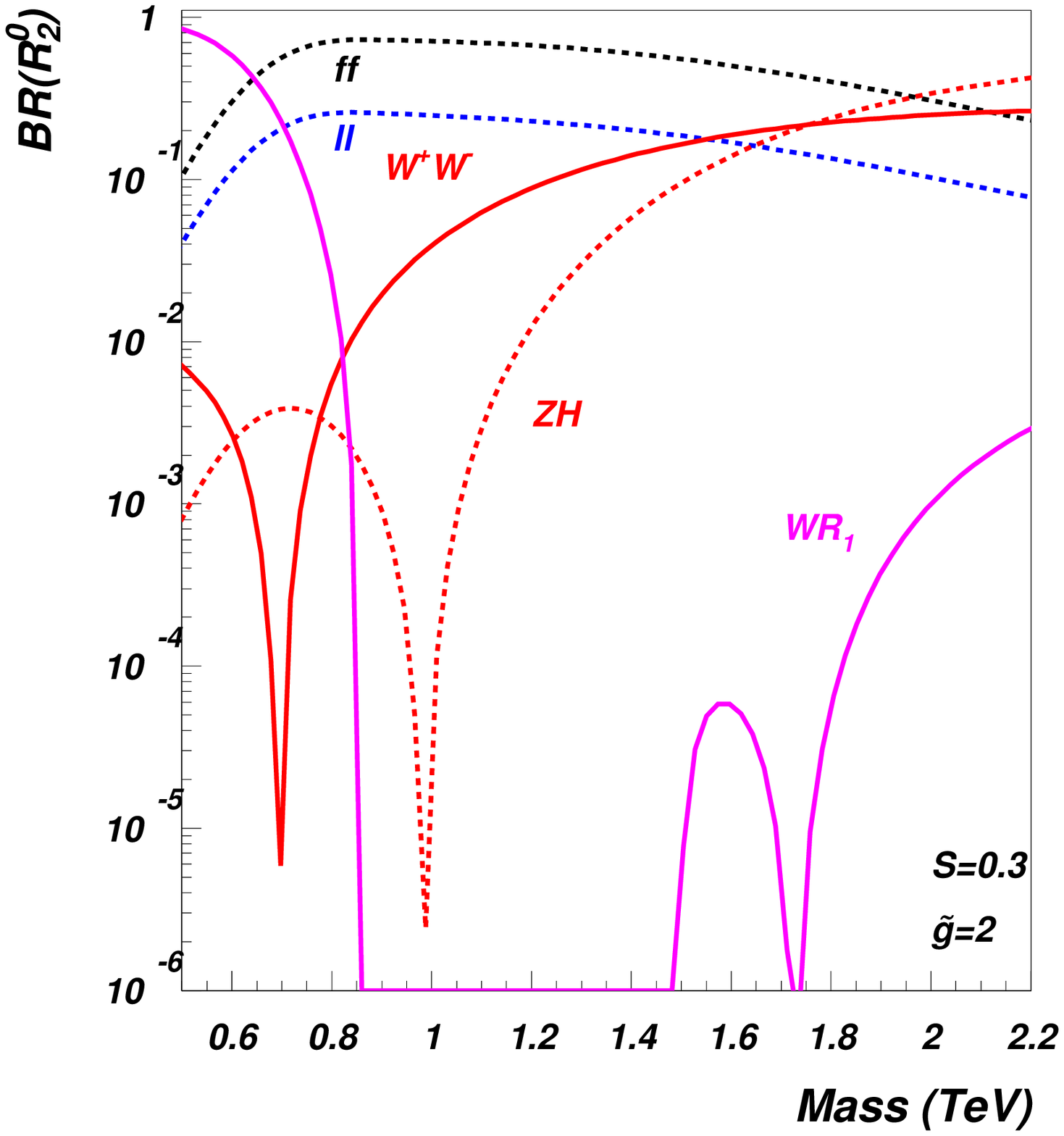}%
 \includegraphics[width=0.45\textwidth,height=0.35\textwidth]{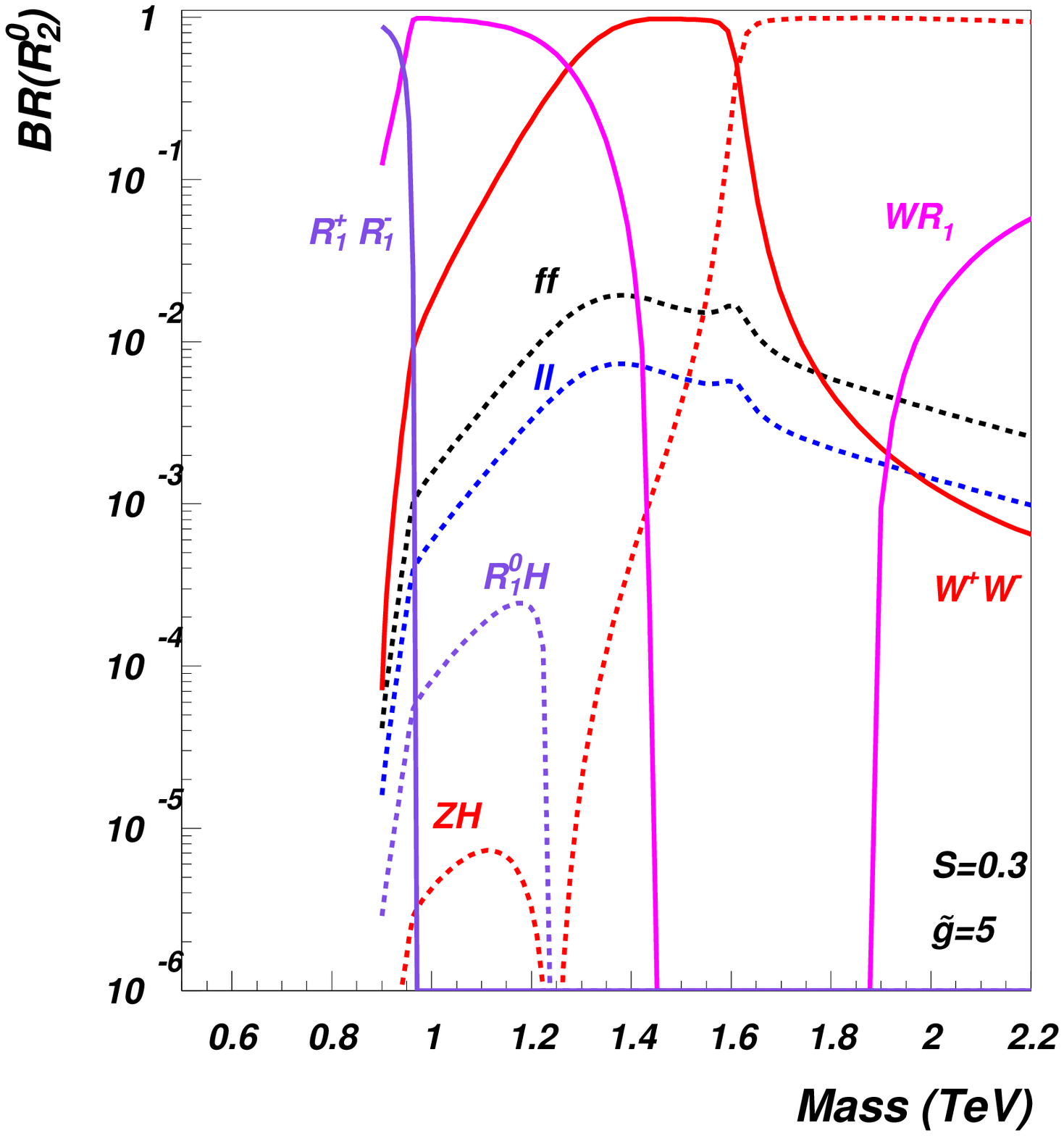}
 \vskip -0.2cm
\caption{Branching ratios of the charged (first row) and neutral (second row) $R_2$ resonance for $S=0.3$ and $\tilde{g}=2,5$ .  
We take $M_H = 0.2 \ \textrm{TeV}, s=0$}\label{fig:BRR2}
\end{figure} 
 
Now we consider the $R_2$ BR's displayed in Fig.~\ref{fig:BRR2}. Being $R_2$ heavier than $R_1$ by definition, new channels like $R_2\rightarrow 2R_1$ and $R_2\rightarrow R_1 X$ show up, where $X$ denotes a SM boson. Notice that there is a qualitative difference in the $R_2$ decay modes for small and large values of $\tilde{g}$. First, for small $\tilde{g}$ the $R_2-R_1$ mass splitting is not large enough to allow the decays $R_2\rightarrow 2R_1$ and $R_2\rightarrow R_1 H$, which are instead present for large $\tilde{g}$. Second, for small $\tilde{g}$ there is a wide range of masses for which the decays to $R_1$ and a SM vector boson are not possible, because of the small mass splitting. 
The BR's to fermions do not drop at the inversion point, because {the $R_2-\widetilde{W}$ mixing does not vanish}.

\subsection{Drell-Yan Production: $p,p \to R_{1,2}$}

\begin{figure}[tbhp]
\vskip -0.2cm
\includegraphics[width=0.43\textwidth]{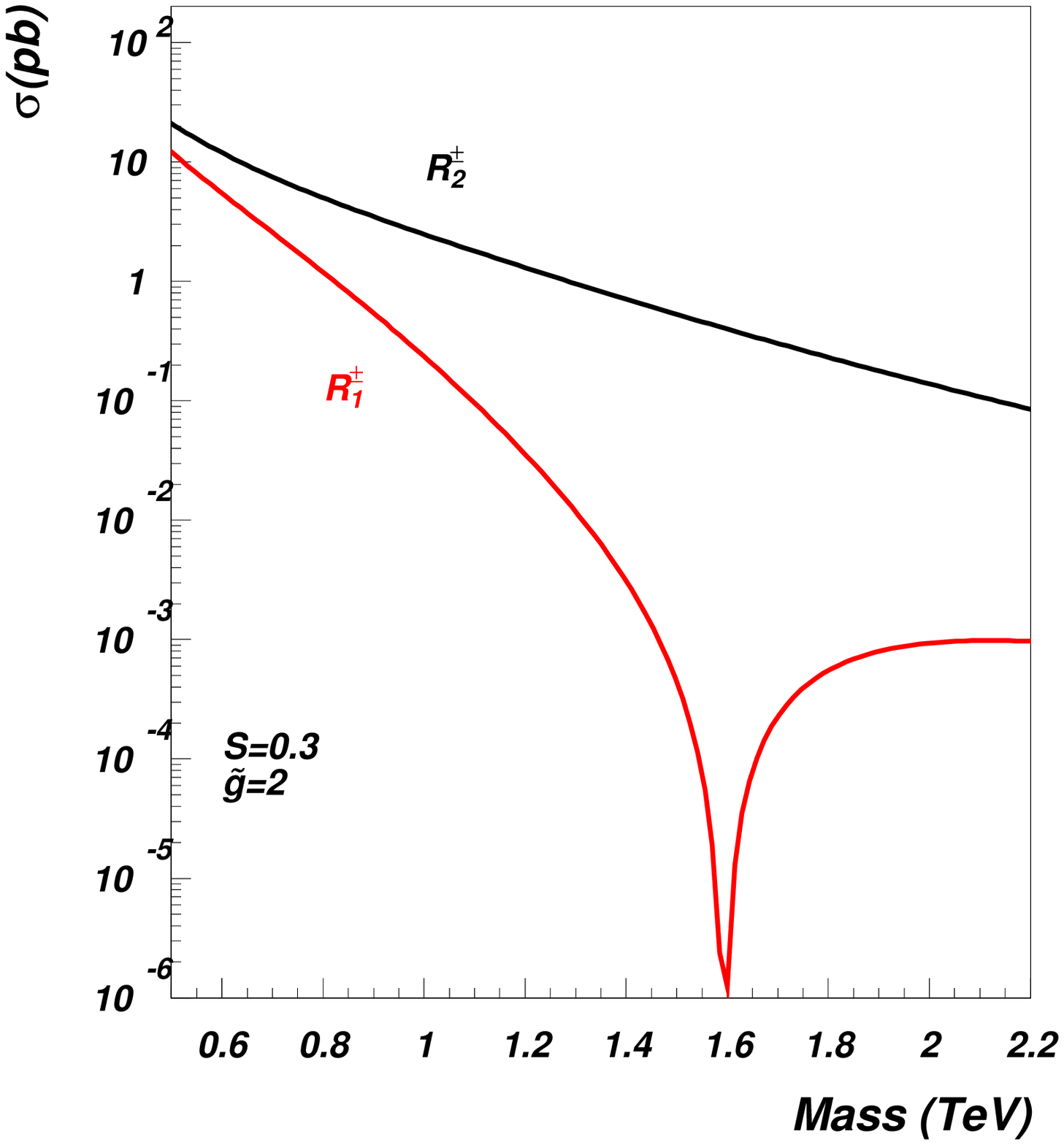}%
\includegraphics[width=0.43\textwidth]{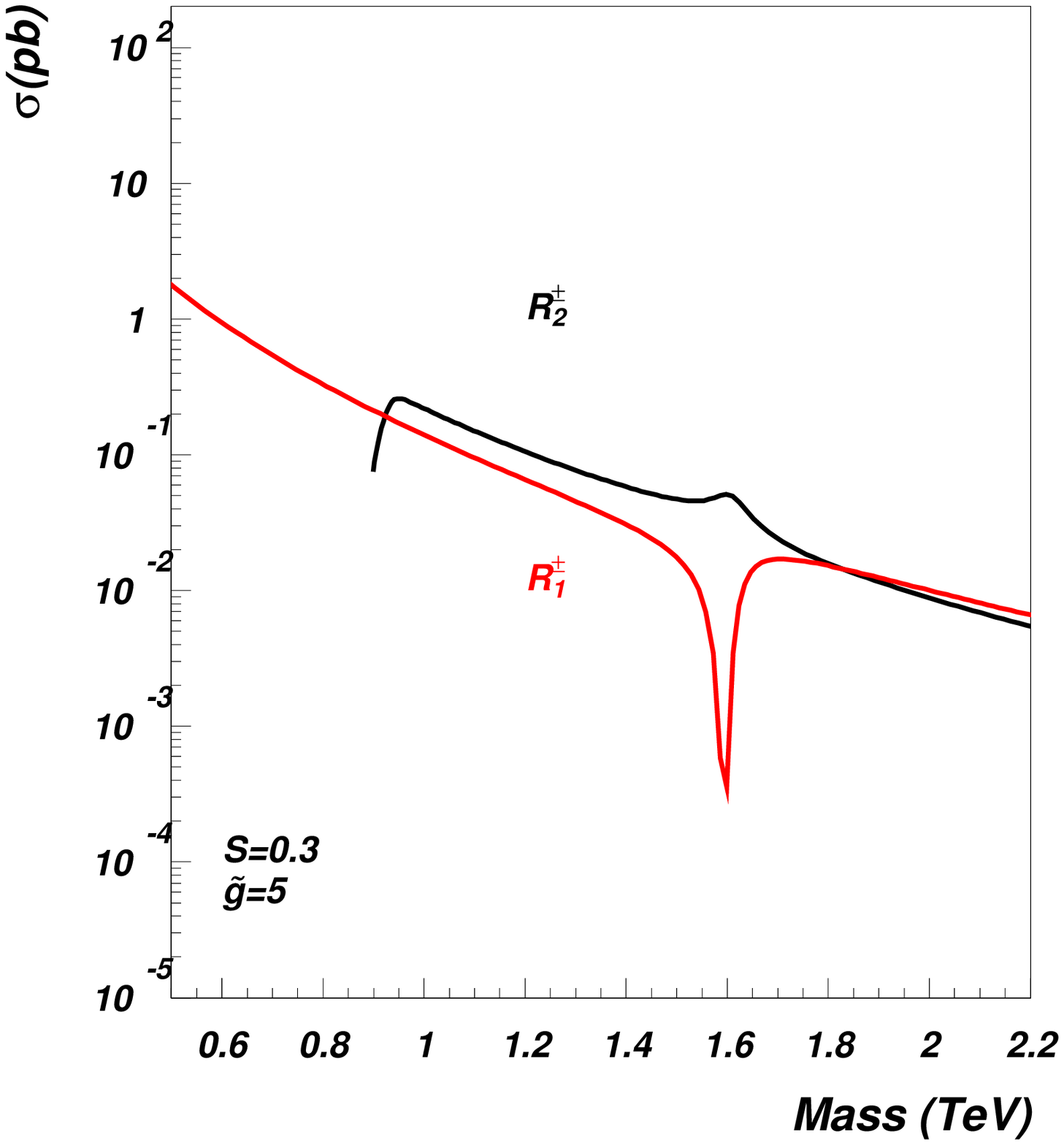}
\vskip -0.2cm
\includegraphics[width=0.43\textwidth]{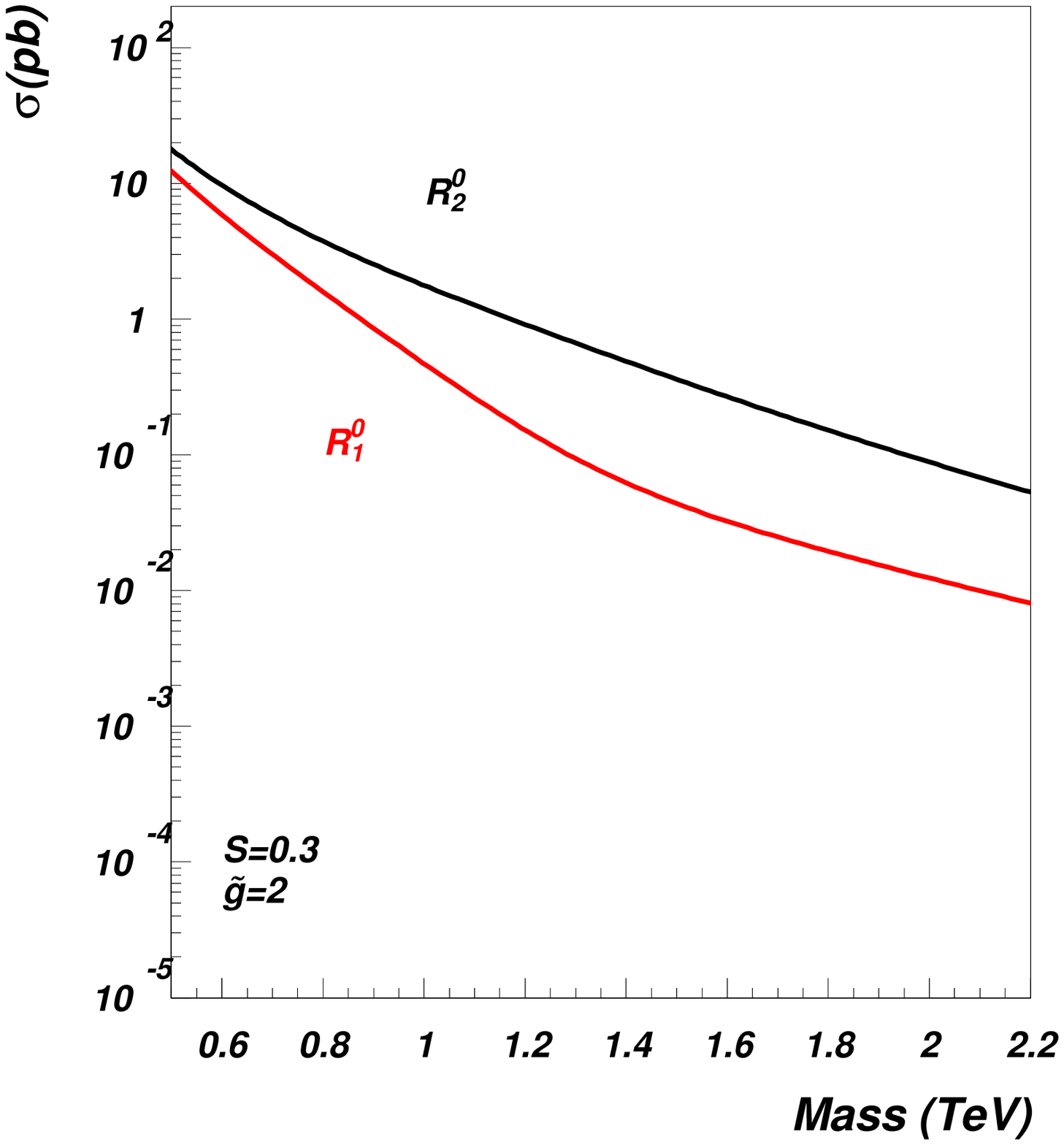}%
\includegraphics[width=0.43\textwidth]{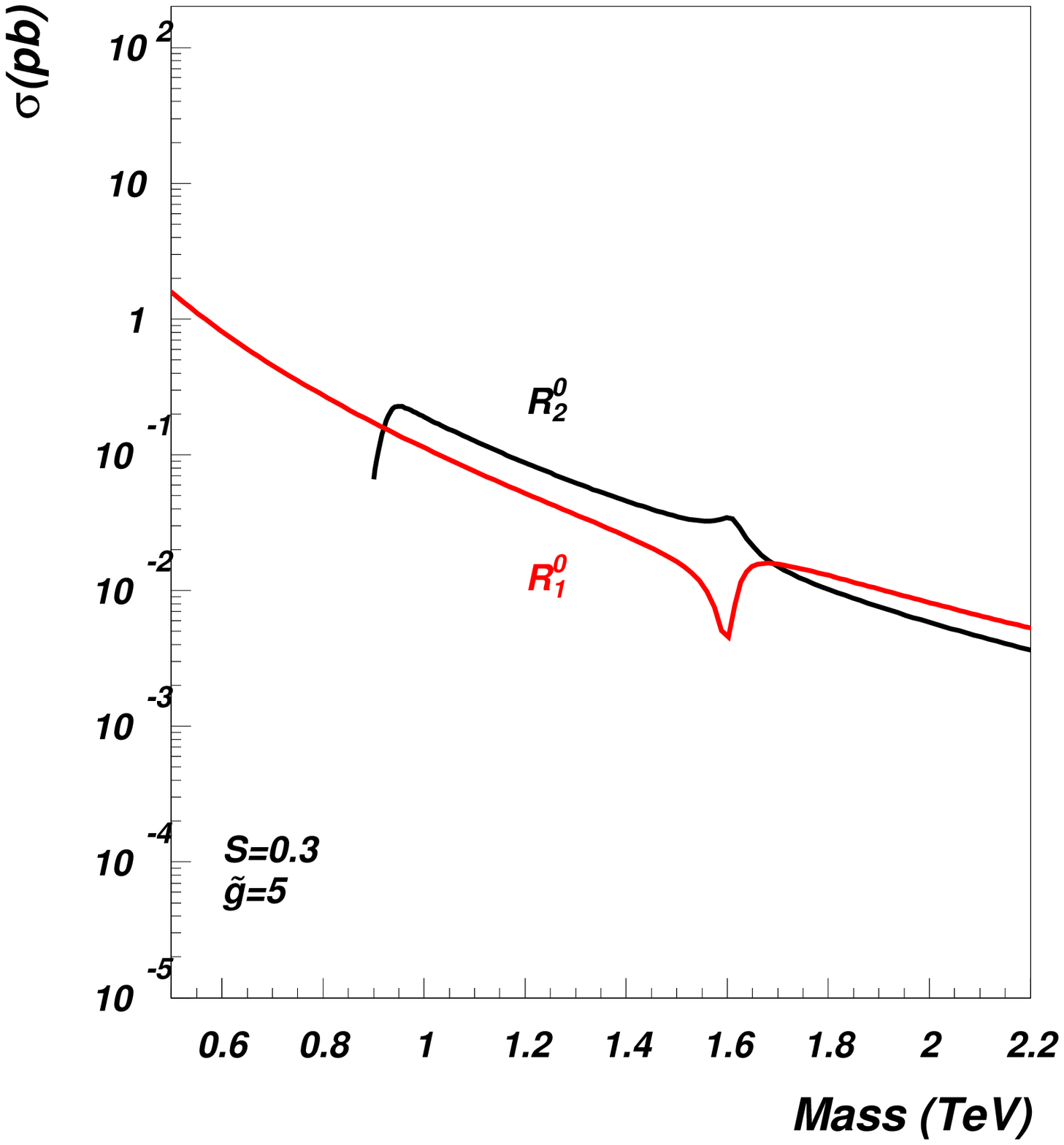}
\caption{Drell-Yan production of the $R_{1,2}^\pm$ (first row) and $R_{1,2}^0$ (second row) resonances, with $S=0.3$ and $\tilde{g}=2,5$.}
\label{fig:DY}
\end{figure}

Spin one resonances can be produced at LHC through the DY processes
$pp \to R_{1,2}$. The corresponding cross sections are shown in
Fig.~\ref{fig:DY}. Consider first the production of $R_2^\pm$, since the
latter is less affected than $R_1^\pm$ by the presence of the mass
inversion point. The cross section decreases as $\tilde{g}$ grows,
because of the reduced $R_2-\widetilde{W}$ mixing. In going from
$\tilde{g}=2$ to $\tilde{g}=5$ the decrease in the production
cross-section of $R_2^\pm$ is roughly one or two orders of magnitude. This is
expected since the leading order contribution to the coupling between
$R_2^\pm$ and fermions is explicitly proportional to $\tilde{g}^{-2}$, as
it is in the D-BESS model~\cite{Casalbuoni:2000gn}. 

As explained in Sec.~\ref{sec:decay} the $R_1^\pm$ resonance becomes
fermiophobic at the inversion point, causing the corresponding DY production
to drop.  In our model the new vectors are fermiophobic only at the mass inversion point differentiating it from
 a class of Higgsless model in which the charged $W^\prime$ resonance is taken to have
strongly suppressed  couplings to  the light fermions for any value of the vector masses.

To estimate the LHC reach for DY production of the $R_{1,2}^0$ and $R_{1,2}^\pm$
resonances we study the following lepton signatures:
\begin{itemize}
\item[(1)]
$\ell^+\ell^-$
signature from the process $pp\to R^{0}_{1,2}\to \ell^+\ell^-$
\item[(2)]
$\ell + \met$
signature from the process $pp\to R^{\pm}_{1,2}\to \ell^\pm\nu$
\item[(3)]
$3\ell + \met$
signature from the process $pp\to R^{\pm}_{1,2}\to ZW^\pm\to 3\ell \nu$
,
\end{itemize}
where $\ell$ denotes a charged lepton -- electron or muon and $\met$ is the missing transverse energy.
We apply detector acceptance cuts of
$|\eta^\ell|<2.5$ and
$p_T^\ell>15$ GeV on the rapidity and transverse momentum of the 
leptons.
For
signature (1)
we use the di-lepton invariant mass distribution $M_{\ell \ell}$
to separate the signal from the background. 
For signatures (2) and (3) 
we use instead the transverse mass variables
$M^T_\ell$ and $M^T_{3\ell}$~\cite{bagger}:
%\begin{equation}
% (M^{T}_\ell)^2=
% \left[\sqrt{M^2(\ell)+p_T^2(\ell)}
%    +|p_T^{\rm miss}|\right]^2-|p_T^{}(\ell)+p_T^{\;\rm miss}|^2
%\end{equation}
%\begin{equation}
% (M^{T}_{3\ell})^2=
% \left[\sqrt{M^2(\ell\ell\ell)+p_T^2(\ell\ell\ell)}
%    +|p_T^{\rm miss}|\right]^2-|p_T^{}(\ell\ell\ell)+p_T^{\;\rm miss}|^2 \ .
%\end{equation}

\begin{equation}
  (M^{T}_\ell)^2=
  [\sqrt{M^2(\ell)+p_T^2(\ell)}
     +|\mpt|]^2-|\vec{p}_T^{}(\ell)+\mptv|^2
\end{equation}
\begin{equation}
  (M^{T}_{3\ell})^2=
  [\sqrt{M^2(\ell\ell\ell)+p_T^2(\ell\ell\ell)}
     +|\mpt|]^2-|\vec{p}_T^{}(\ell\ell\ell)+\mptv|^2
\end{equation}

We also add a cut on the transverse
missing energy $\met>15$ GeV.
We consider the  representative parameter space points
$\tilde{g}=2,5$ and $M_A=0.5, 1, 1.5, 2$ TeV
for our plots and discussion.

The  invariant mass and transverse 
mass distributions for signatures (1)-(3) are shown in 
Figs.~\ref{fig:sig1}-\ref{fig:sig3}.
\begin{figure}[tbhp]
\vskip -0.2cm 
\includegraphics[width=0.5\textwidth]{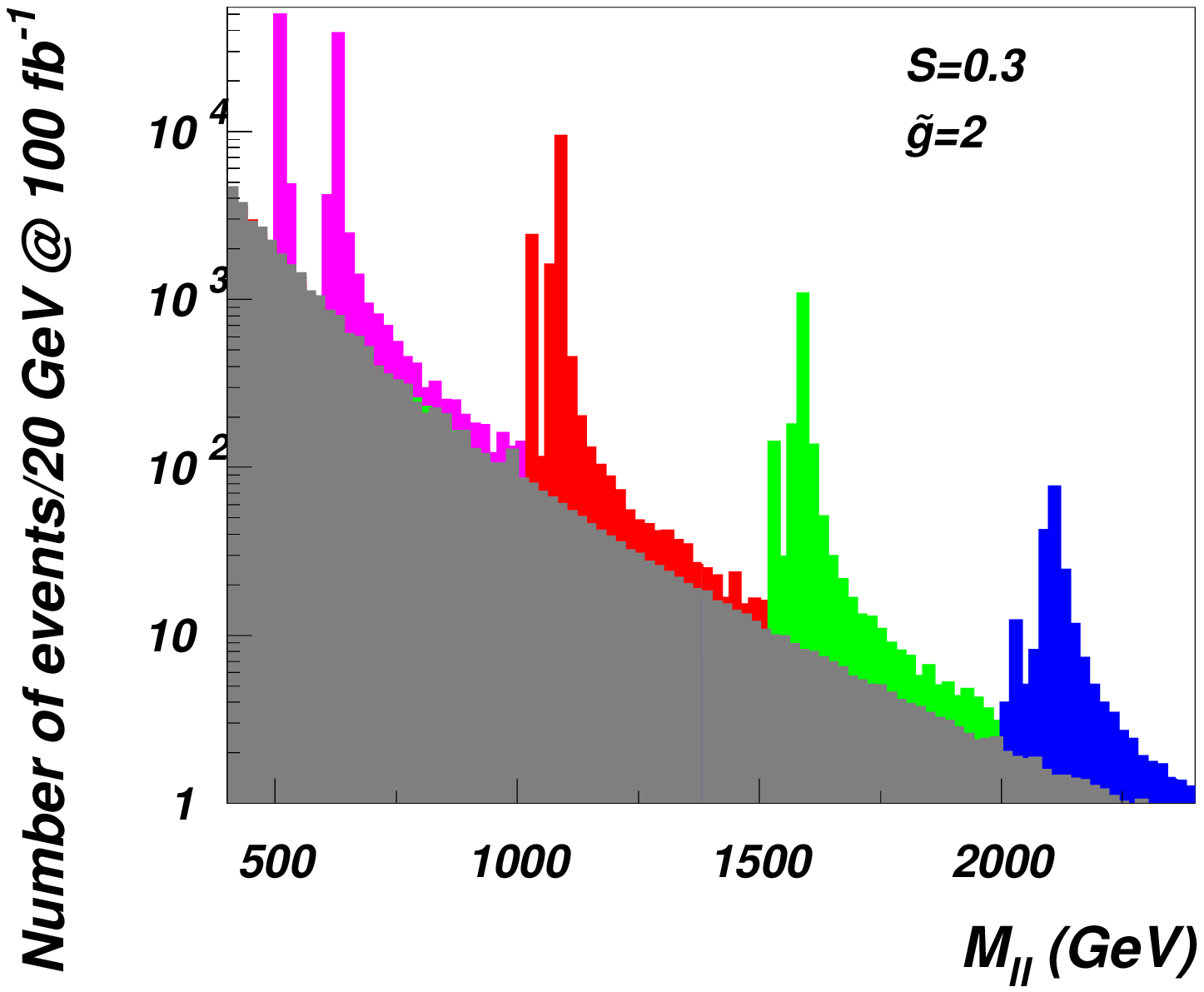}%
\includegraphics[width=0.5\textwidth]{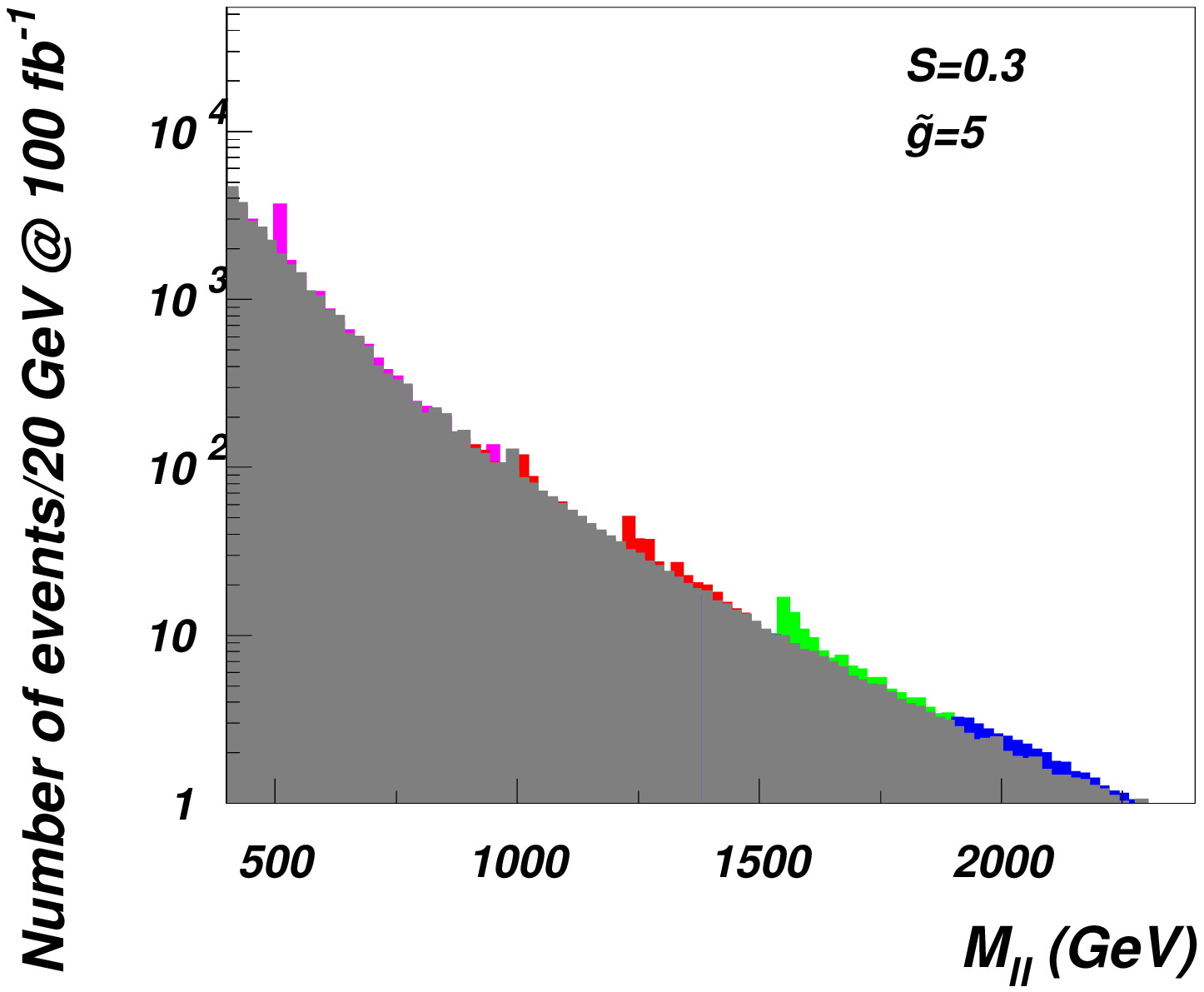}%
\caption{Dilepton invariant mass distribution
$M_{\ell \ell}$ for $pp\to R^{0}_{1,2}\to \ell^+\ell^-$
signal and background processes.
We consider $\tilde{g}=2,5$ respectively
and masses $M_A=0.5$ Tev (purple), $M_A=1$ Tev (red), 
$M_A=1.5$ Tev (green) and $M_A=2$ Tev
(blue). }
\label{fig:sig1}
\end{figure}
\begin{figure}[tbhp]
\vskip -0.2cm
\includegraphics[width=0.5\textwidth]{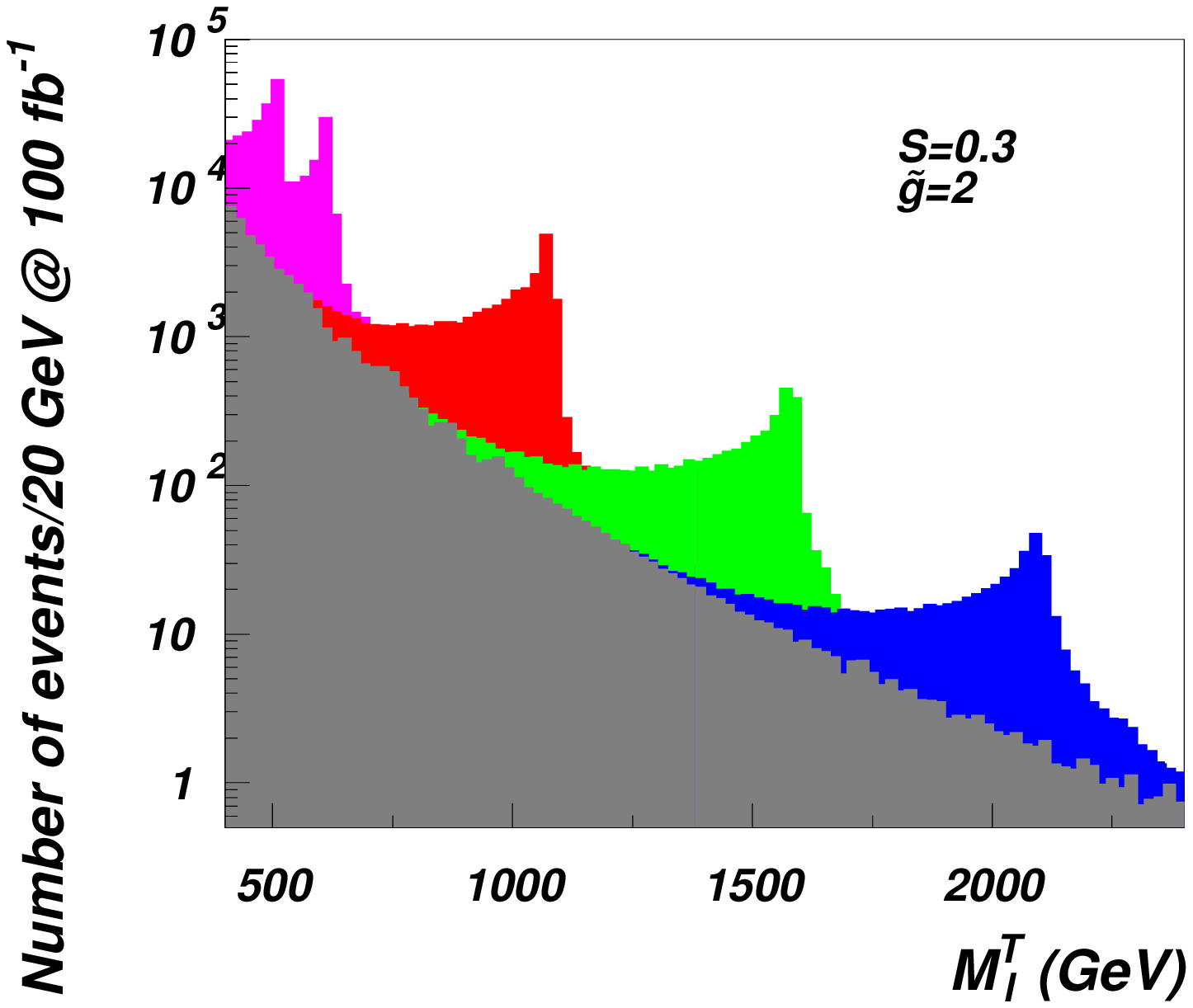}%
\includegraphics[width=0.5\textwidth]{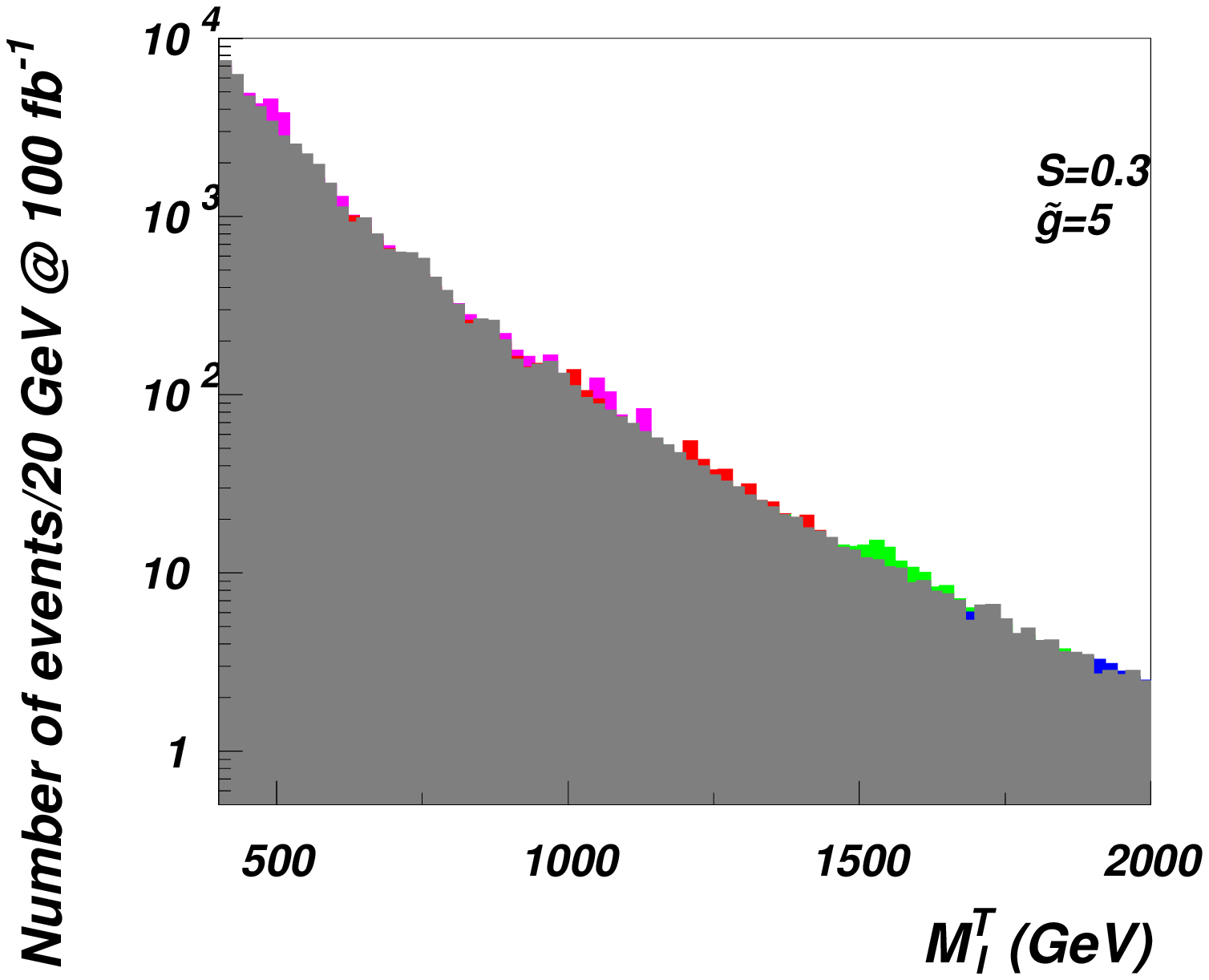}%
\caption{$M^T_\ell$ mass distribution
for $pp\to R^{\pm}_{1,2}\to \ell^\pm\nu$
signal and background processes.
We consider $\tilde{g}=2,5$ respectively
and masses $M_A=0.5$ Tev (purple), $M_A=1$ Tev (red), 
$M_A=1.5$ Tev (green) and $M_A=2$ Tev
(blue). 
\label{fig:sig2}}
\end{figure}
\begin{figure}[tbhp]
\vskip -0.2cm
\includegraphics[width=0.5\textwidth]{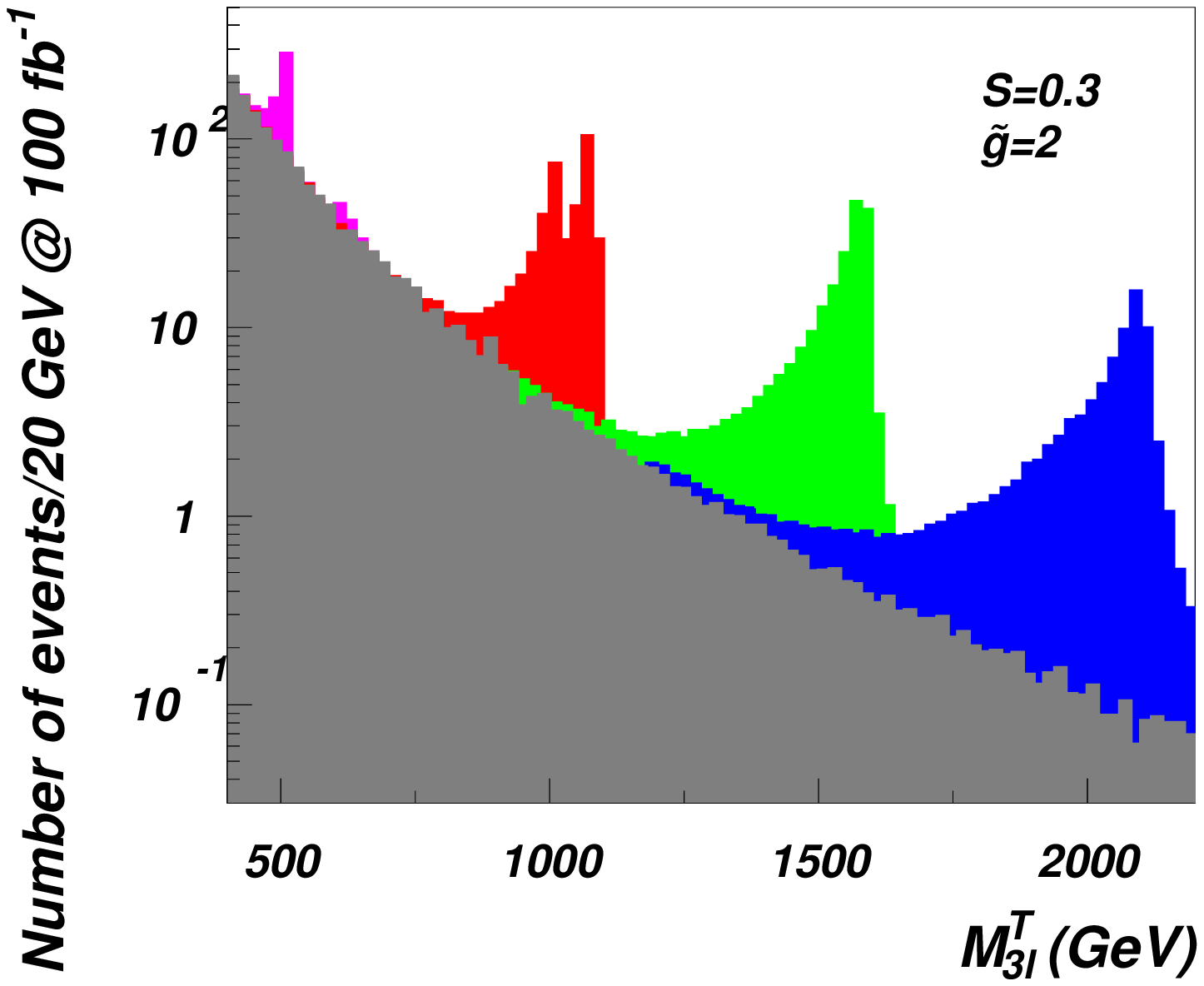}%
\includegraphics[width=0.5\textwidth]{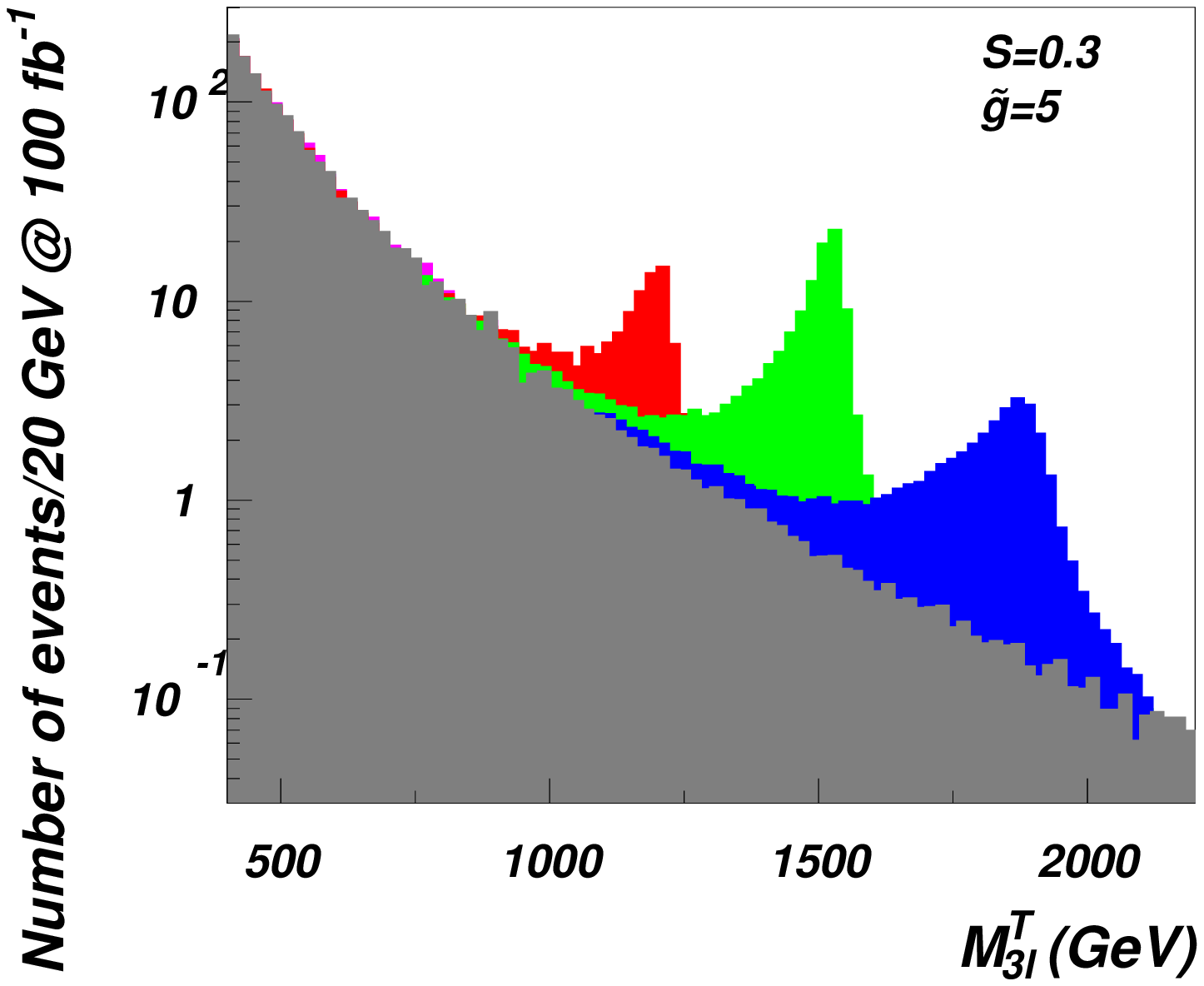}
\caption{$M^T_{3\ell}$ mass distribution
for $pp\to R^{\pm}_{1,2}\to ZW^\pm\to 3\ell\nu$
signal and background processes.
We consider $\tilde{g}=2,5$ respectively
and masses $M_A=0.5$ Tev (purple), $M_A=1$ Tev (red), 
$M_A=1.5$ Tev (green) and $M_A=2$ Tev
(blue). 
\label{fig:sig3}}
\end{figure}
In the left frames of Figs.~\ref{fig:sig1} and ~\ref{fig:sig2},
corresponding to $\tilde g=2$, clear signals from the leptonic decays of 
$R^{0}_{1,2}$ and  $R^{\pm}_{1,2}$ are seen even for 2
TeV resonances. Moreover Fig.~\ref{fig:sig1} demonstrates that
for $\tilde g=2$  both peaks
from  $R^{0}_{1}$ and $R^{0}_{2}$ may be resolved. 
The lepton energy resolution effects should not visibly  affect the
presented distributions. In the case of signature (2) a double-resonance peak is also seen at low mass, but the
transverse mass distribution $M^{T}_\ell$ is not able to resolve the
signal pattern as well as the $M^{\ell\ell}$ distribution for
signature (1), because of the presence of missing transverse momenta
from the neutrino. This analysis must be improved via a full-detector simulation.  However, for larger masses only a single resonance is visible because the $R^\pm_{1}$ coupling to fermions is strongly suppressed.
This is a distinguishing footprint of the NMWT model at higher masses closer to the inversion point: only a single peak from the $R_2^\pm$ will appear in the single lepton channel while a double peak should be visible in the di-lepton channel.

Let us now turn to the case of $\tilde g=5$ in the right
frames of   Figs.~\ref{fig:sig1} and \ref{fig:sig2}. For large $\tilde g$ the $Rff$ couplings are suppressed, so observing signatures (1) and (2) could be problematic  (quantative results for the LHC reach for all signatures
are presented below). {}However, for large
$\tilde g$, the triple-vector coupling 
is enchanced, 
so one
can observe a clear signal in the $M^{T}_{3\ell}$ distribution presented in Fig.~\ref{fig:sig3}. At low masses 
the decays of the heavy vector mesons to SM gauge bosons are suppressed  and the signal disappears.  This mass range can, however, be covered with signatures (1) and (2) 
as we demonstrate below.
\begin{figure}[tbhp]
\vskip -0.2cm
\includegraphics[width=0.43\textwidth]{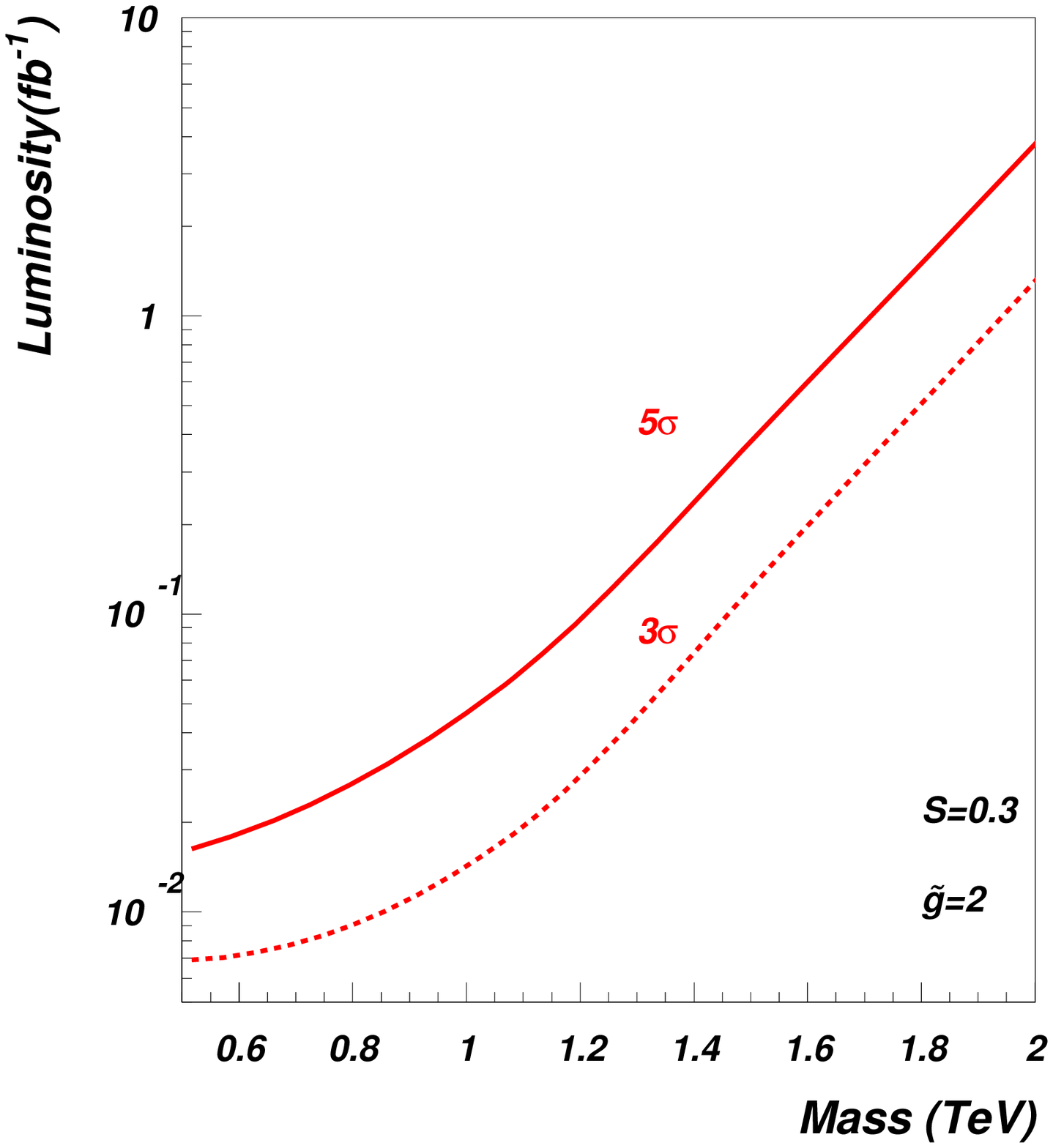}%
\includegraphics[width=0.43\textwidth]{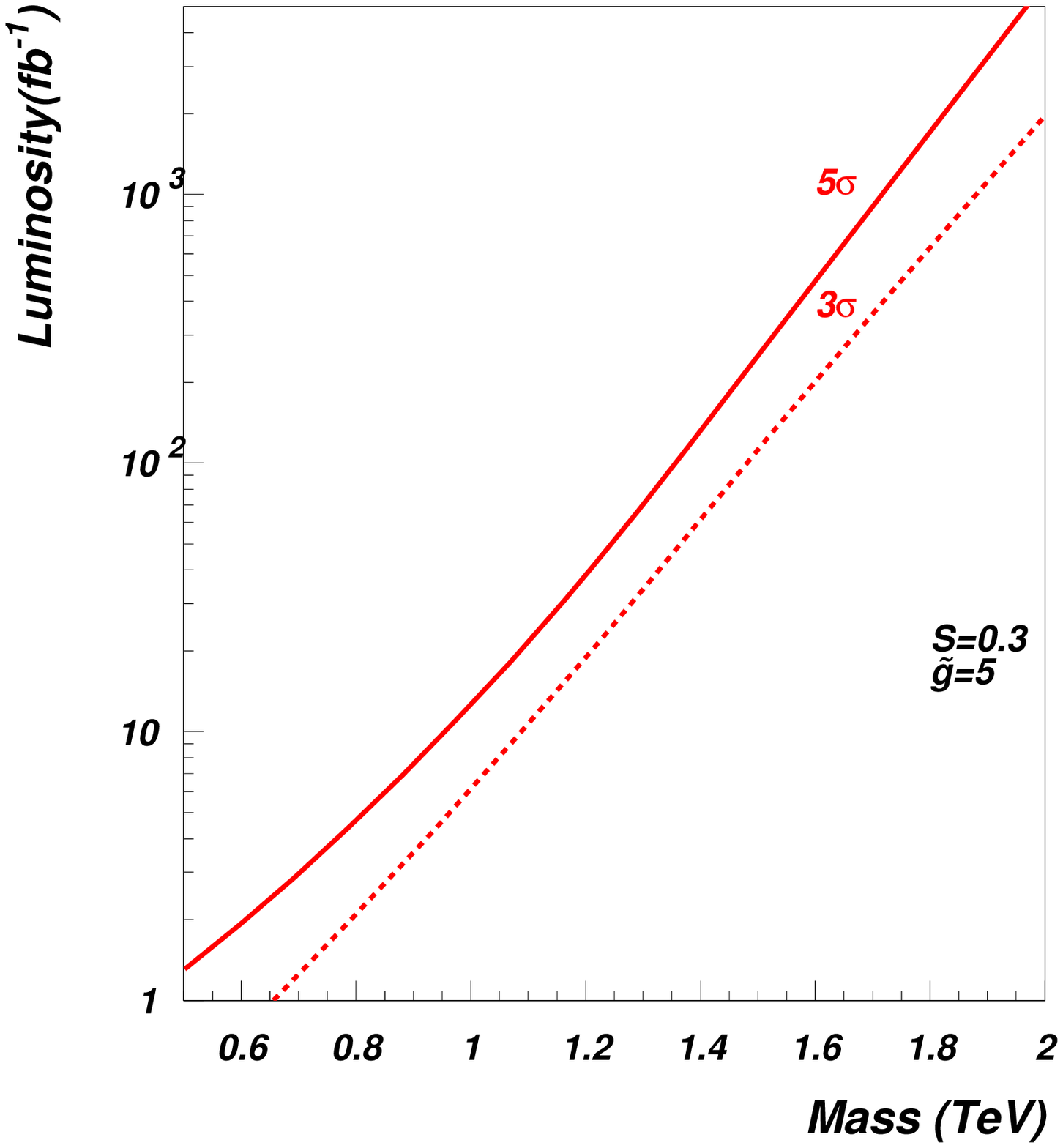}
\vskip -0.2cm
\includegraphics[width=0.43\textwidth]{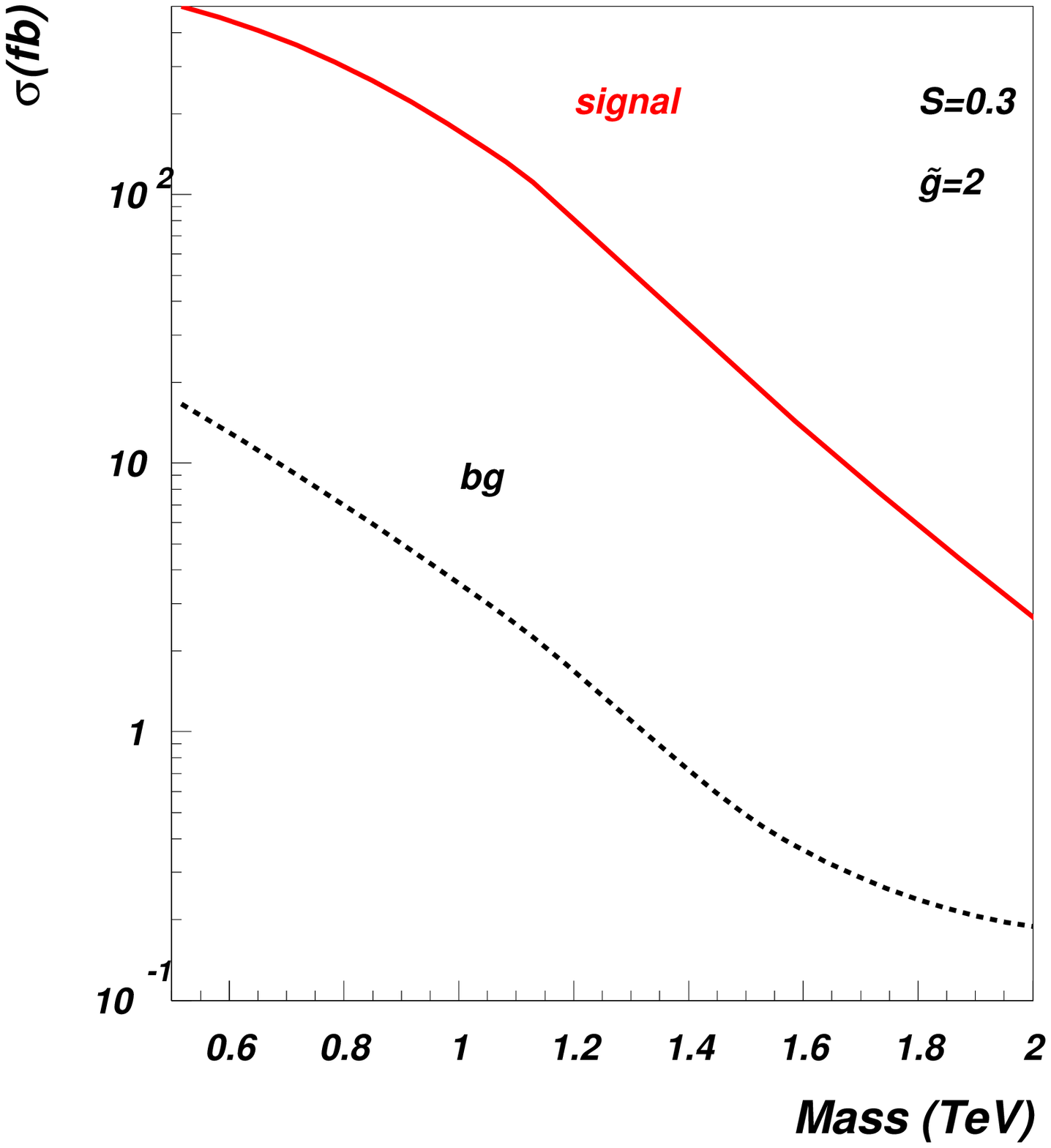}%
\includegraphics[width=0.43\textwidth]{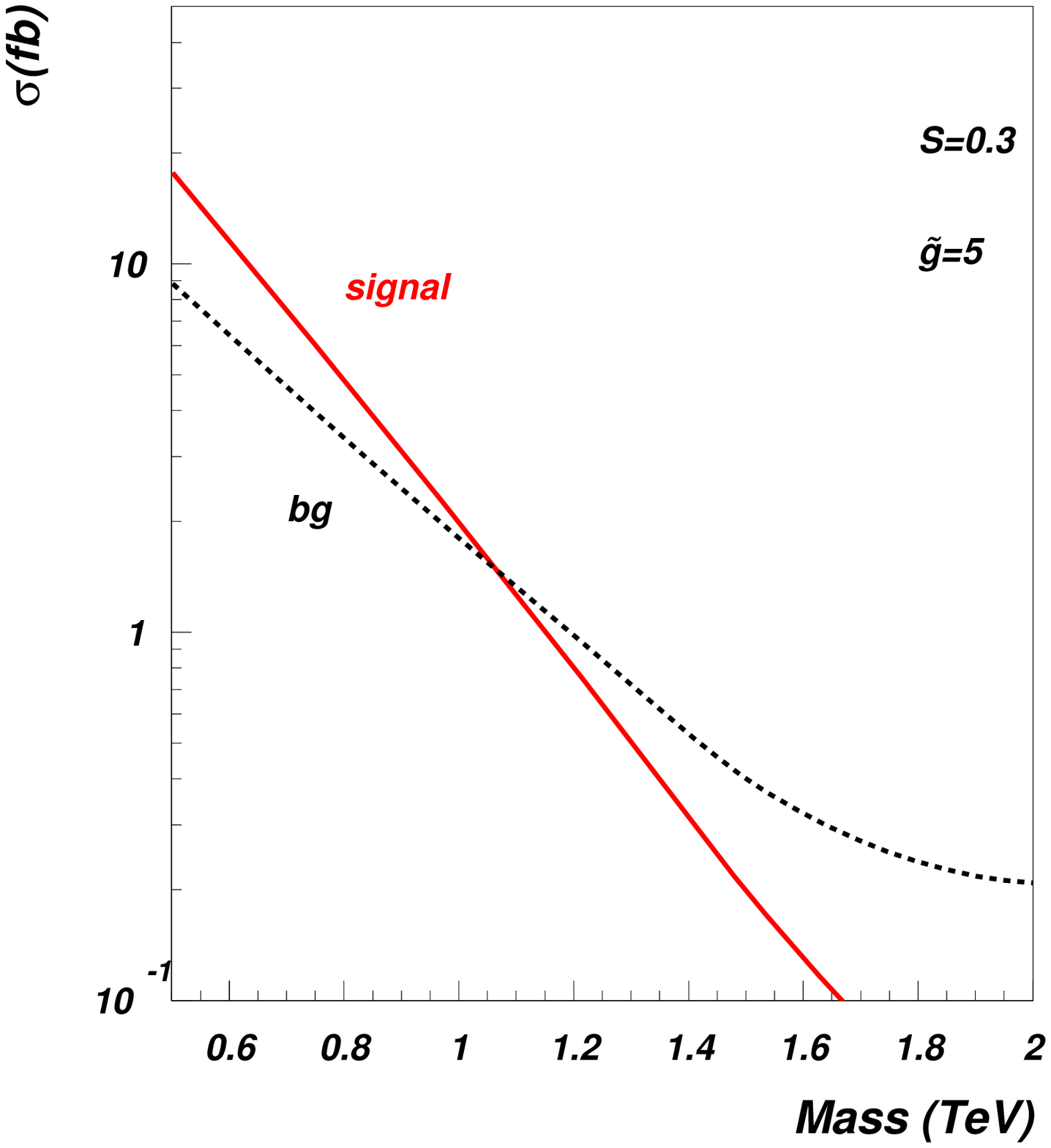}
\caption{LHC reach for the $R^0_{1,2}$ resonance signal in $pp\to R^0_{1,2} \to e^+ e^-$ with $S=0.3$ and
$\tilde{g}=2,5$. 
Top row: the required integrated luminosisty for
achieveing a 5$\sigma$ and a 3$\sigma$ significance. Bottom row:  the
signal and background cross sections for this process.}
\label{fig:signif1}
\end{figure}
 
\begin{figure}[tbhp]
\vskip -0.2cm
\includegraphics[width=0.43\textwidth]{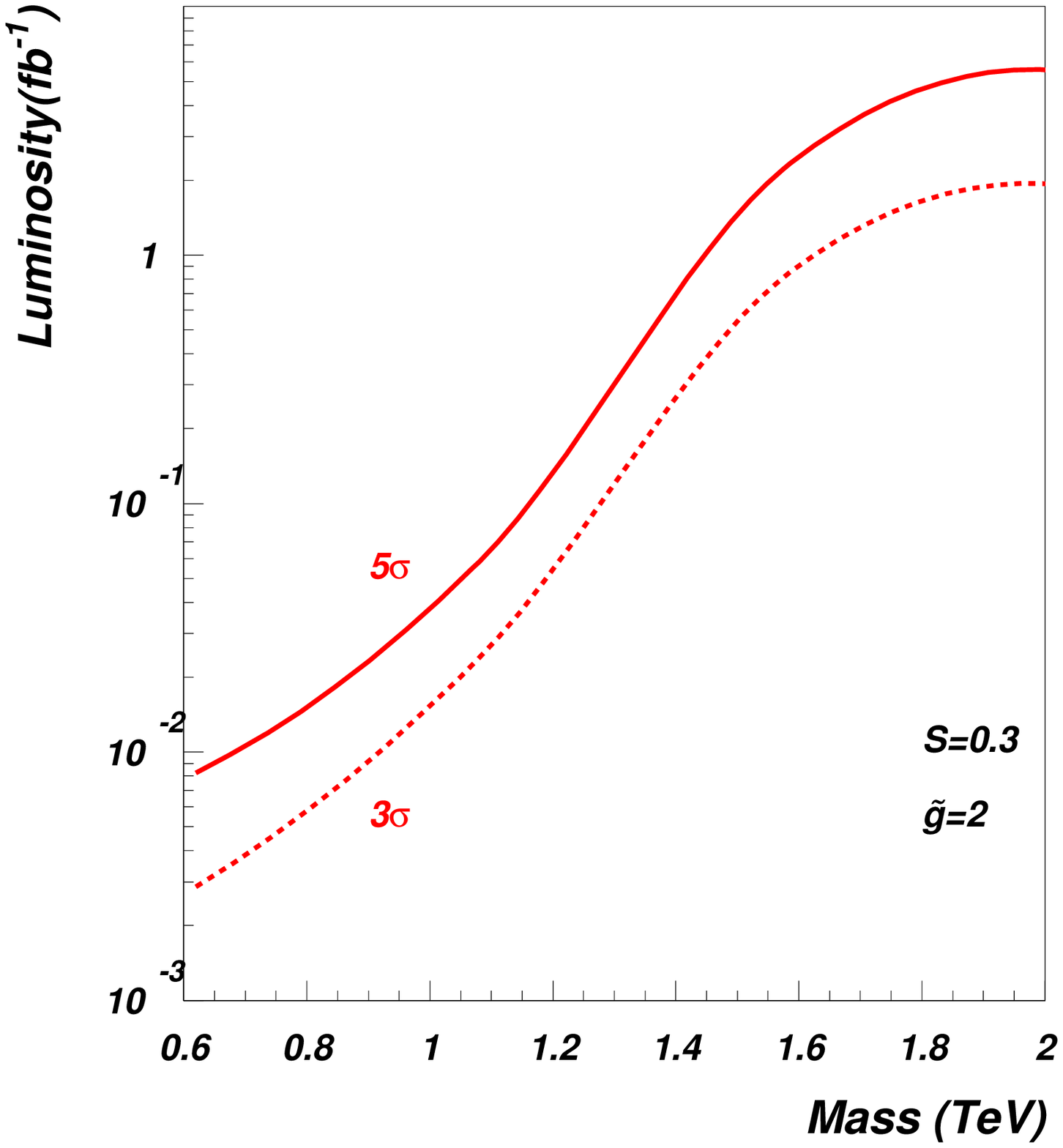}%
\includegraphics[width=0.43\textwidth]{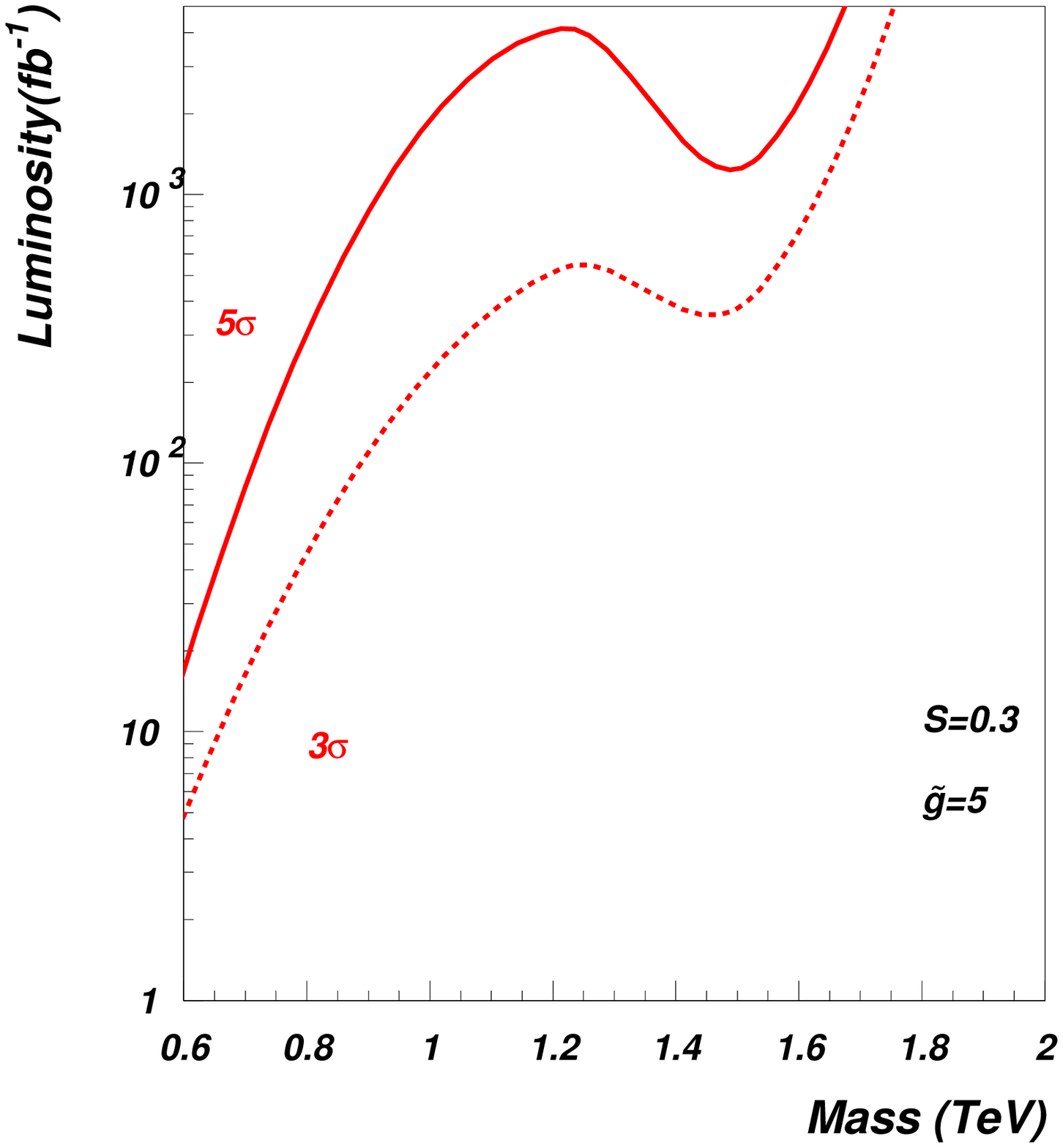}
\vskip -0.2cm
\includegraphics[width=0.43\textwidth]{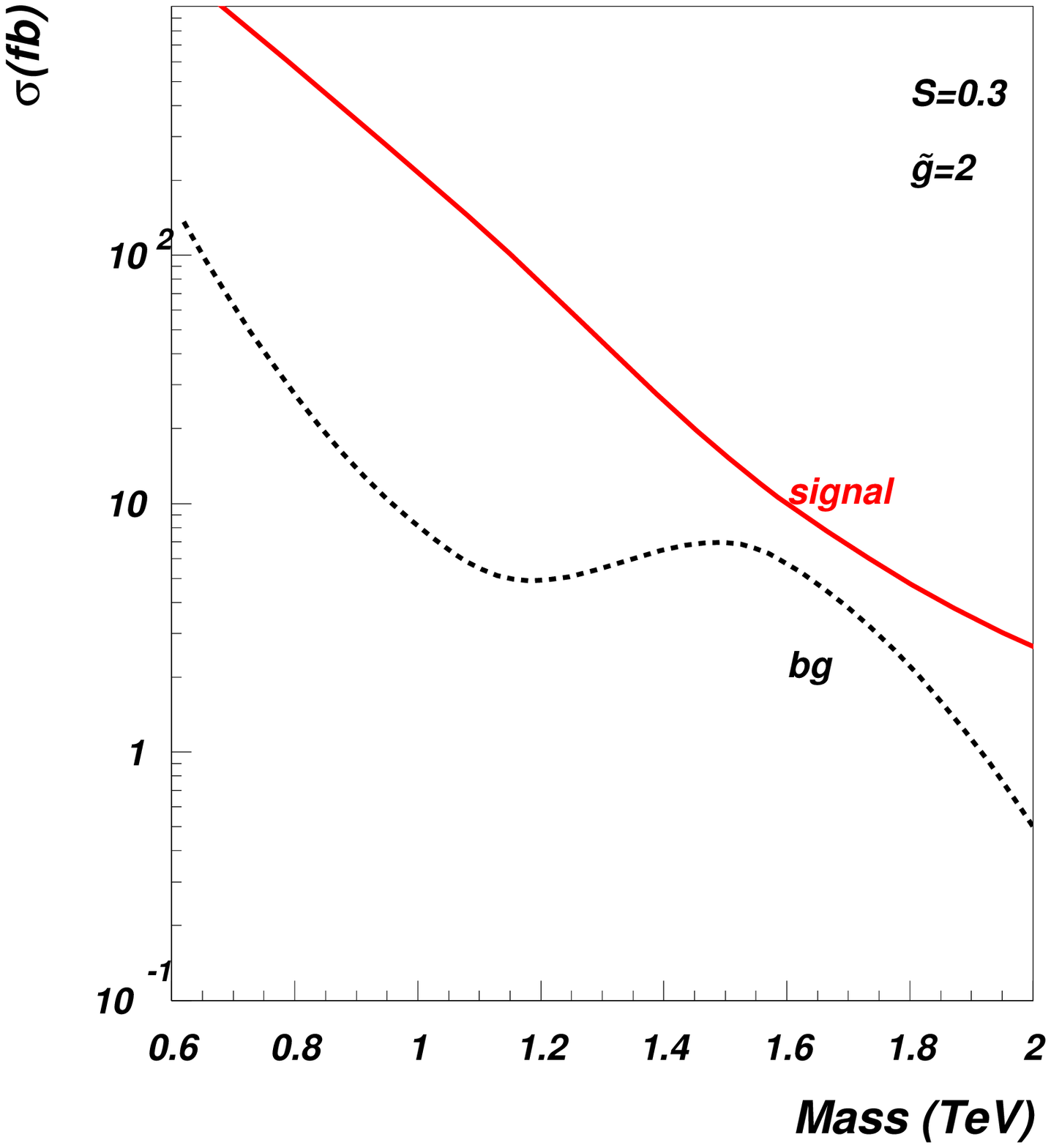}%
\includegraphics[width=0.43\textwidth]{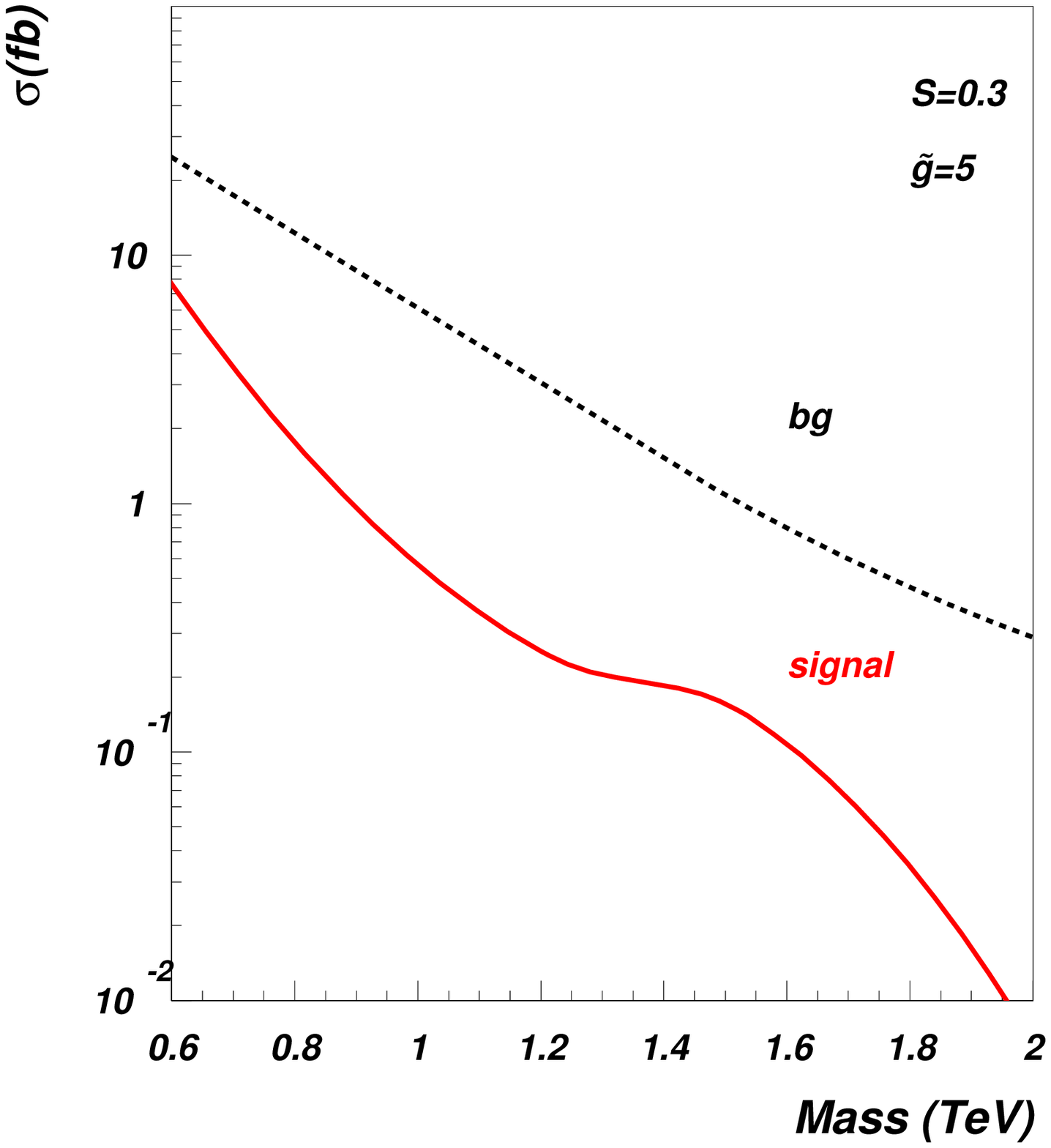}
\caption{LHC reach for the $R^\pm_{1,2}$ resonance signal in  $pp\to R^\pm_{1,2} \to \ell\met$ signature with $S=0.3$
 and
$\tilde{g}=2,5$. 
Top row: the required integrated luminosisty for
achieveing a 5$\sigma$ and a 3$\sigma$ significance. Bottom row:  the
signal and background cross sections for this process.
\label{fig:signif2}}
\end{figure}

We end this section by quantifying the LHC reach for signatures (1)-(3) in terms of the luminosity required to observe the $R_{1,2}$ mass peaks at a significance of 3 and 5 sigma. 
To do so we define the signal as the difference between the
NMWT cross section and the SM cross section in a certain mass window around the peak.
We optimize the invariant or
transverse  mass window  cuts, on a case by case basis, for each signature and parameter space point. 
For example,
signatures (2) and (3) require assymetric mass window
cuts since the transverse one- and tri-lepton mass distribution have
low-end tails. We single out the most significant peak when applying the mass window cut. The significance of the signal is then defined as the number
of signal events divided by the square root of the number of
background events when the number of events is large, while a
Poisson distribution is used when the number of events is small.

\begin{figure}[tbhp]
\vskip -0.2cm
\includegraphics[width=0.43\textwidth]{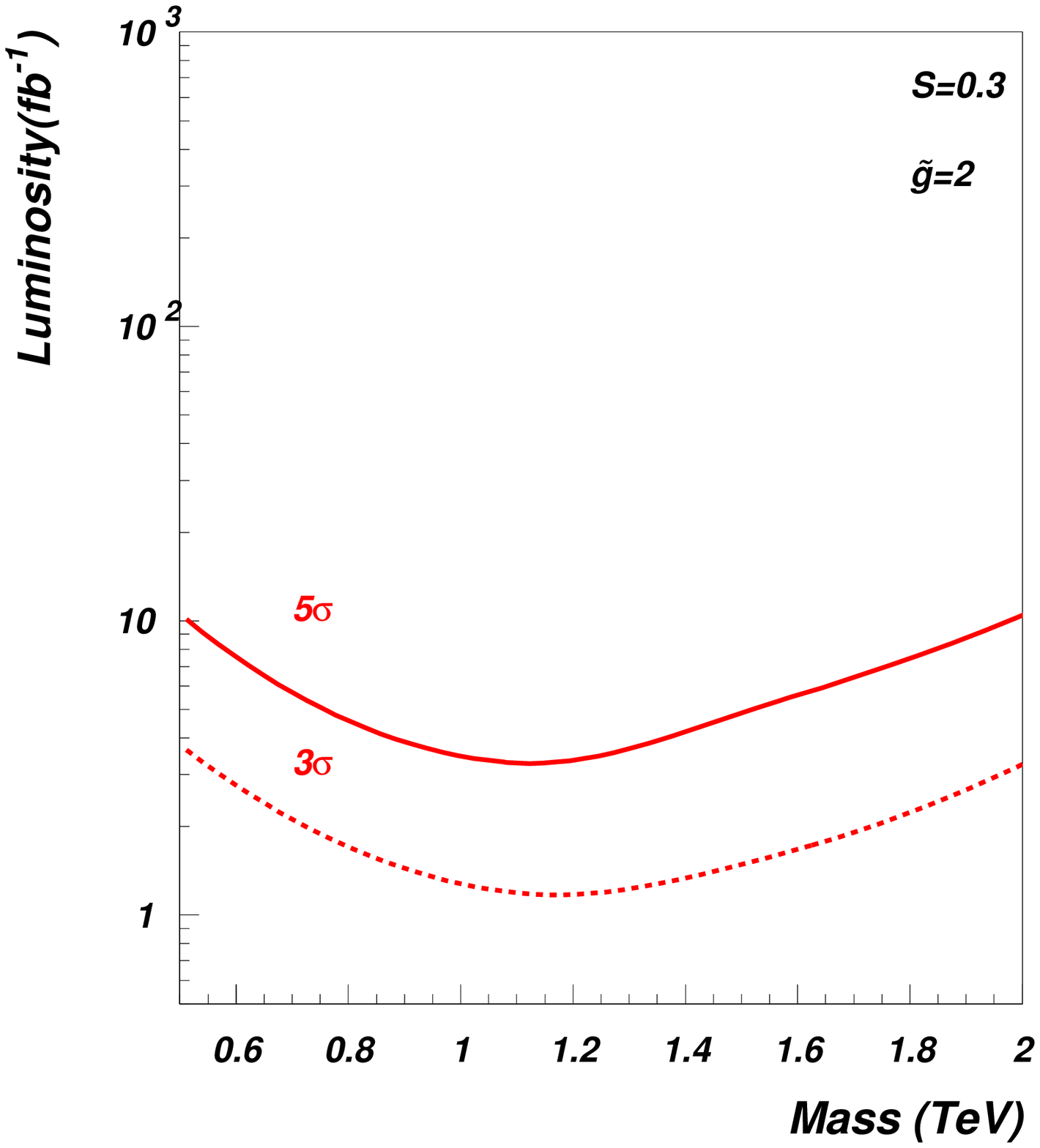}%
\includegraphics[width=0.43\textwidth]{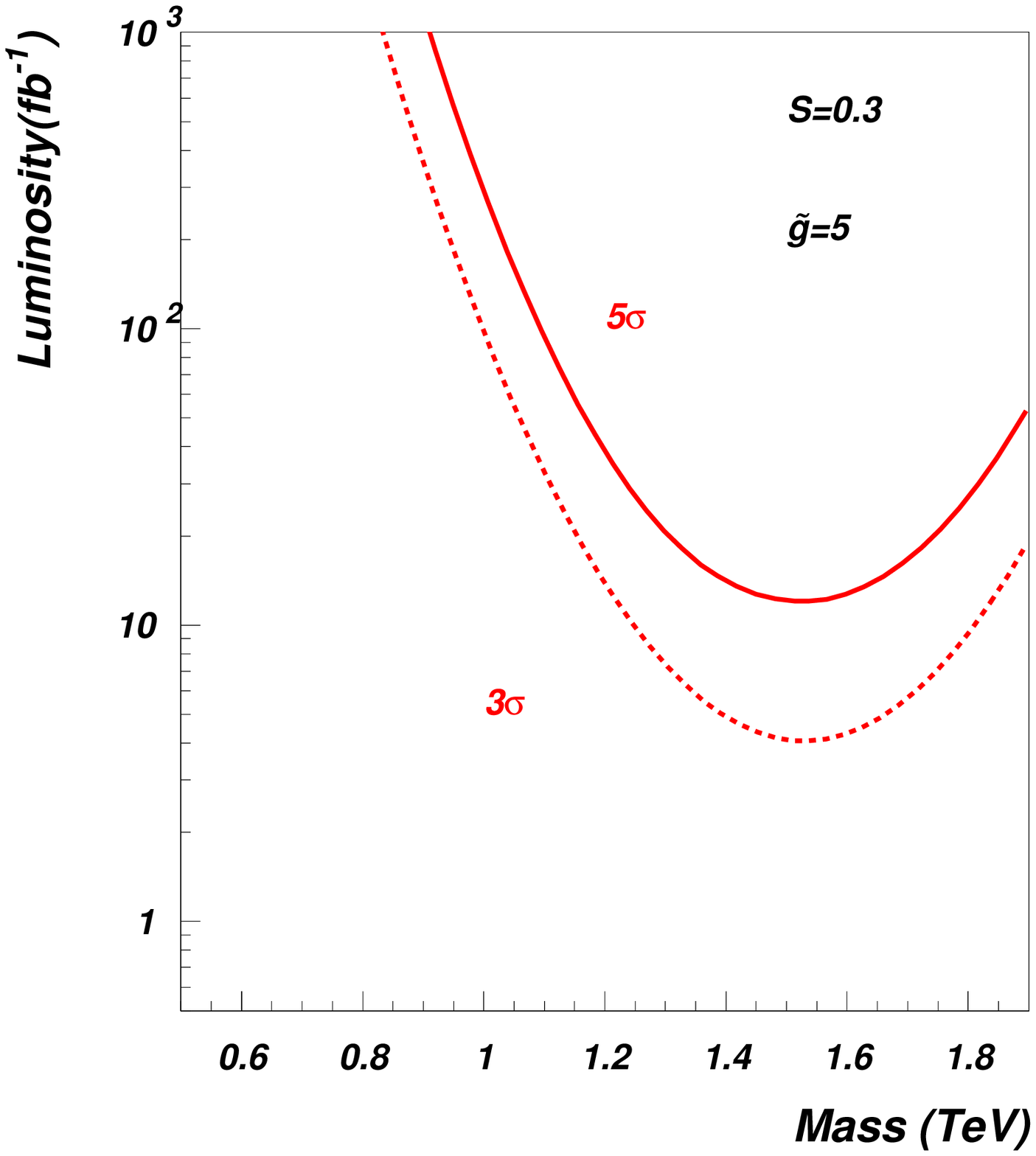}
\vskip -0.2cm
\includegraphics[width=0.43\textwidth]{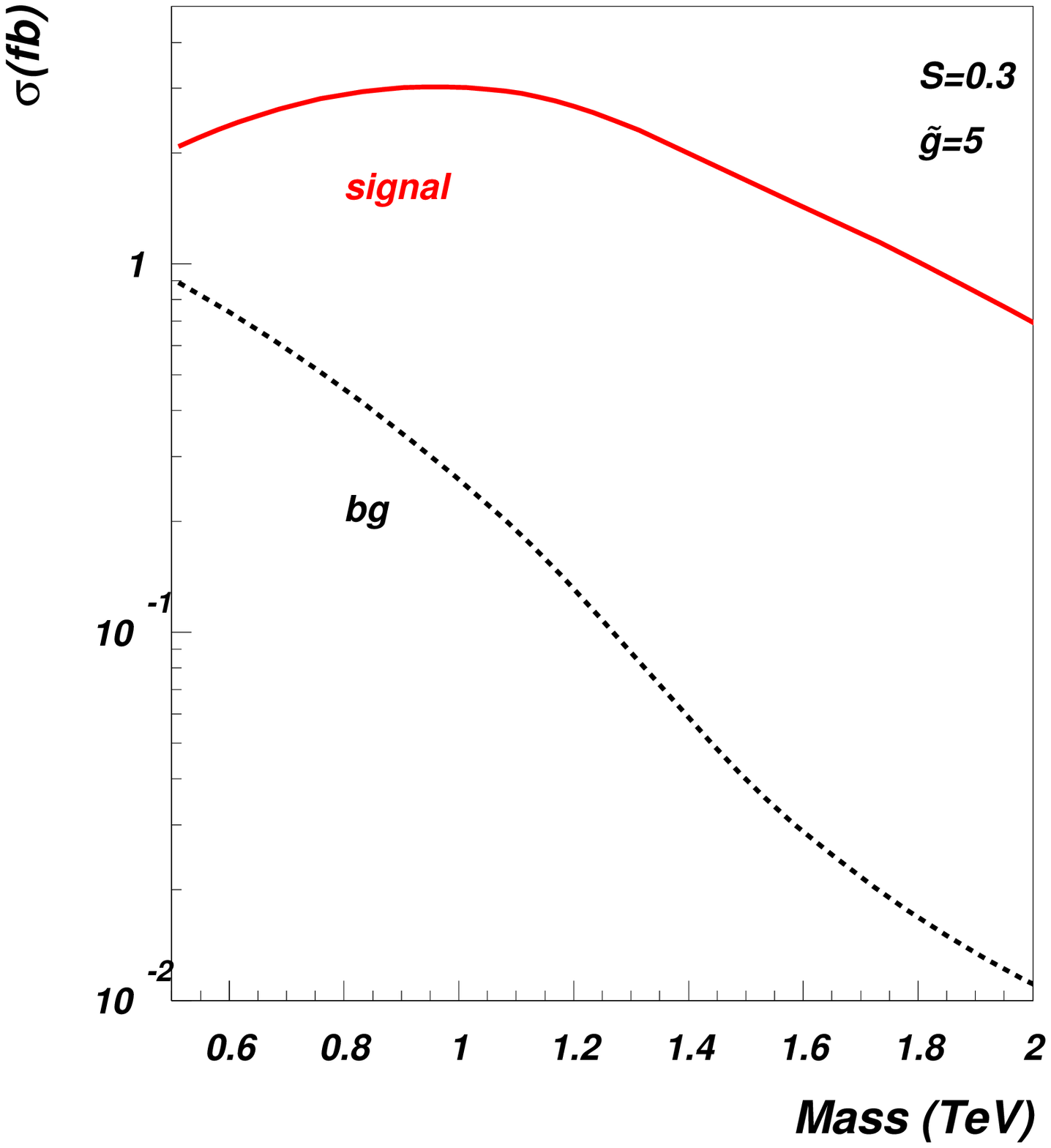}%
\includegraphics[width=0.43\textwidth]{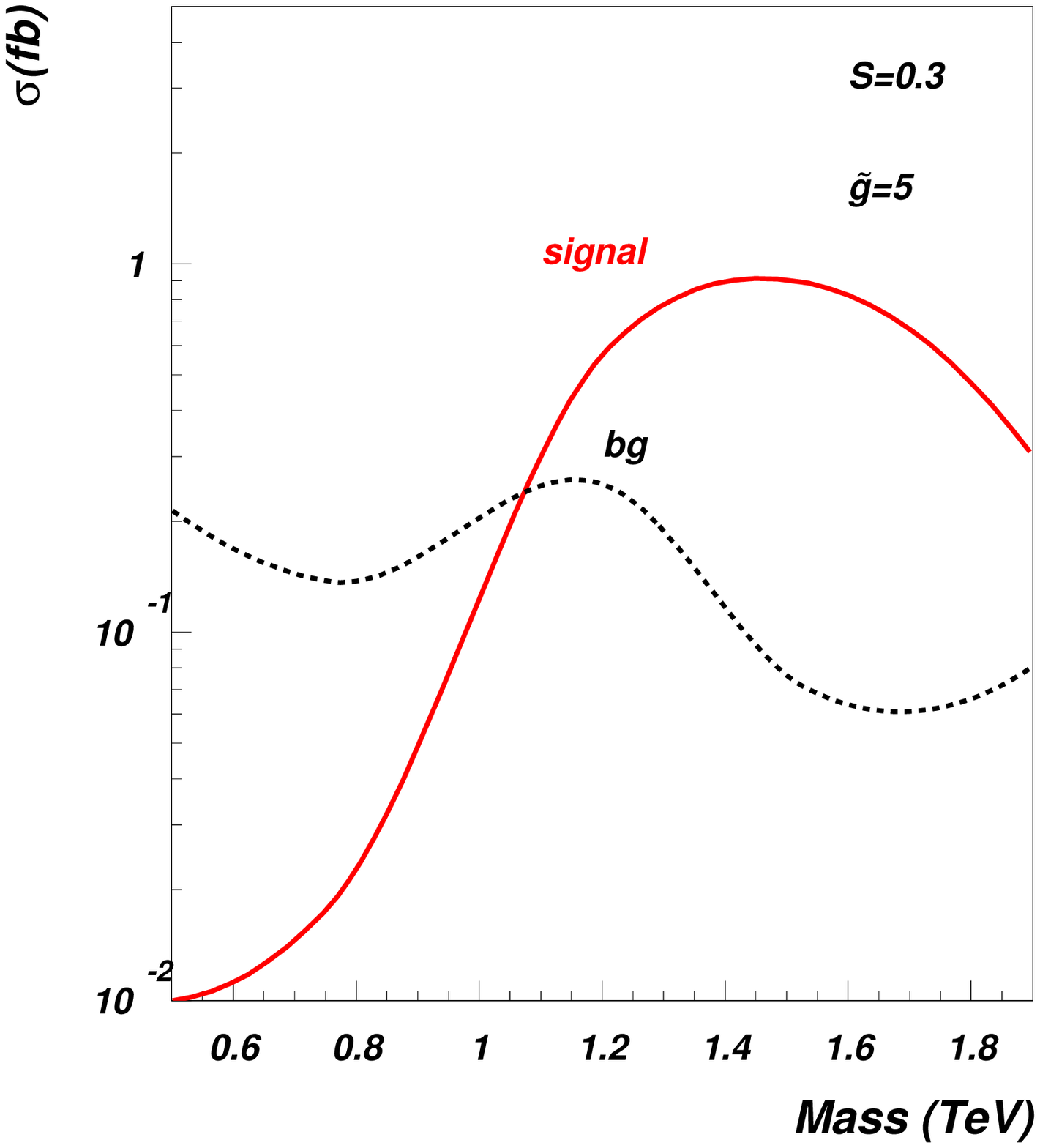}
\caption{LHC reach for the $R^\pm_{1,2}$ resonance signal in $pp\to R^\pm_{1,2}\to W^\pm Z \to 3\ell\met$ 
signature
with $S=0.3$ and
$\tilde{g}=2,5$. 
Top row: the required integrated luminosisty for
achieveing a 5$\sigma$ and a 3$\sigma$ significance. Bottom row:  the
signal and background cross sections for this process.
\label{fig:signif3}}
\end{figure}
The luminosity required
for 5$\sigma$ and 3$\sigma$ significance for signature (1) is
shown in the first row of plots in Fig.~\ref{fig:signif1} as a function
of the mass of the resonance
while the signal and background cross sections are shown in
the second row of plots. For $\tilde{g}=2$ one can see that
even for 5~fb$^{-1}$  of integrated luminosity LHC will 
observe  vector mesons up to 2 TeV mass through signature (1).
On the other hand, for $\tilde{g}=5$
even with 100~fb$^{-1}$  integrated luminosity
one would not be able to observe vector mesons
heavier than 1.4 TeV in this channel. The reach of the LHC for signature (2)
is quite similar to the one for signature (1) for $\tilde{g}=2$ but less promising for $\tilde{g}=5$ as one can see in Fig.~\ref{fig:signif2}.

The LHC reach for signature (3) is presented in
Fig.~\ref{fig:signif3}.
For $\tilde{g}=2$ the
LHC will cover the whole mass range under study
with 10~fb$^{-1}$ of integrated luminosity
through signature (3).
For  $\tilde{g}=5$
it will be able to cover the large mass region
inaccessible to signatures (1) and (2) through signature (3)
with an integrated luminosity of 10-50~fb$^{-1}$
while the low mass region could be covered 
by signatures (1) and (2)
with an integrated luminosity of 10-100~fb$^{-1}$. Thus signature (3) is, in a very important way, complementary to signatures (1) and (2).

\subsection{Vector Boson Fusion production: $p,p \to R_{1,2},j,j$}

VBF is potentially an important channel for vector meson production,
especially in theories in which the vector resonances are quasi
fermiophobic.  

We consider VBF production of the charged $R_{1}$ and  $R_2$  vectors. 
We impose the following kinematical cuts on 
{ the jet transverse
momentum $p^j_T$, energy $E^j$, and rapidity gap $\Delta \eta^{jj}$, 
as well as rapidity acceptance}
$|\eta^j|$~\cite{He:2007ge,Birkedal:2005yg}:
\begin{eqnarray}
|\eta^j|<4.5\ , \quad  p^j_T> 30 \mbox{ GeV} \ , \quad   E^j> 300 \mbox{ GeV} \ , \quad  \Delta \eta^{jj} > 4 \ .
\label{eq:vbfcuts}
\end{eqnarray}
\begin{figure}[tbhp]
\vskip -0.2cm
\includegraphics[width=0.43\textwidth]{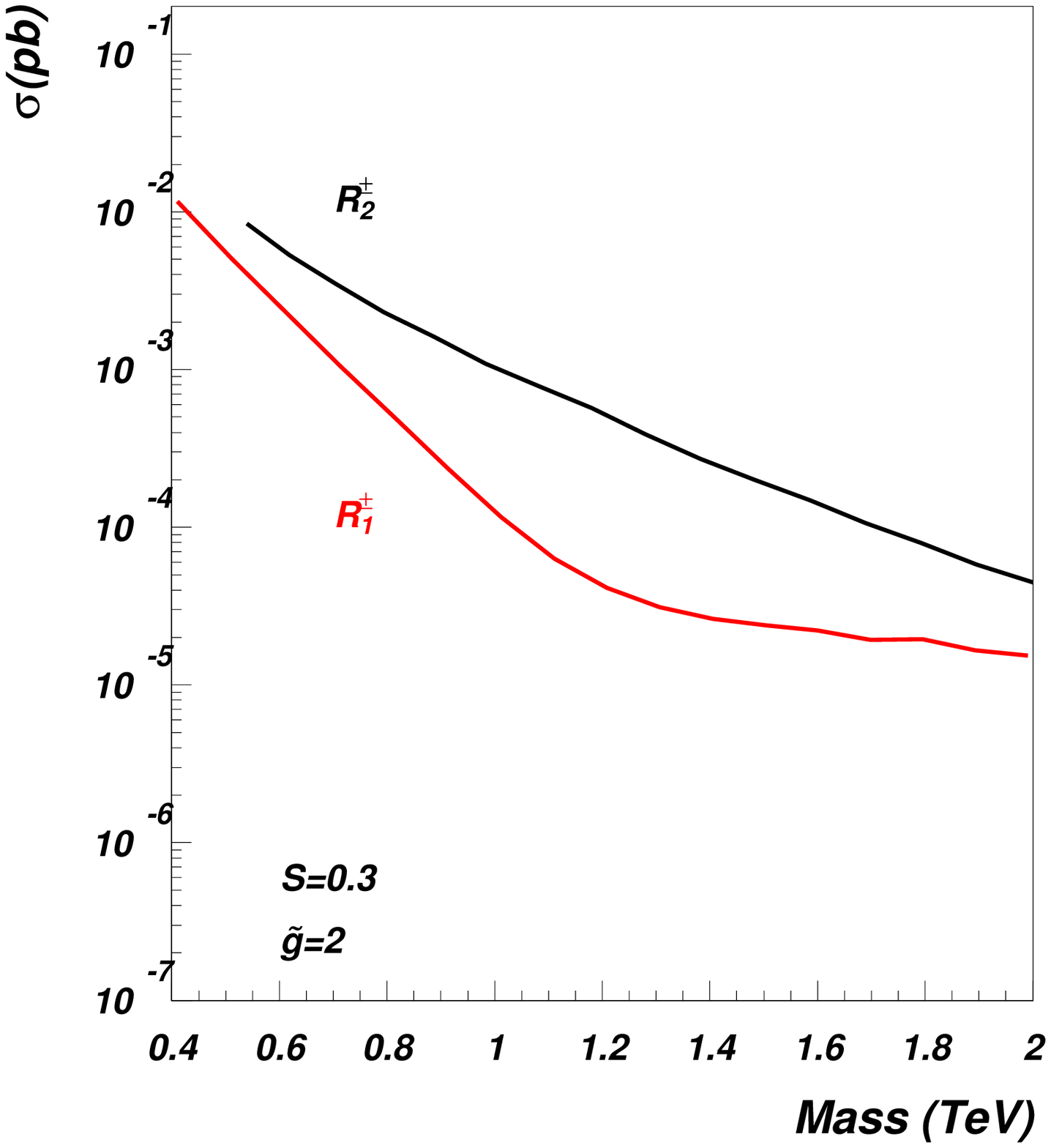}%
\includegraphics[width=0.43\textwidth]{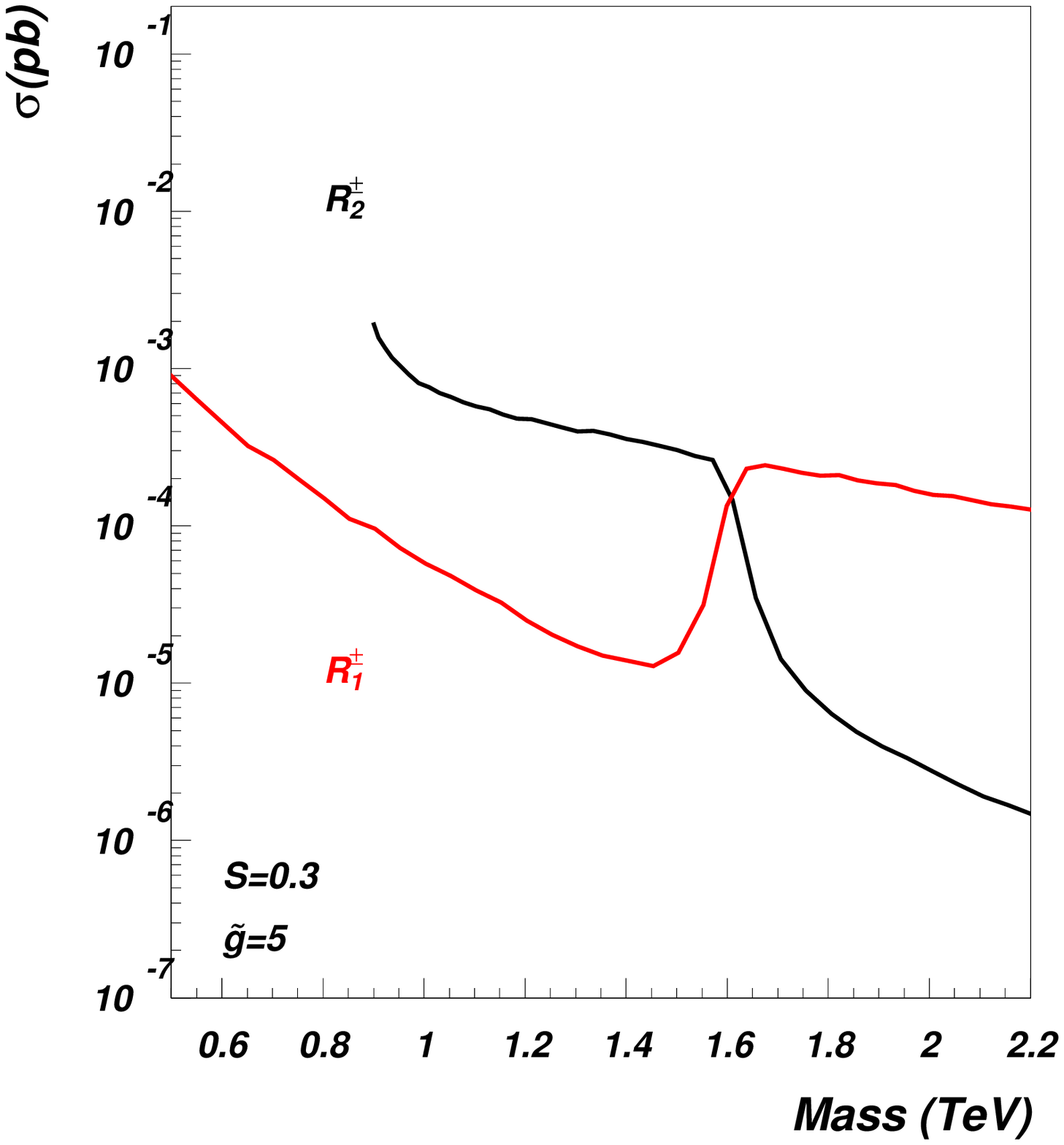}
\vskip -0.2cm
\caption{Vector boson fusion production cross sections for the $R_{1,2}^\pm$ resonances, with $S=0.3$ and $\tilde{g}=2,5$. The jet cuts are: $|\eta^j|<4.5$, $p^j_T>30$ GeV, $E^j>300$ GeV, $\Delta \eta^{jj}>4$. See text for details.}
\label{fig:VBF}
\end{figure}
The VBF production cross section for the charged $R_1$ and $R_2$ vector
resonances is shown in Fig.~\ref{fig:VBF} for the set of cuts given
by Eq.~\ref{eq:vbfcuts}. An interesting feature of the VBF
production is the observed crossover around the mass degeneracy point for $\widetilde{g}=5$. This is a direct consequence of the fact that the $R_{1,2}$ resonances switch their vector/axial nature at the inversion point.  {}For smal $\tilde{g}$ the crossover does not occur due to the interplay between the electroweak and the Technicolor corrections. 
In D-BESS VBF processes are not very relevant, since there are no direct
interactions between the heavy mesons and the SM vectors. However in fermiophobic
Higgsless models VBF is the main production channel of the heavy resonances.  {Since the production rate of $R^\pm_{1,2}$
is below 1 fb VBF is not a
promising channel at the LHC.
}

\subsection{Composite Higgs Phenomenology}

The composite Higgs phenomenology is interesting due to its interactions with the new massive vector bosons and their mixing with SM gauge bosons. We first analyze the Higgs coupling to the $W$- and $Z$-
gauge bosons. In Fig.~\ref{fig:ghww} (left) we present the $g_{HWW}/g_{HWW}^{SM}$
ratio as a function of $M_A$. The behaviour of the $g_{HZZ}$ and $g_{HWW}$ 
couplings are identical.  We keep fixed $S=0.3$ 
and consider two values of $\tilde g$,  2 (solid line) and 5 (dashed line). 
We repeat the plots for three choices of the $s$ parameter
$(+1,0,-1)$ depicted in black, blue and green colors respectively.
The deviation of  $g_{HWW}$ from $g_{HWW}^{SM}$  increases with $M_A$  due to the
fact that we hold the $S$ parameter fixed. 
\begin{figure}
\includegraphics[width=0.5\textwidth]{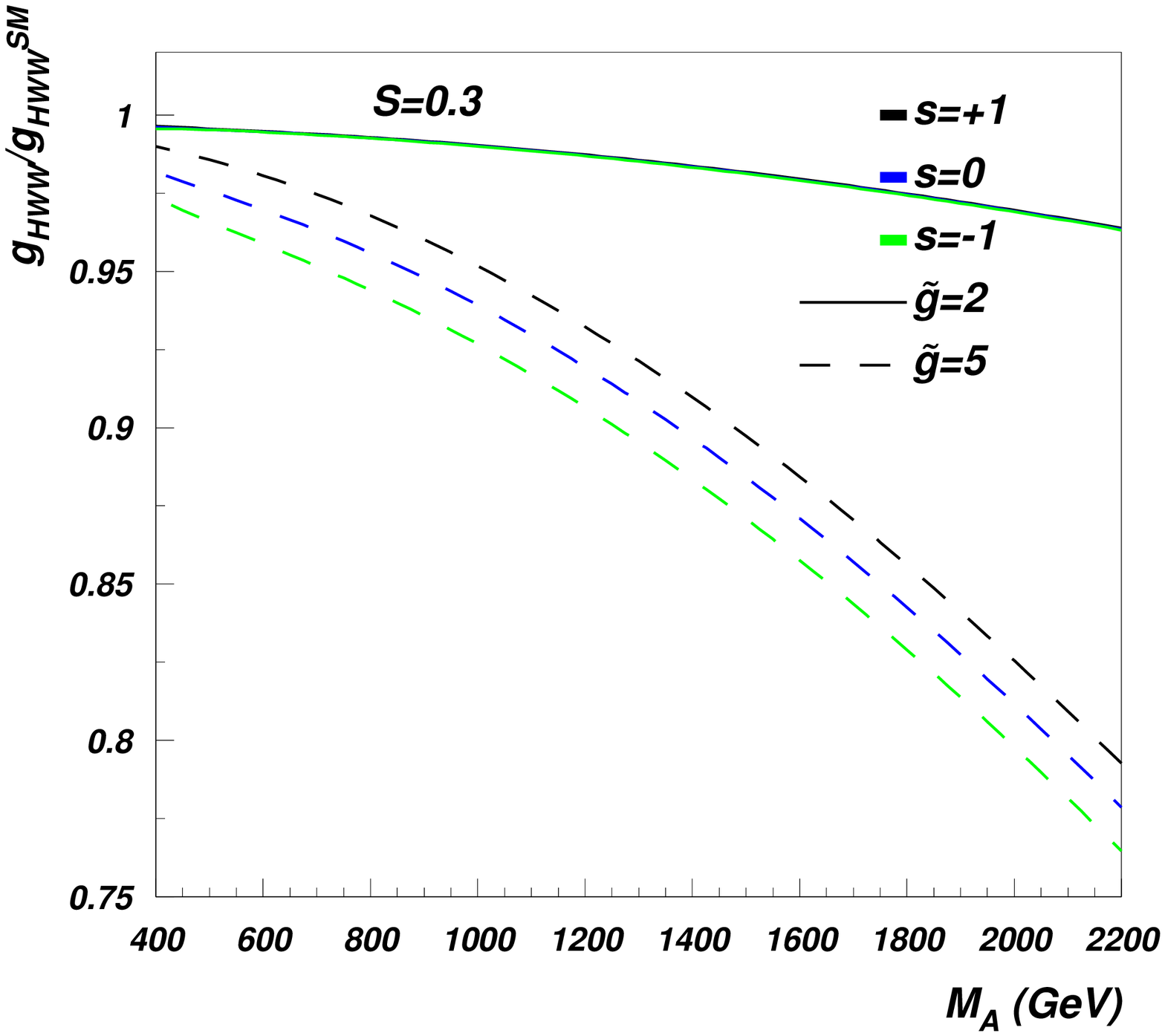}%
\includegraphics[width=0.5\textwidth]{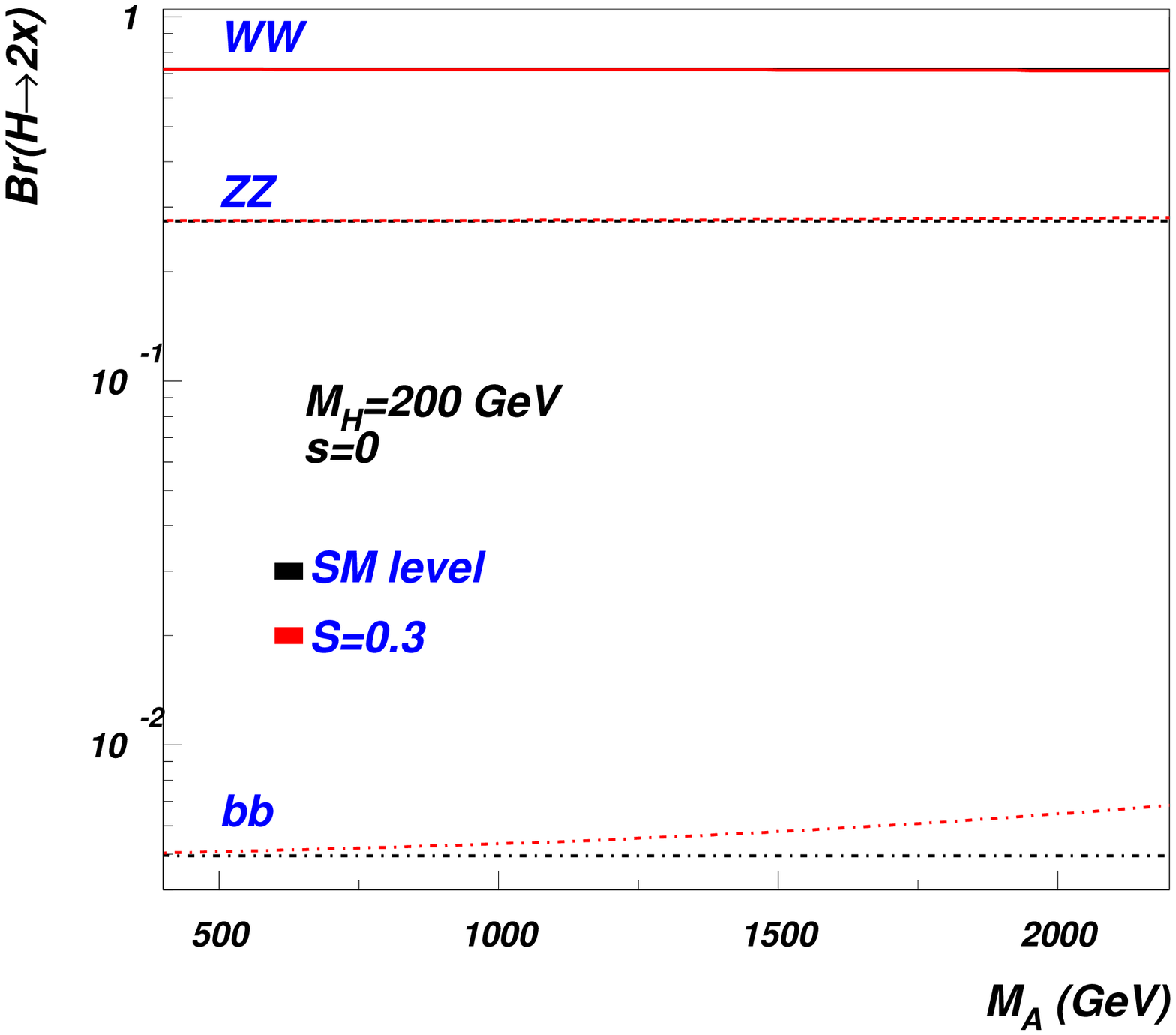}%
\caption{\label{fig:ghww} 
Left: $g_{HWW}/g_{HWW}^{SM}$
ratio as a function of $M_A$. The behavior of the $g_{HZZ}$ coupling
is identical to the $g_{HWW}$ one.
Results are presented for $S=0.3$,
$\tilde g$=2 and 5 (solid and dashed lines respectively),
 and for $s=(+1,0,-1)$ 
 (black, blue and green colors respectively).
 \\
 Right: branching ratios of the composite Higgs (red) and SM Higgs (black) as function  $M_A$ with $s=0$.
$M_H$= 200 GeV.
}
\end{figure}
One reaches deviations from the SM couplings of 20$\%$ when $M_A\simeq 2$~TeV.
This is reflected in the small deviations of the Higgs branching ratios
when compared with the SM ones 
as shown in  Fig.~\ref{fig:ghww} (right). Here we used as reference point $s=0$.

The presence of the heavy vectors is prominent in the associate production of the composite Higgs with SM vector bosons, as first pointed out in \cite{Zerwekh:2005wh}. 
Parton level Feynman diagrams for  the $pp\to WH$ and $pp\to ZH$ processes are shown in Fig.~\ref{fig:whdiag} (left)  and
Fig.~\ref{fig:whdiag} (right) respectively.
\begin{figure}
\includegraphics[width=0.7\textwidth]{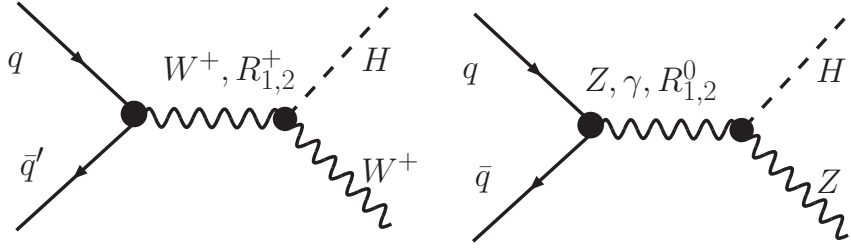}%
\caption{\label{fig:whdiag}Feynman
 diagrams for the composite Higgs production 
 in association with  SM gauge bosons. }
\end{figure}
\begin{figure}
\includegraphics[width=0.5\textwidth]{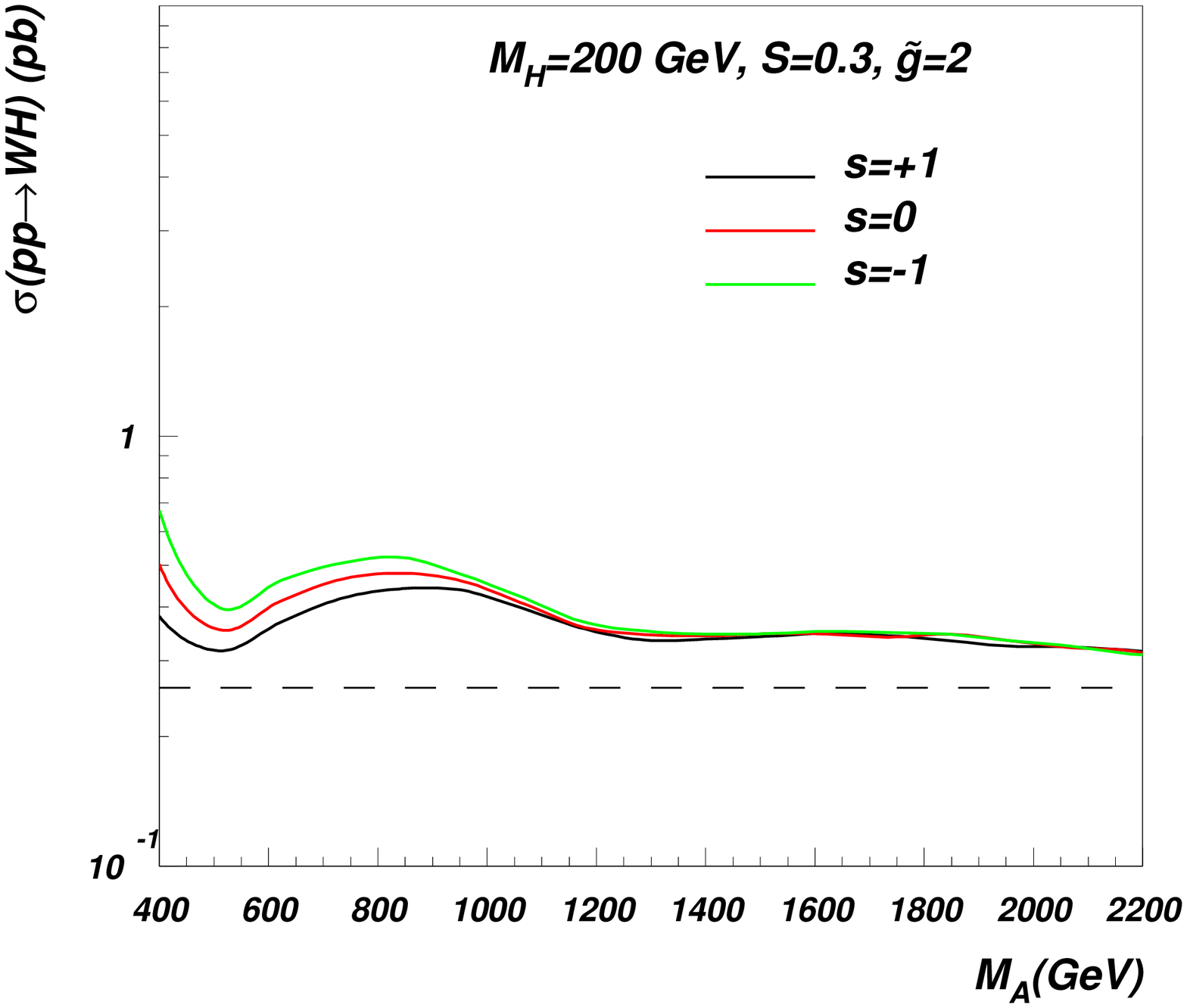}%
\includegraphics[width=0.5\textwidth]{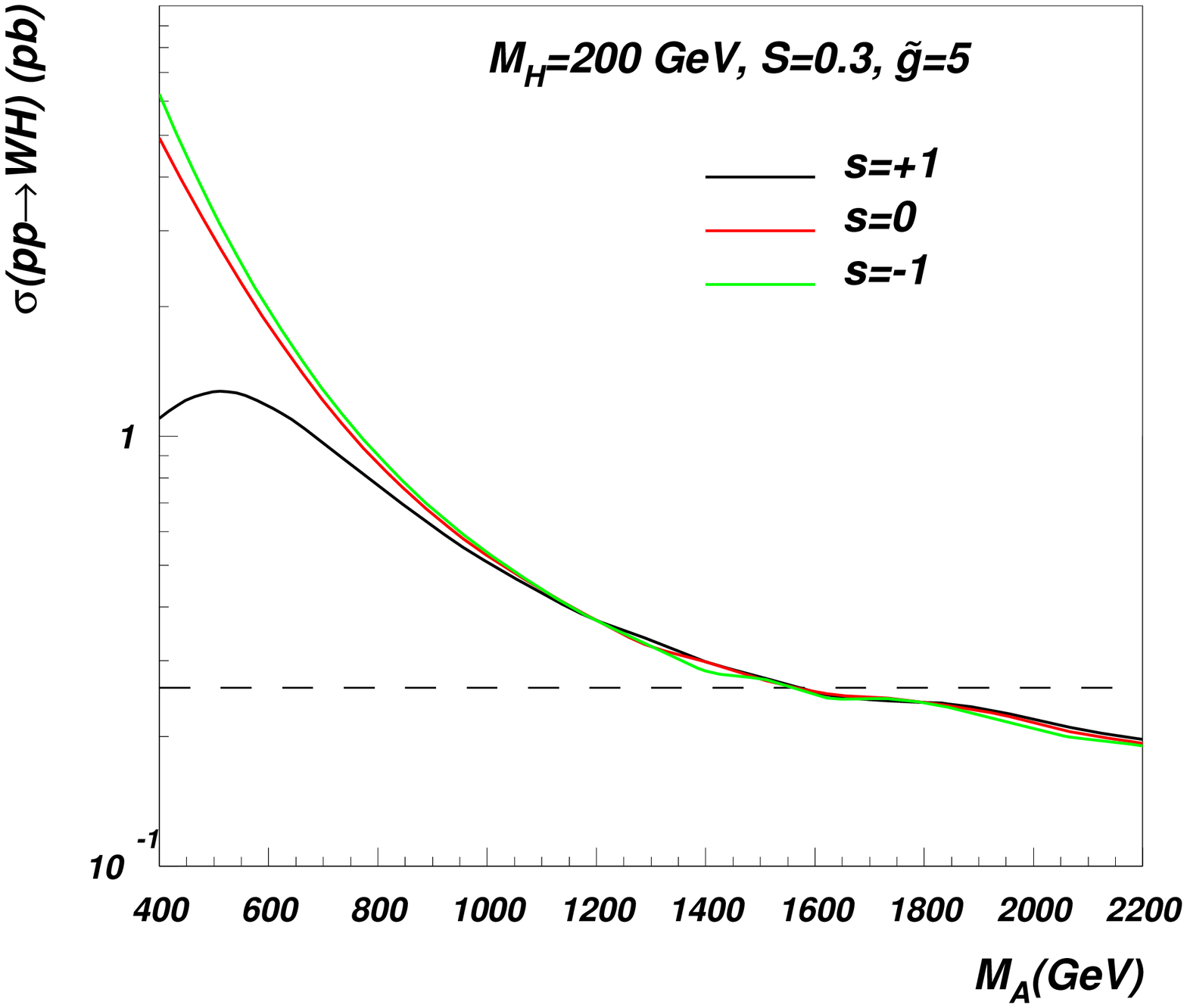}%
%\vskip -0.2cm
%\includegraphics[width=0.5\textwidth]{figures/zh.eps}%
%\includegraphics[width=0.5\textwidth]{figures/vn.eps}%
\caption{\label{fig:vh}
The cross section for $pp\to WH$ production
($W^+H$ and $W^-H$ modes are summed up)
versus $M_A$ for  $S=0.3$, $s=(+1,0,1)$
and $\tilde g =2$ (left) and $\tilde g=5$ (right).
The dashed line at the bottom indicates the SM cross section level.}
\end{figure}
The resonant production of heavy vectors can enhance $HW$ and
$ZH$ production by a factor 10 as one can see in Fig.~\ref{fig:vh} (right). This enhancement occurs for low values of the vector meson mass
and large values of $\tilde g$. This behavior is shown  in Fig.~\ref{fig:vh} (right) for $\tilde{g}=5$. These are values of the parameters not excluded  by Tevatron data (see Fig.~\ref{fig:bounds}).

%The $pp\to ZH$ rate is half
%the $pp\to WH$ rate, but both have an identical behavior as a function of $M_A$.
%Fig.~\ref{fig:vh}(right) also shows 
%that for large $M_A$ values the cross section of $pp\to ZH$ and $pp\to WH$ 
%can be slightly suppressed relative to the SM.
%This happens because
%for large $M_A$ the associate Higgs production
%takes place mainly via exchanges of $Z$ and $W$ bosons,
%whose couplings with the Higgs are suppressed, as Fig.~\ref{fig:ghww}(left) shows.
%In case of small $\tilde g$, the deviations of the 
%$pp\to WH$ production cross section relative to the SM is very
%moderate, even in case of low $M_A$ 
%as Fig.~\ref{fig:vh}(left) shows.
%However the region of low $M_A$ and low $\tilde g$
%is already excluded by Tevatron data and precision EW data.

The contribution from heavy vector to $pp\to VH$ ($V=W^{\pm},Z$ can be clearly identified from the peaks
in the invariant mass distributions of $WZZ$ or $ZZZ$ 
presented in Fig.~\ref{fig:mvvv}.
\begin{figure}
\includegraphics[width=0.7\textwidth,height=0.5\textwidth]{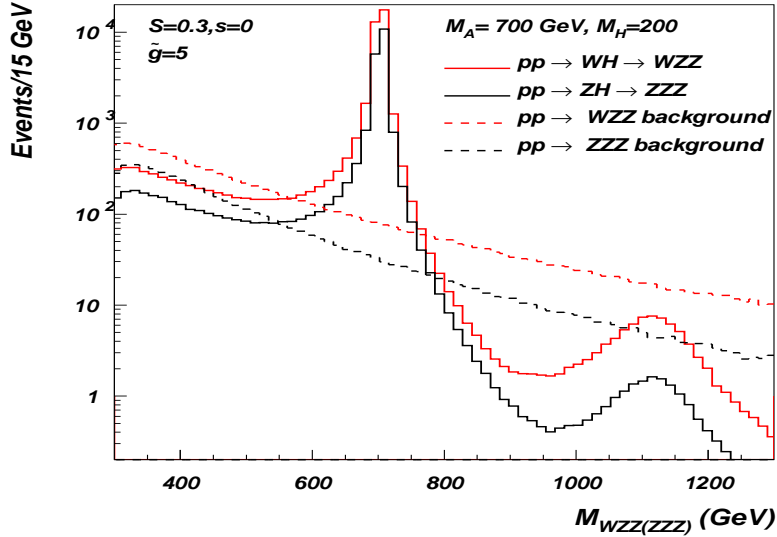}
\caption{\label{fig:mvvv}
Invariant mass distributions of $WZZ$ or $ZZZ$
for signal (solid lines) and background (dashed lines)
for 100 fb$^{-1}$.
}
\end{figure}
 One should consider these distributions as qualitative ones, since at the
experimental level $WZZ$ or $ZZZ$ invariant masses will be reconstructed from 
leptons and jets in the final state with appropriate acceptance cuts applied.
However, one can eventually expect that visibility of the signal will remain.  Taking into account 
leptonic branching ratios of the two $Z$-bosons 
{and the hadronic branching ratios for the third gauge boson (W or Z)}
we
estimate about 40 clean events under the peak with negligible background.
The second broader vector peak will not be observed.
\begin{figure}
\includegraphics[width=0.7\textwidth,height=0.5\textwidth]{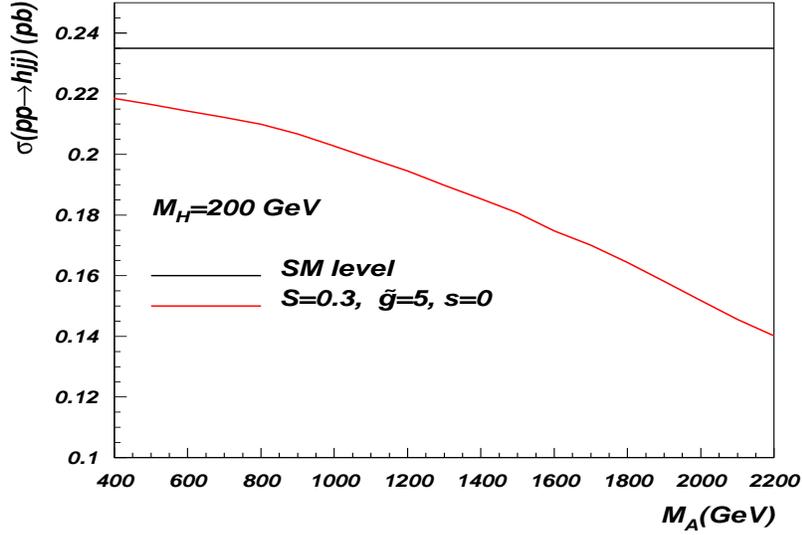}%
\caption{\label{fig:hjj}
Composite Higgs production cross section via the VBF mechanism.}
\end{figure}
We have also analyzed the composite Higgs  production
in vector boson fusion processes $pp\to Hjj$. We find that it is not enhanced with respect to the corresponding process in the SM as it is clear from Fig.~\ref{fig:hjj}. The behavior of the $\sigma(pp\to Hjj)$ as function of $M_A$  traces the one of the Higgs-gauge bosons coupling shown in Fig.~\ref{fig:ghww}.

\subsection{Extending the Parameter Space}\label{sec:narrow}
In Sec.~\ref{sec:decay} we saw that the vector resonances are very narrow. The only exception occurs for the $R_2$ meson, when the $R_2\to R_1,X$ and $R_2\to 2R_1$ channels are important ($X$ denotes a SM gauge boson). The origin of this can be attributed to the fact that in writing Eq.~(\ref{eq:boson}) we used only renormalizable operators. Consider for example the operator \cite{Kaymakcalan:1984bz,Sannino:2008ha}:
\begin{eqnarray}
-\frac{2\gamma}{v^2} {\rm Tr}\left[F_{{\rm L}\mu\nu} M F_{\rm L}^{\mu\nu} M^\dagger\right] \ .
\label{eq:gamma}
\end{eqnarray}
This operator was introduced many years ago by Kaymakcalan and Schechter in \cite{Kaymakcalan:1984bz} and appeared for the first time in DEWSB effective Lagrangians in \cite{Sannino:2008ha}. The effects of a similar Lagrangian term has been also recently discussed, in the context of four-site Higgsless model, by Chivukula and Simmons \cite{Chivukula:2008gz}. This term
affects several couplings. For example the $g_{V\pi\pi}$ coupling reads
\begin{eqnarray}
g_{V\pi\pi} = \frac{F_V M_V}{2 F_\pi^2}\left[1-\frac{1+\gamma}{1-\gamma}\frac{F_A^2}{M_A^2}\frac{M_V^2}{F_V^2}\right] \ .
\label{Vpp}
\end{eqnarray}
Taking either the $M_A\to\infty$ or the $\gamma\to -1$ limit returns the known formula for $g_{\rho\pi\pi}$ in QCD. To better appreciate the physical content of this term we combine Eq.~(\ref{eq:gamma}) and  Eq.~(\ref{eq:boson}) yielding the following kinetic terms for the vector and axial states:
\begin{eqnarray}
-\frac{1+\gamma}{4}(\partial_\mu V_\nu - \partial_\nu V_\mu)^2-\frac{1-\gamma}{4}(\partial_\mu A_\nu - \partial_\nu A_\mu)^2
\end{eqnarray}
From this it follows that requiring the vector mesons to be non-tachyonic propagating fields implies $-1<\gamma<1$. Moreover it is unreasonable to take $\gamma$ too close to either $-1$ or $1$, because this would naturally lead to infinitely large masses for the vector mesons. 

Taking a value of $\gamma$ not close to $\pm 1$ but different from zero has anyway a large impact on the meson widths. In Fig.~\ref{fig:widthgamma}
the $R_1$ and $R_2$ widths are shown for both $\gamma=0$ (solid lines) and $\gamma=-0.5$ (dashed lines). The widths can increase by two orders of magnitude.
\begin{figure}[tbhp]
 \vskip -0.4cm
 \includegraphics[width=0.45\textwidth,height=0.35\textwidth]{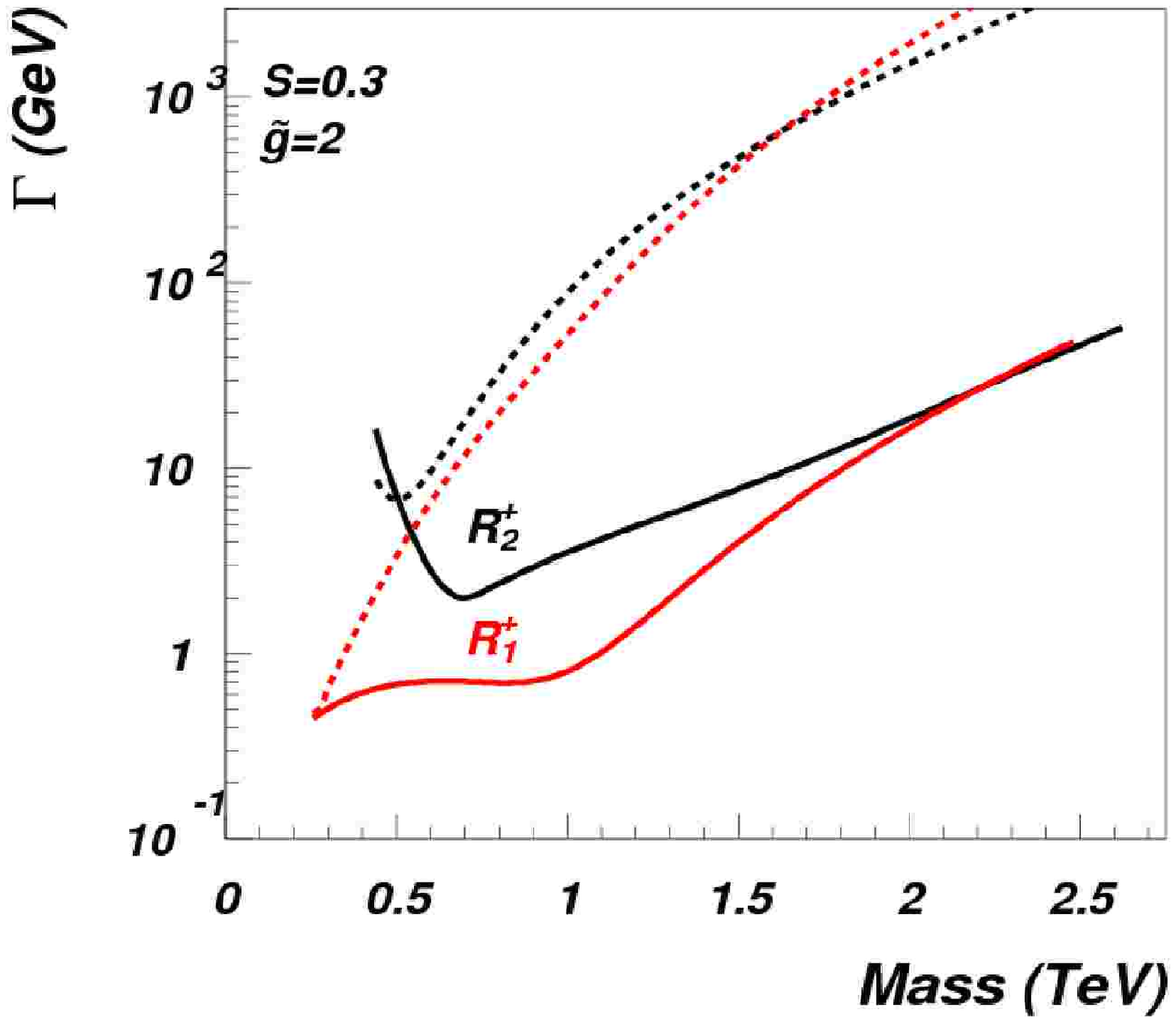}%
 \includegraphics[width=0.45\textwidth,height=0.35\textwidth]{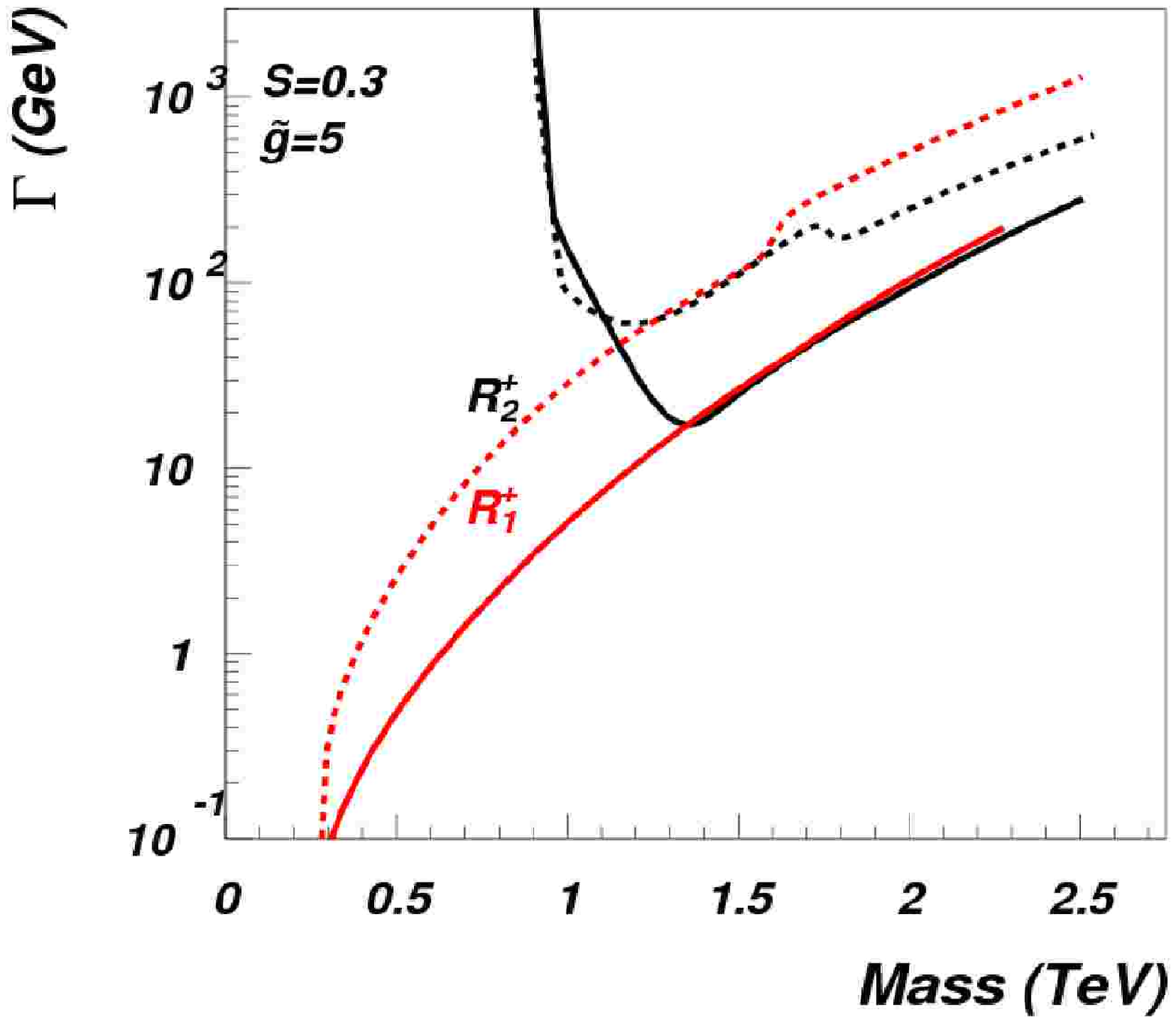}
 \vskip -0.4cm
 \includegraphics[width=0.45\textwidth,height=0.35\textwidth]{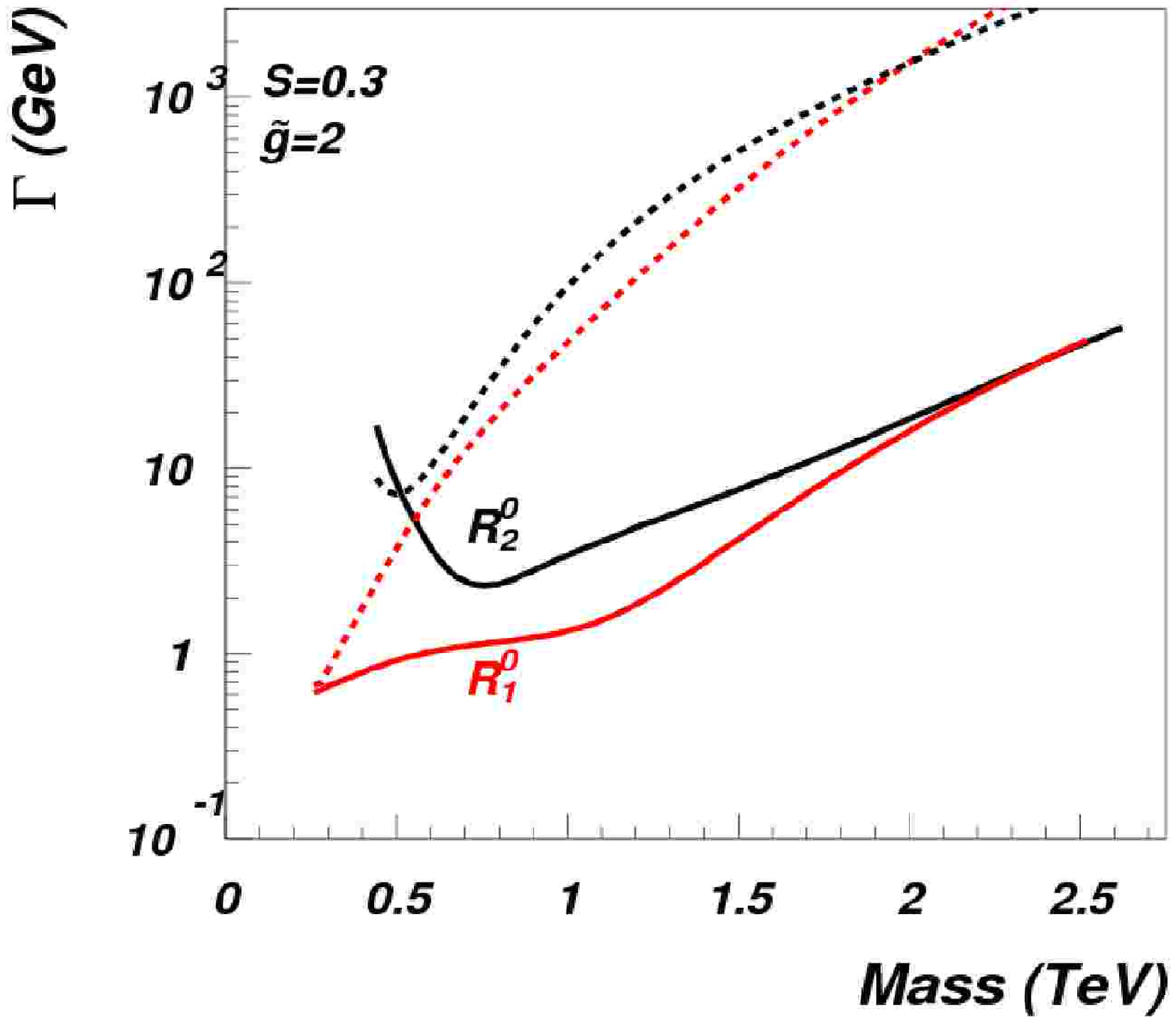}%
 \includegraphics[width=0.45\textwidth,height=0.35\textwidth]{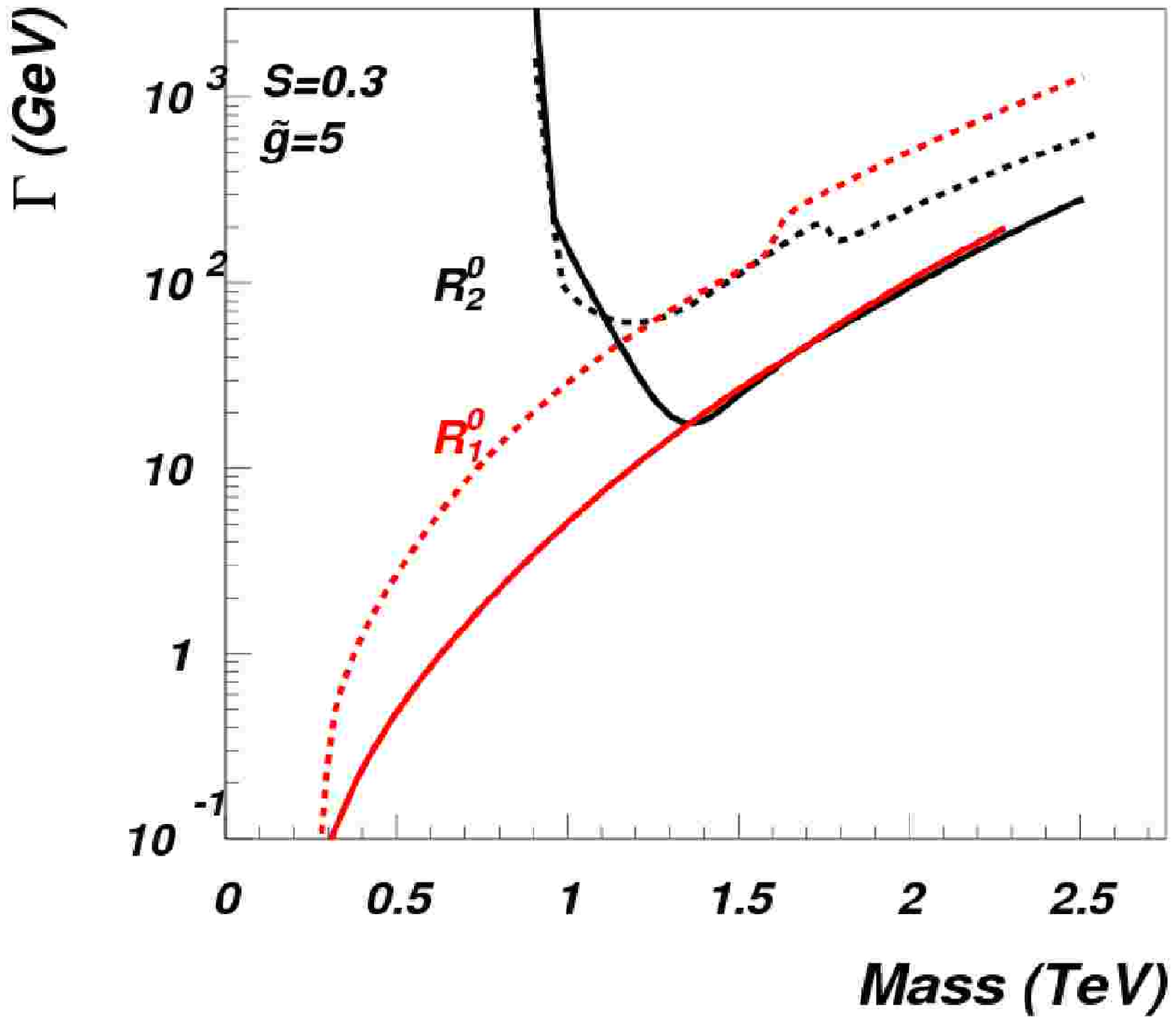}
 \vskip -0.4cm
\caption{Decay width of the charged (first row) and neutral (second row) 
vector mesons for $S=0.3$ and $\tilde{g}=2,5$. The solid lines correspond to $\gamma=0$, the dashed lines correspond to $\gamma=-0.5$. We take $M_H = 0.2 \ \textrm{TeV},\ s=0$.}
\label{fig:widthgamma}
\end{figure}
The DY production of the heavy vectors is unaffected by $\gamma$, at the tree level, since the fermion couplings to the vector mesons do not depend on it. We have also checked that the contributions from this term do not substantially affect the other results. 

\section{Conclusions}
We have analyzed the potential of the Large Hadron Collider (LHC) to observe 
signatures of phenomenologically viable Walking Technicolor  models. 
We studied and compared the Drell-Yan (DY) and
Vector Boson Fusion (VBF)  mechanisms for the production of composite
heavy vectors. The DY production mechanism constitutes the most promising way to detect and study the technicolor spin one states. 

We have compared, when possible, with earlier analysis and
shown that our description reproduces all of the earlier results while
extending them by incorporating basic
properties of walking dynamics such as the  mass relation between the vector
and axial spin one resonances.

LHC can be sensitive to spin one states as heavy as 2 TeV. One TeV spin one states can  be observed already with 100~pb$^{-1}$ integrated luminosity in the di-lepton channel. 
The VBF production of heavy mesons is, however, suppressed 
and will not be observed. The enhancement of the composite Higgs production is another promising signature.

We identified distinct DY signatures which allow to cover at the LHC, in a complementary way, a great deal of the model's parameter space. 

\acknowledgments
We gladly thank Neil Christensen for providing us an improved CalcHEP batch interface and Elena Vataga for discussions related to the experimental signatures. We thank Dennis D. Dietrich for discussions. The work of R.F., M.T.F., M.J. , and F.S. is supported by the Marie Curie Excellence Grant under contract MEXT-CT-2004-013510.

\end{document}